\documentclass[a4paper,12pt]{article}

\pdfoutput=1

\usepackage{amsfonts}
\usepackage{amsmath,amssymb,amsthm}
\usepackage{hyperref}
\usepackage{graphicx}
\usepackage{color}
\usepackage[latin1]{inputenc}

\newcommand{\beq}{\begin{equation}}
\newcommand{\eeq}{\end{equation}}
\newcommand{\bea}{\begin{eqnarray}}
\newcommand{\eea}{\end{eqnarray}}
\def\no{\nonumber}
\newcommand{\half}{\frac{1}{2}}

\newcommand{\p}{\partial}

\def\N{{\cal N}}

\def\qth{\tau}

\newcommand{\rf}[1]{(\ref{#1})}

\numberwithin{equation}{section}

\title{{\Large IIB  Duals of $D=3$ $\N=4$ Circular Quivers }}

\author{ \\
{\large Benjamin~Assel,$^{\natural,1}$
  Costas~Bachas,$^{\natural,2}$
    John~Estes,$^{*,\flat,3}$ and
    Jaume~Gomis$^{\dagger,4}$}}

\date{ }

\begin{document}

\maketitle

\vskip -6.5cm
\rightline{LPTENS-12/34}
\rightline{Imperial/TP/2012/JE/02}
\vskip 5.0cm

\centerline{$^\natural$ Laboratoire de Physique Th\'eorique de l'\'Ecole Normale Sup\'erieure, }
\centerline{24 rue Lhomond, 75231 Paris cedex, France}

\vskip 4mm

\centerline{$^*$ Blackett Laboratory, Imperial College,}
\centerline{London, SW7 2AZ, United Kingdom }

\vskip 4mm

\centerline{$^\flat$ Instituut voor Theoretische Fysica, KULeuven, }
\centerline{Celestijnenlaan 200D B-3001 Leuven, Belgium}

\vskip 4mm

\centerline{$^\dagger$ Perimeter Institute for Theoretical Physics, }
\centerline{Waterloo, Ontario, N2L2Y5, Canada}

\vspace{0.3cm}

\centerline{\tt  $^1$benjamin.assel@lpt.ens.fr, ${}^2$bachas@lpt.ens.fr,}
\centerline{\tt ${}^3$johnaldonestes@gmail.com, $^4$jgomis@perimeterinstitute.ca}

\vskip7mm

\abstract{ \normalsize{
We   construct the type-IIB $AdS_4\ltimes K$ supergravity solutions which are dual to the  three-dimensional  ${\cal N}=4$
superconformal field theories that arise as infrared fixed points of circular-quiver gauge theories.  
 These superconformal field theories are labeled by a triple $(\rho,\hat\rho,L)$ subject to   constraints, where $\rho$ and $\hat\rho$ are two partitions of  the same number $N$,
 and $L$ is a positive integer. We show that in the limit of large $L$ the localized five-branes  in our solutions are effectively smeared, and these type-IIB  solutions
 are dual to the near-horizon geometry of   M-theory   M2-branes at a $\mathbb{C}^4/(Z_k\times Z_{\hat k})$ orbifold singularity. 
 Our IIB solutions resolve the  singularity into localized five-brane throats,  without
 breaking the conformal symmetry.
The  constraints satisfied by 
 the triple  $(\rho,\hat\rho,L)$,
 together with  the enhanced non-abelian flavour symmetries of the superconformal field theories,  
are precisely reproduced by the supergravity solutions.  
As a bonus, we uncover a novel type of ``orbifold equivalence" between different quantum field theories and  provide quantitative evidence for this equivalence.
}}

\vfill\eject
\tableofcontents
\section{Introduction}

Conformal field theories  play a distinguished role in the space of all quantum field theories.
They reside at fixed points of the renormalization group,  which generates a flow in the space of quantum field theories. Any non-conformal field theory can be reached under renormalization group evolution by perturbing a conformal  theory with  a suitable operator.
The central role that conformal field theories play in our understanding of quantum field theory is
 one of the reasons why they  have been a subject  of enduring interest.

 A large class of conformal field theories can be obtained by perturbing a  Gaussian fixed-point theory in the ultraviolet 
 by a relevant operator,
 and following the renormalization group flow in the infrared.
Infrared fixed points which arise after ``long" renormalization group flows are, however,  inherently strongly coupled, and therefore 
not amenable to study with  conventional field theory  techniques. 
The discovery of the AdS/CFT correspondence \cite{Maldacena:1997re,Gubser:1998bc,Witten:1998qj} has opened 
a new window into the world of strongly-coupled conformal field theories,
  turning the study of some of them  (in the large $N$ limit) to the study of string theory in asymptotically anti-de-Sitter backgrounds.

 In this paper, we provide the type-IIB string theory backgrounds dual to a very large class of strongly coupled, three dimensional ${\cal N}=4$ superconformal field theories. These backgrounds  have a warped $AdS_4 \ltimes K$  geometry with very specific five-brane sources. They are dual to the non-trivial infrared fixed points of  three-dimensional ${\cal N}=4$ supersymmetric gauge theories corresponding to {\it circular quiver} gauge theories depicted in figure \ref{fig1a}.
These quiver gauge theories admit an elegant realization in terms of the low energy limit of intersecting brane configurations  \cite{Hanany:1996ie}. The dual type-IIB backgrounds we construct encode the backreacted near-horizon geometry of these brane configurations.

 The circular quiver gauge theories that give rise to irreducible superconformal field theories in the infrared are labeled by the triple
 \beq
 (\rho,\hat\rho,L)\,,
\nonumber
 \eeq
where $\rho$ and $\hat\rho$ are two ordered partitions of $N$ and $L$ is a positive integer. In order for the ultraviolet quiver gauge theory to be well defined, the Young tableaux corresponding to the partitions $\rho$ and $\hat\rho$ must satisfy a set of  constraints which can be summarized, as we will explain, by the inequality
\beq
 \rho^T +L >    \hat\rho\,.
 \label{constraintss}
\eeq
Besides being invariant under the $OSp(4|4)$ superconformal symmetry group, these infrared fixed-point theories 
have a rich pattern of global symmetries that go beyond those present in the theory in the ultraviolet. For a superconformal field theory emerging from a circular quiver labeled by partitions $\rho$ and $\hat\rho$,  the infrared   theory acquires an enhanced global symmetry
\beq
H_\rho\times H_{\hat\rho}\,,
\label{globalsymm}
\eeq
where $H_\sigma$
 is the commutant of $SU(2)$ in $U(N)$ for the embedding
$
 \sigma: SU(2)\rightarrow  U(N)
$
 characterized by the partition $\sigma$ of $N$. The explicit  $AdS_{4}\ltimes K$ type-IIB string backgrounds we construct provide a  concrete realization  of the
constraints \rf{constraintss} and reproduce the precise pattern of global symmetries \rf{globalsymm}.

The construction of these string backgrounds extends  previous work in \cite{ABEG:2011,Aharony:2011yc}, where the bulk description of the
 strongly coupled, three dimensional ${\cal N}=4$ superconformal field theories that arise as infrared fixed points of {\it linear}
  quiver gauge theories was presented. All these solutions emerge  from the analysis of the equations derived in \cite{Gomis:2006cu} (see also \cite{Lunin:2006xr}) 
  which   determine the most general $OSp(4|4)$ invariant type-IIB supergravity backgrounds. 
  These equations can be, in turn, 
   elegantly solved  
 \cite{DEG1,DEG2} in terms of two harmonic functions on an open  Riemann surface.
The Riemann surface relevant for the  linear quivers is a strip, whereas the one  relevant for circular quivers is an annulus. 
 Pleasingly, the data of a circular quiver gauge theory which flows to a superconformal field theory in the infrared is encoded in special  ``punctures"  at the boundary of the annulus.
 
 \smallskip
 
A class of three-dimensional ${\cal N}=4$ superconformal field theories that arise from circular quivers are known to admit an
 M-theory description in terms of  orbifolds of the seven-sphere  \cite{Hosomichi:2008jd,Benna:2008zy,Imamura:2008ji}. 
By taking a certain (large $L$) smearing limit of our solutions, T-dualizing the periodic coordinate of the annulus 
 and   lifting the resulting type-IIA background to eleven dimensions,  we reproduce the relevant M-theory geometries  ${\rm AdS}_4\times {\rm S}^7/(\mathbb{Z}_k\times
\mathbb{Z}_{\hat k})$. In  the process one looses however  the dependence on 
  the full quiver data $(\rho,\hat\rho,L)$.  This data
  can be in principle encoded  in the non-contractible 3-cycles of the compact space
   and the associated 3-form fluxes  \cite{Imamura:2008ji, Witten:2009xu,Aharony:2009fc,Dey:2011pt}.  
   The  3-cycles  degenerate however  in the orbifold limit,   and  we are not aware of any solutions of eleven-dimensional supergravity that resolve   
  the singularity   on the M-theory side. By contrast in our IIB solutions, the full  data  is encoded in the positions of  five-brane throats along the annulus circle,
  and the singularity is effectively 
  resolved.\footnote{Even the simpler problem of T-dualizing pure NS5-branes  is notoriously subtle
\cite{Gregory:1997te, Tong:2002rq, Harvey:2005ab,Okuyama:2005gx}. As explained in these references, 
the contributions of  world-sheet instantons are 
 responsible for the localization of the NS5-branes on the type-IIB side \cite{Tong:2002rq},  and
for creating the dual throats in winding space on the type-IIA side \cite{Harvey:2005ab}.  It could be  interesting to extend this
analysis to the present, more complicated context.}

 
\smallskip

 An interesting outcome of our investigations is  a new type of ``orbifold equivalence" between different quantum field theories. 
 Based on the $SL(2,\mathbb{R})$ symmetry of type-IIB supergravity we arrive at the conjecture that gauge theories living on brane configurations 
 which are related by $SL(2,\mathbb{Q})$ transformations are equivalent in a certain large $N$ limit. 
  Theories related by  $SL(2,\mathbb{Z})$ transformations are of course exactly equivalent,  or mirror-symmetric, while  
  more general  $SL(2,\mathbb{Q})$ transformations can be regarded as orbifoldings of the F-theory torus. 
  A similar generalization
  of the T-duality group $O(d,d, \mathbb{Z})$ to a semigroup extension of $O(d,d, \mathbb{Q})$ has been analyzed recently in \cite{Bachas:2012bj}. 
  We provide quantitative evidence for  this new orbifold equivalence  by explicitly computing  the partition function 
   on $S^3$ of two different theories, which are related by $SL(2,\mathbb{Q})$, and showing that these partition functions match exactly.

 \smallskip

 The plan of the rest of the paper is as follows. In section \ref{sec:quivers} we introduce  the linear and circular quivers, which are the main objects of study in this paper. We also recall the conditions under which an ultraviolet quiver gauge theory is expected to flow in the infrared to an irreducible superconformal field theory and provide the data and constraints characterizing irreducible superconformal field theories, both for linear and circular quivers. In section \ref{gravitysolns} we present the supergravity solutions corresponding to the fixed points arising from circular quivers, also introducing the main features of the solutions corresponding to linear quivers that are needed for our analysis. In section \ref{sec:ads} we establish the dictionary between the infrared SCFTs and our supergravity solutions and find perfect agreement. Section \ref{sec:lim} contains the analysis of various interesting limits of the solutions and discusses their interpretation. This includes an interesting ``smeared" limit that results in M-theory geometries describing M2-branes at orbifold singularities. In section \ref{sec:sl2r} we explain how our supergravity solutions can be used to yield theories that are equivalent under a novel type of ``orbifold equivalence".
 We provide quantitative evidence for this by computing the large $N$ partition function of a proposed dual pair and we find perfect agreement. We have relegated to the Appendices some details and computations.


\section{Quivers, Infrared Fixed Points and Branes}
\label{sec:quivers}

\subsection{Linear and circular quivers}

The three-dimensional $\N=4$ superconformal field theories considered in this paper arise as
non-trivial infrared  fixed points of three-dimensional  quiver gauge theories with
$\N=4$ supersymmetries. Their field content and  their microscopic
Lagrangians are  succinctly summarized by a quiver diagram \cite{Douglas:1996sw}.
In our case  the diagrams will have  either linear or circular topology (see figures 1 and 2). We  refer to
 the corresponding quivers as linear and circular respectively.
\begin{figure}[h]
\centering
\includegraphics[height=6cm,width=8.5cm]{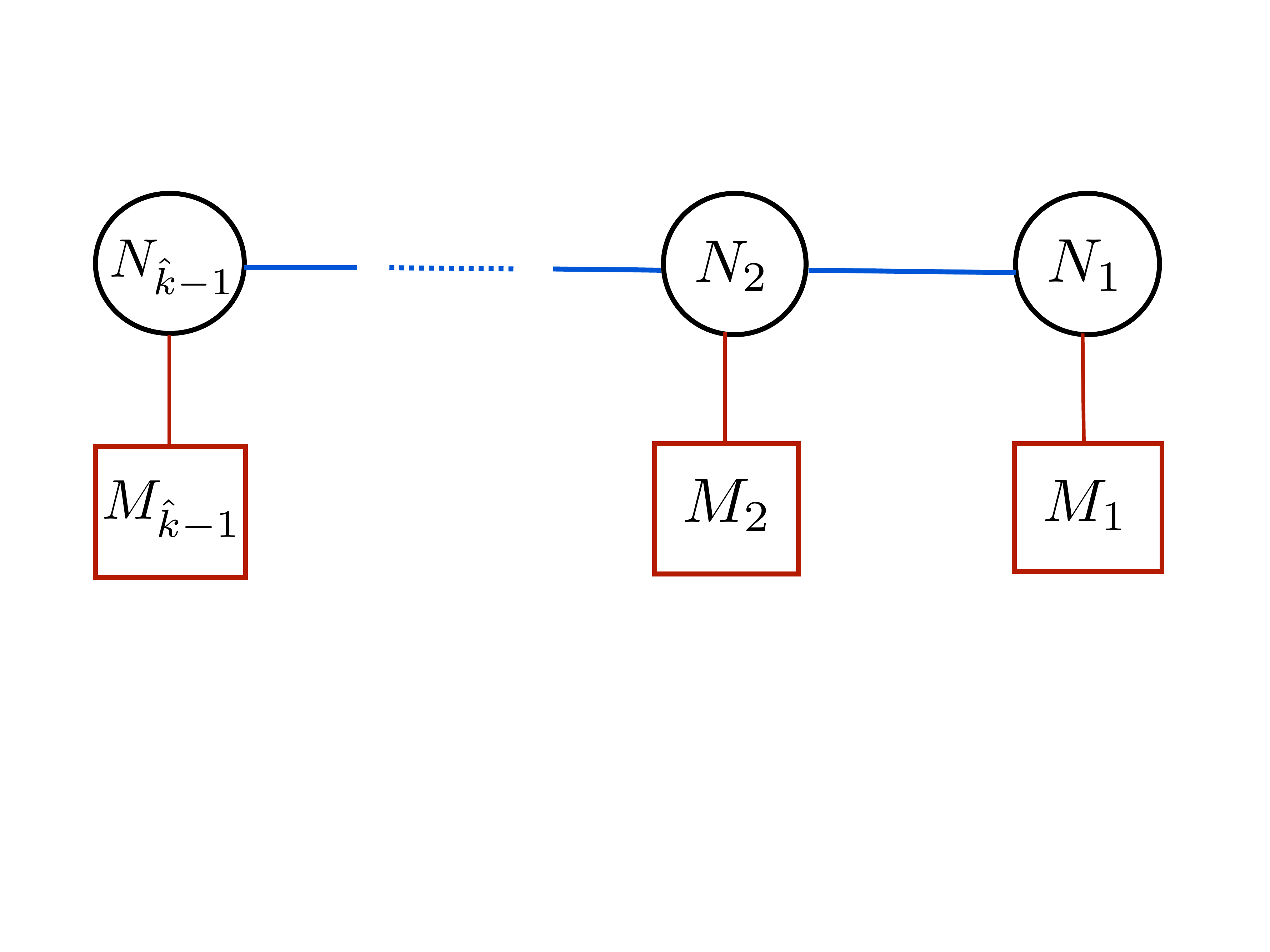}
\vskip -1cm
\caption{\footnotesize A  linear quiver with $\hat k-1$ gauge-group factors $U(N_1)\times U(N_2)\times\cdots$. The red boxes indicate the
numbers of  hypermultiplets in the fundamental representation of each gauge-group  factor.}
\label{fig1}
\end{figure}
\begin{figure}[h]
\centering
\includegraphics[height=6cm,width=8.5cm]{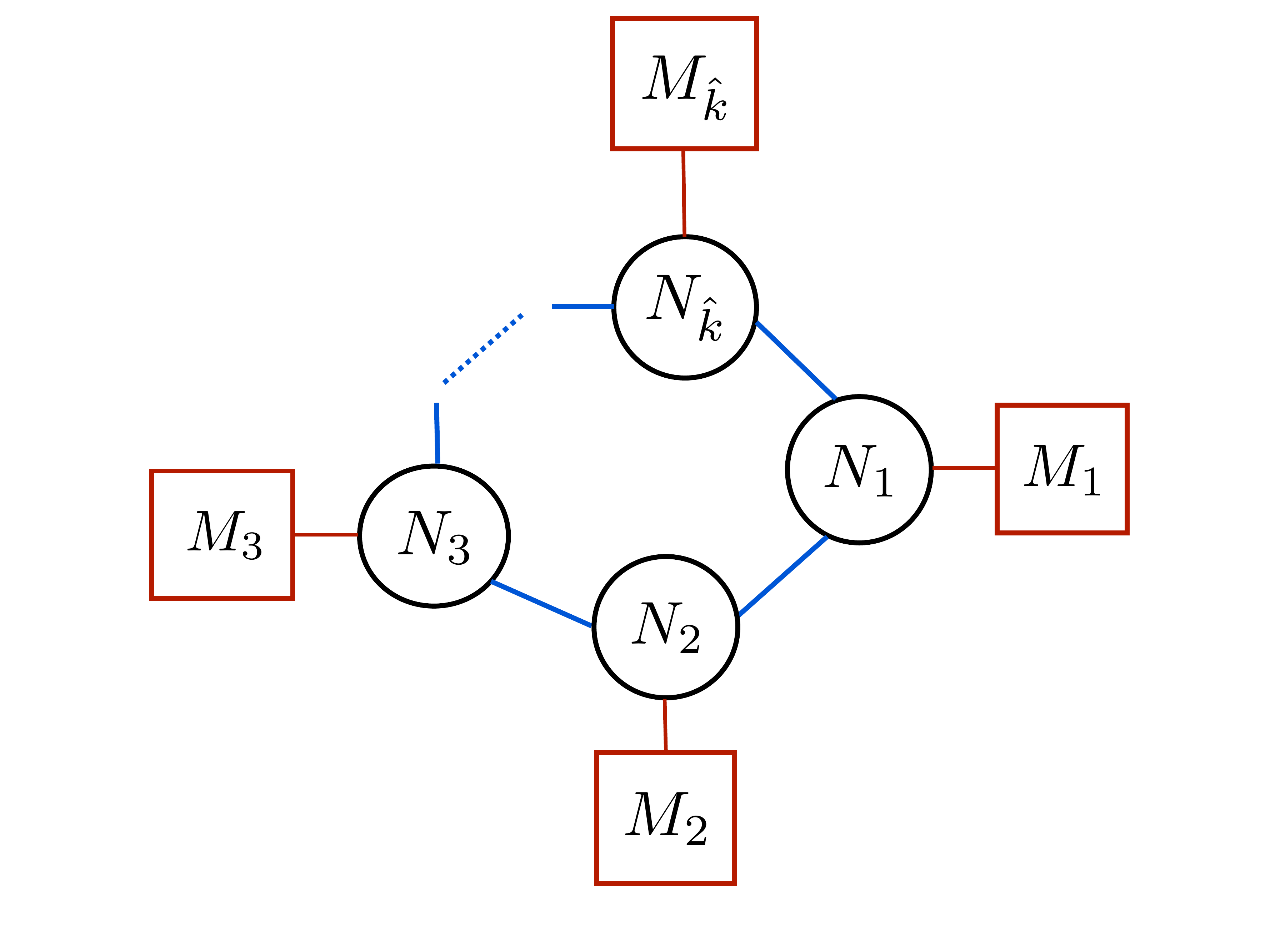}
\caption{\footnotesize A circular quiver with $\hat k$ gauge-group factors. The $U(N_i)$  theories
  interact via  bifundamental hypermultiplets (the blue lines)
which form a circular chain, as opposed to the linear chain of figure 1.}
\label{fig1a}
\end{figure}

\noindent

The  vector multiplets of these  quiver gauge-field  theories    transform in the adjoint representation of the gauge group
\beq
 U(N_1) \times U(N_2) \times ... \  U(N_i) \times ... \, .
\eeq
 Moreover, these theories contain a hypermultiplet transforming in the bifundamental representation of each consecutive pair of gauge groups $U(N_i)\times U(N_{i+1})$. For linear quivers $1\leq i \leq \hat k-2$, while for the circular quivers $1\leq i\leq \hat k$  with the convention that  $U(N_{\hat k+1})\equiv U(N_1)$.
 Finally, there are $M_i$ hypermultiplets in the fundamental representation of the group $U(N_i)$.\footnote{In the special case $\hat k=1$ the circular quiver has a single
   gauge-group factor,  $U(N_1)$,  and the  bifundamental hypermultiplet  is a hypermultiplet in  the adjoint representation of   $U(N_1)$. }
\smallskip

A central question about the dynamics of these gauge theories is  whether they flow to a non-trivial fixed point of the renormalization group  in the  infrared.
 Since massive fields decouple in the infrared, we will assume that  hypermultiplet masses  and  Fayet-Iliopoulos terms are set to zero.
 We also assume  for now vanishing Chern-Simons terms -- we will consider however such terms later in the paper.
 The  quiver data and the  extended $\N=4$  supersymmetry   specify  then completely  the  microscopic,
  renormalizable Lagrangian.

   We are interested in  ``irreducible superconformal field theories", i.e.  theories  not containing a decoupled sector with free vector multiplets and/or free 
   hypermultiplets.\footnote{On general grounds, we do not expect the bulk dual of a strictly free field theory to be describable by supergravity. Such a theory would require, due to the existence of  higher spin conserved currents,   higher spin fields propagating in the bulk.}
  It has been conjectured by Gaiotto and Witten \cite{Gaiotto:2008ak}  that a
  necessary and sufficient condition for a  gauge theory to flow to an irreducible superconformal field theory is
  \beq
N_{F,i}\geq 2N_i\,.
\label{Higgsing}
\eeq
In words, each  gauge-group  factor $U(N_i)$ should have at  least   $2N_i$ hypermultiplets transforming  in the fundamental representation.
A more refined irreducibility condition will be discussed below.
Recall  that a hypermultiplet in the fundamental and anti-fundamental representation are equivalent.
 Therefore, for a quiver gauge theory,  the above requirement of irreducibility in the infrared  imposes the following inequalities on the quiver data
 \beq
M_i+N_{i-1}+N_{i+1}\geq 2N_i\,.
\label{condiSCFT}
\eeq

One way to argue for  the above
 conditions is that when they are obeyed the  gauge group can be completely Higgsed  \cite{BHOOY:1996},
 and there exists a singularity at the origin of  the Higgs branch, from which the Coulomb branch emanates.
 A non-trivial superconformal field theory  appears in the infrared limit of the gauge theory around that vacuum.
 Conversely, when complete Higgsing is not possible, decoupled multiplets remain in the infrared, thus yielding non-irreducible theories.

 \smallskip

The quiver data that characterizes the irreducible superconformal field theories can be repackaged in a convenient way
in terms of two partitions,  $\rho$ and $\hat\rho$,  of the same number N  (this is explained below).  As usual,  one can associate
a Young tableau to each partition. The quiver theory can be described by the following data

\medskip\smallskip
$\bullet\ $   for linear quivers :   $(\rho,\hat\rho)$
 subject to the  constraints
\beq
    \rho^T > \hat \rho\  ;
  \label{fixedpoint}
\eeq

\medskip\smallskip
 $\bullet\ $ for   circular quivers:   $(\rho,\hat\rho,  L )$   subject to the constraints
 \medskip\smallskip
  \beq
  \rho^T \geq     \hat\rho\ , \qquad  L>0
   \,  .
  \label{fixedpointcirc}
\eeq
Here $\rho$ and $\hat\rho$ denote the two  partitions of  $N$,  and $L$  is a  positive  integer.
The inequality  $\rho  > \sigma$ between partitions means that the total number of boxes
 in the first $n$ rows of $\rho$ exceeds the same number in $\sigma$,  for all $n$.
 Transposition interchanges the columns and rows of a Young tableau.  The inequality \eqref{fixedpoint} has
 appeared previously  in different contexts related to solutions of Nahm's equations, see e.g. \cite{Kronheimer:1990ay,BHP}.

\smallskip

We denote the linear-quiver   theory  associated to  $(\rho,\hat\rho)$  by
$T_{\hat\rho}^\rho(SU(N))$,   and the circular-quiver theory with data  $(\rho,\hat\rho,  L )$
 by  $C_{\hat\rho}^\rho(SU(N),  L )$.
  It turns out  that the above  Young-tableaux constraints  are automatically satisfied
 if  the ranks of all the gauge groups of the ultraviolet  theories are positive, that is if  all $N_i>0$.
 If  some Young-tableaux inequalities were saturated for a linear quiver, the quiver would break down to decoupled
 quivers plus free hypermultiplets. Circular quivers,  on the other hand,
  degenerate to linear quivers when $L=0$.

\smallskip
 As we shall see, this   data also completely encodes the field content of the ultraviolet mirror pair \cite{Intriligator:1996ex} of quiver gauge theories which flow to the
same fixed point  in the infrared.
Mirror symmetry  for this class of quiver gauge theories is realized very simply by the exchange of the two partitions
 \beq
\text{mirror symmetry}:\qquad \rho\longleftrightarrow \hat\rho\,   .
\eeq
 Therefore, $T_{\hat\rho}^\rho(SU(N))$ and $T_{\rho}^{\hat\rho}(SU(N))$ are mirror linear-quiver gauge theories,
  while $C_{\hat\rho}^\rho(SU(N), L)$
 and $C_{\rho}^{\hat\rho}(SU(N),  L)$  are mirror circular quivers. The
 Young tableaux constraints  are symmetric under the exchange of $\rho$ and $\hat\rho$,
 see appendix \ref{sec:ineq},  and are therefore  consistent with  mirror symmetry.

\smallskip

These infrared  superconformal field theories are believed to have  a rich pattern of global symmetries,
  inherited from the symmetries acting on the Higgs and Coulomb branch of the
    quiver gauge theory from which the fixed point is reached in the infrared.  Since mirror symmetry exchanges
     the Higgs and Coulomb branches of mirror pairs, we conclude that the
global symmetry at the fixed point is
\beq
H \times \hat H \,,
\eeq
where
   \beq
    H =\prod_i U(M_i)  \qquad {\rm and}\qquad  \hat H = \prod_i U(\hat M_i) \,.
    \eeq
$H$  is the symmetry  that  rotates  the fundamental hypermultiplets    of
$T^{\rho}_{\hat \rho}(SU(N))$ or $ C_{\hat\rho}^\rho(SU(N), L)$, while $\hat H$ rotates the fundamental hypermultiplets of  their
 mirror duals. The two symmetries coexist at the superconformal fixed point.

\smallskip

In this paper we find the bulk  gravitational description of   the irreducible three dimensional $\N=4$ superconformal theories to which circular quiver gauge theories
of the above type  flow in the infrared.  We already presented the supergravity description of the superconformal  theories associated to linear quivers in \cite{ABEG:2011}.


 \subsection{Brane Realization}
\label{sec:branes}

The above three-dimensional $\N=4$ supersymmetric linear and circular quiver gauge theories admit an elegant realization as the low-energy limit of brane configurations in type-IIB string theory  \cite{Hanany:1996ie}.  The brane configuration consists of an array of D3, D5 and NS5 branes oriented as
shown in the table.\footnote{For more details of these brane constructions
 see  \cite{Hanany:1996ie}\cite{Gaiotto:2008ak}.}
\smallskip\smallskip

\begin{table}[h]
\label{tab:probeconfig}
\begin{center}
\begin{tabular}{|c||c|c|c|c|c|c|c|c|c|c||}
  \hline
      & 0 & 1 & 2 & 3 & 4 & 5 & 6 & 7 & 8 & 9 \\ \hline
  D3  & X & X & X & X &   &   &   &   &   &   \\
  D5  & X & X & X &   & X & X & X &   &   &   \\
  NS5 & X & X & X &   &   &   &   & X & X & X \\ \hline
\end{tabular}
\caption{\footnotesize Brane array for three-dimensional quiver gauge theories}
\end{center}
\end{table}

\noindent For linear quivers, the D3 branes span a finite interval along the $x^3$ direction and terminate on the five-branes,
 while for circular quivers $x^3$ parametrizes a circle.

\medskip
\noindent{\it Linear Quivers}
\medskip

\begin{figure}[th]
\centering
\includegraphics[height=6cm,width=11cm]{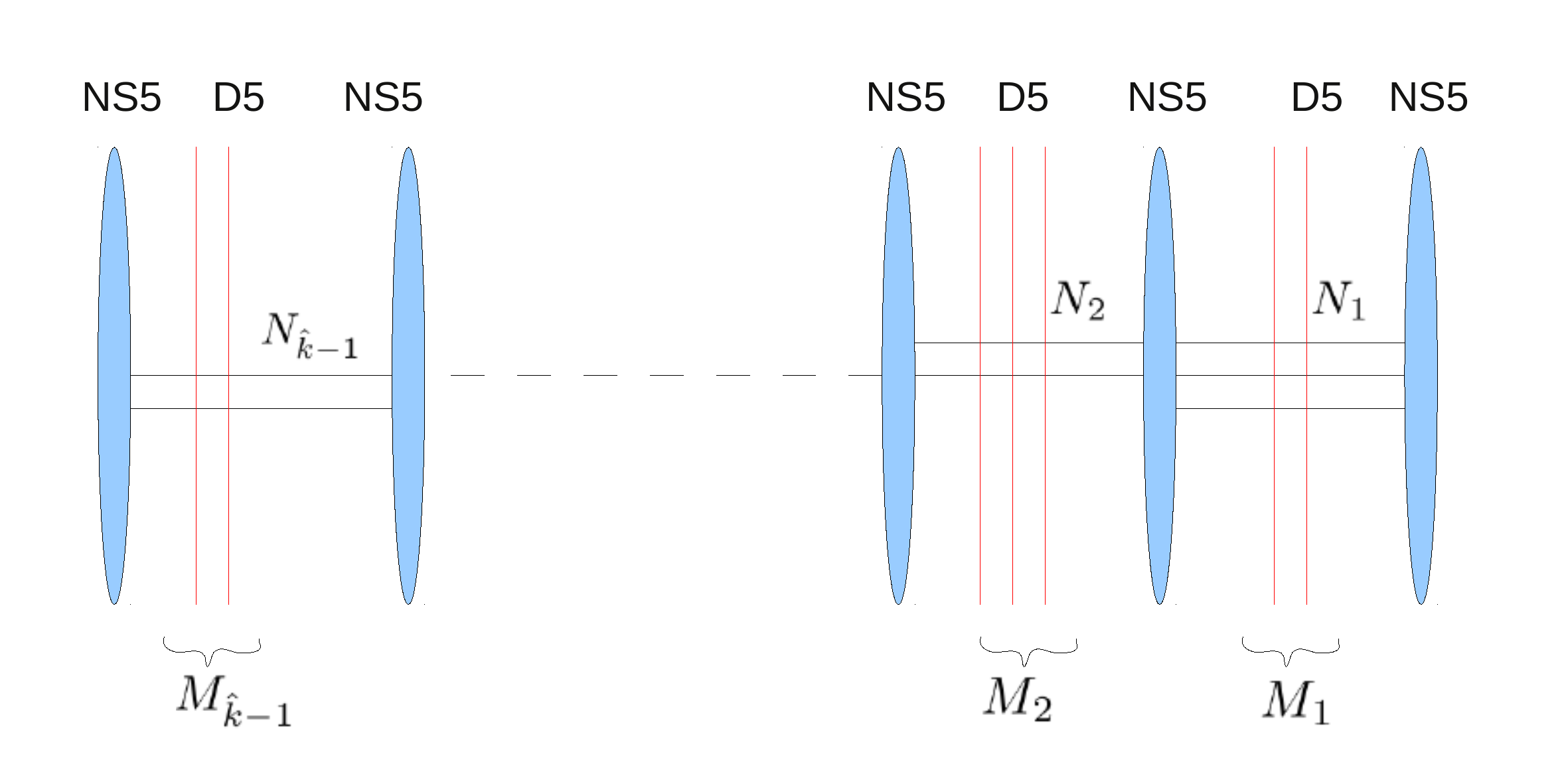}
\caption{\footnotesize Brane realization of linear quivers}
\label{fig2}
\end{figure}

The brane configuration corresponding to the linear quiver gauge theory of Figure \ref{fig1} is depicted in Figure \ref{fig2}.
An  invariant way of encoding a brane configuration -- and the corresponding quiver gauge theory -- is by specifying the {\it linking numbers} of the five-branes.
They can be defined as follows
\begin{align}\
l_i &= - n_i + R_i^{\rm NS5} \qquad (i=1,...,k) \cr
\hat l_j &= \hat n_j + L_j^{\rm D5} \qquad (j=1,...,\hat k) \ ,
\label{defnlinking}
\end{align}
where $n_i$ is the number of D3 branes ending on the $i$th D5 brane from the right minus the number ending from the left, $\hat n_j$ is the same quantity for the $j$th NS5 brane, $R_i^{\rm NS5}$ is the number of NS5 branes lying to the right of the $i$th D5 brane and $L_j^{\rm D5}$ is the number of D5 branes lying to the left of the $j$th NS5 brane.  These numbers  are {\it invariant} under Hanany-Witten moves \cite{Hanany:1996ie},   when a D5 brane crosses a NS5 brane. Since the extreme infrared limit is expected to be  insensitive to these moves,  it is   convenient to label the infrared dynamics in terms of the linking numbers.

\begin{figure}[th]
\centering
\includegraphics[height=6cm,width=11cm]{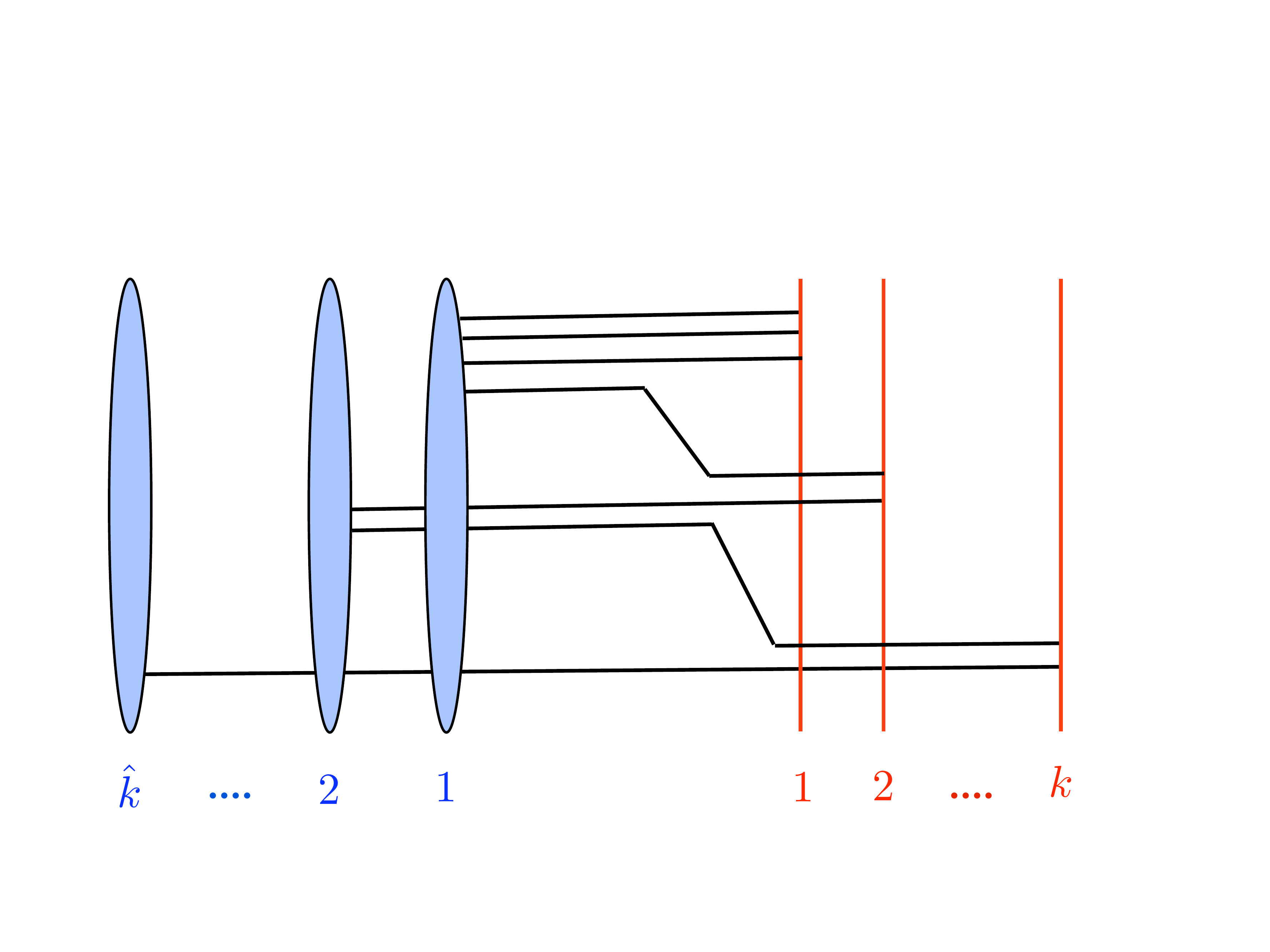}
\caption{\footnotesize Pushing all D5-branes to the right of all NS5-branes makes it easy to read the linking numbers, as
the net number of D3-branes ending on each five-brane.
In this example  $\rho = (3,2, \cdots 2)$ and $\hat\rho = (4, 2, \cdots 1)$. }
\label{separate}
\end{figure}

The brane construction of the linear quivers shown in Figure \ref{fig2} is characterized by the following linking numbers
\bea
l_i &=& j \quad \textrm{for a D5 brane between the $j$-th and $(j+1)$-th NS5 brane,} \\
 && \textrm{with a labeling such that $l_1 \geq .. \geq l_k$} ; \nonumber\\
\hat l_j &=& N_{j-1} - N_{j} + \sum_{s=j}^{\hat k-1} M_s \quad \textrm{for} \quad j=1,..,\hat k. \quad (N_{0}=N_{\hat k}=0)\,.\label{linkNS}
\eea
We may move all the NS5 branes to the left and all the D5 branes to the right, noting that a new D3  brane is created every time that a
D5  crosses a NS5. In the end, all the D3 branes will be suspended between  a  NS5 brane on the left and
 a D5 brane on the right, so  that the linking numbers satisfy the sum rule
\beq\label{sumrule}
\sum_{i=1}^{k} l_i = \sum_{j=1}^{\hat k} \hat l_j    \equiv N \,,
\eeq
where $N$ is the total number of  suspended D3 branes. This implies that the
  two sets of five-brane linking numbers define two partitions of $N$
   \bea\label{partitions}
  \rho: \qquad  ~N &=&  l_1+\ldots+l_{k}  \no  \\
 &=&   \underbrace{1+\ldots+1}_{M_1}\, +\, \underbrace{2+\ldots+2}_{M_2}\, +\, \ldots+ \ldots \\
  \hat \rho: \qquad  ~N &=& \hat l_1+\ldots +\hat l_{\hat k}\no  \\
  &=&   \underbrace{1+\ldots+1}_{\hat M_1}\, +\, \underbrace{2+\ldots+2}_{\hat M_2}\, +\, \ldots+ \ldots \,.
 \eea
 This is the repackaging of the quiver data in terms of partitions of $N$, mentioned above.
 It is illustrated by Figure  \ref{separate}.

 \vskip 1mm

In the original configuration of Figure  \ref{fig2}
 the
 D5 brane linking numbers are, by construction, positive and
     non-increasing, i.e.  $l_1\geq \cdots \geq  l_i\geq l_{i+1}\cdots \geq l_k>0$,
but this is not automatic for the linking numbers of the NS5 branes.   Requiring that the
     NS5 brane linking numbers be non-increasing, that is $\hat l_1 \geq \cdots \geq \hat l_i\geq \hat l_{i+1}\cdots \geq \hat l_{\hat k} = N_{\hat k - 1}$,
 is equivalent,  as follows from \rf{linkNS},   to
\beq\label{goodtheories}
M_i+N_{i-1}+N_{i+1}\geq 2N_i\,.
\eeq
This is the same as  \rf{condiSCFT}, the
 necessary and sufficient conditions for the corresponding (`good') quiver gauge theories to flow to an irreducible superconformal
  field theory in the infrared. Notice that if  these
   conditions  are not obeyed the linking numbers of the NS5 branes  need not even be positive integers.
   Furthermore, for the good theories that obey \eqref{goodtheories},
     it  follows from the expressions \rf{linkNS}  that the
   Young tableaux conditions  $  \rho^T >  \hat \rho$
are  automatically satisfied,  as long as the rank  of each  gauge-group factor  in the quiver diagram is positive.
\smallskip

In the configuration of  Figure \ref{separate} on the other hand, the
 meaning of the above conditions  changes. The ordering and positivity of all  linking numbers is  now automatic (more precisely, it can
 be trivially arranged by  moving 5-branes of the same type past each other).
   The constraints $  \rho^T >  \hat \rho$ on the other hand  are non-trivial;  they are the ones that   guarantee
  that a supersymmetric configuration like the one of Figure \ref{fig2} can be reached by a sequence of Hanany-Witten moves
  \cite{ABEG:2011}.  The two types of  configuration shown in the figures are in one-to-one correspondence when {\it all}
  these inequalities are satisfied by the five-brane linking numbers.

  \smallskip

Summarizing, the linear-quiver ${\cal N} = 4$
gauge theories  conjectured in  \cite{Gaiotto:2008ak} to  flow to irreducible fixed points in the infrared,  without extra free decoupled multiplets,
 are labeled in an invariant way by two ordered partitions of $N$, with associated Young tableaux $\rho$ and $\hat\rho$
 subject to the conditions  $  \rho^T >  \hat \rho$.
\smallskip


 \medskip
\noindent{\it Circular Quivers}
\medskip

\begin{figure}[th1]
\centering
\includegraphics[height=7.3cm,width=11cm]{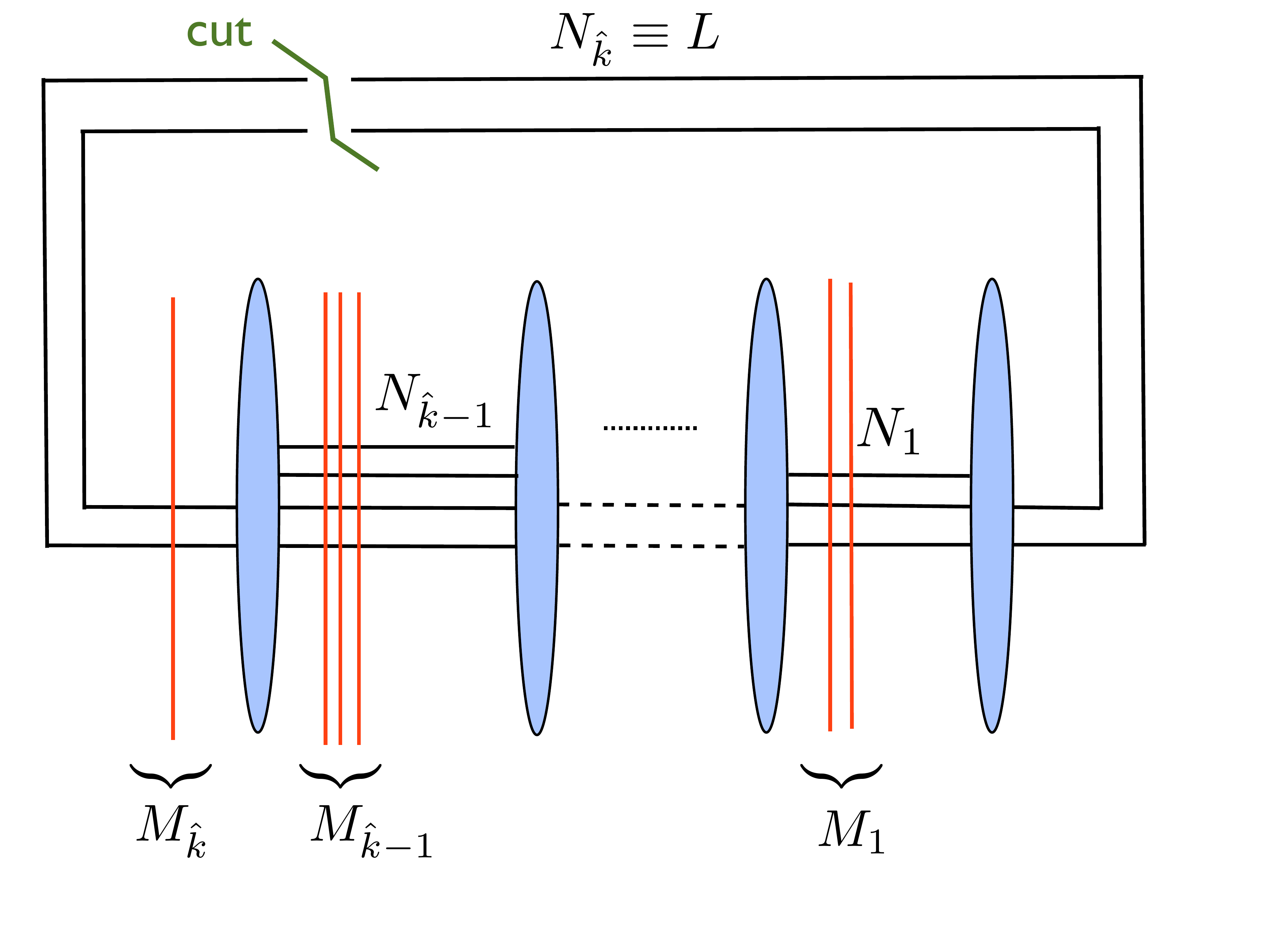}
\caption{\footnotesize Brane realization of circular quivers. To attribute linking numbers to the five-branes we cut open
the $k$-th stack of D3 branes, and place the $k$-th D5 stack at the left-most end.}
\label{fig2a}
\end{figure}

The brane configuration corresponding to the circular-quiver gauge theory of Figure \ref{fig1a} is given  in Figure \ref{fig2a}.
In this case  the $x^3$ coordinate along the D3 branes is periodic.  Compared to the linear case,   there are
$N_{\hat k} >0$   additional D3 branes extended between the first and the $\hat k$th NS5 branes that  close the circle. There can be, as well,
  $M_{\hat k}\geq 0 $  extra D5 branes giving rise to  fundamental hypermultiplets.

We can associate linking numbers to the five-branes by cutting open the circular
quiver  along one of the suspended D3-brane stacks, say the $k$-th stack.
We also choose to place the
 $\hat k$-th stack of D5 branes   at the left-most end of the open chain, as shown  in  Figure \ref{fig2a}.
The linking numbers are gauge-variant quantities, and the above choices amount to fixing partially a gauge.
 In this  gauge   the   linking numbers    read:
\bea
l_i &=& j \quad \textrm{for the $j$-th stack of D5 branes}
 \, ,   \\
\hat l_j &=& N_{j-1} - N_{j} + \sum_{s=j}^{\hat k } M_s\ ,  \quad \textrm{with} \quad j=1,..,\hat k\ . \quad  (N_0 = N_{\hat k})
\label{circularlinking}
\eea
As in the case of linear quivers, we    label the  NS5 branes in order of appearance
 from right to left,  and   the D5 branes  from left to right.

Defined as above, the linking numbers obey the sum rule \eqref{sumrule} with $N\equiv \sum_{s=1}^{\hat k} sM_s$.
Furthermore  the   linking numbers of the D5 branes  are by construction  non-increasing,   positive and bounded by
the number of NS5 branes, i.e.
\beq\label{orderedD5}
\hat k \geq l_1\geq  \cdots l_i \geq l_{i+1} \cdots \geq l_k >0\ .
\eeq
 What  about the linking numbers of NS5 branes? For  linear quivers,
    imposing that  the $\hat l_j$ be non-increasing  was equivalent to the  Higgsing conditions \eqref{Higgsing}
     that singled out  the  `good theories',  i.e. those believed to flow  to  an irreducible
 superconformal fixed point in the infrared.  Now, the Higgsing conditions can be written as
 \bea\label{upperconds}
0 &\leq& N_{j+1} + N_{j-1} - 2 N_j + M_j =
 \hat l_j - \hat l_{j+1}  \quad   \textrm{for} \, \   j = 1,..,\hat k-1 \\
0 &< &  N_{1} + N_{\hat k -1} - 2 N_{\hat k}  + M_{\hat k} =
 \hat l_{\hat k} - \hat l_1 + \sum_{s = 1}^{\hat k} M_s  \,.
 \label{strict}
\eea
The second line, which gives the condition for Higgsing of the $\hat k$-th gauge-group factor,
needs explaining. We have assumed that,  for this factor,  the inequality \eqref{Higgsing} is strict.
A good circular quiver  always has  at least one   such  gauge-group  factor   because,    if all the  inequalities \eqref{Higgsing}
were saturated,  it can be shown  that  all the $N_j$  are equal,  and all $M_j=0$.  So, in this case,  there would be
 only bi-fundamental hypermultiplets, but these  cannot
break completely  the gauge group  since they are neutral  under the diagonal $U(1)$. This
possibility must thus be excluded, i.e. one or more of the inequalities \eqref{Higgsing} must be strict.
 We choose to cut open the circular quiver at a D3-brane stack  for which $N_{F,j} > 2N_j$. Without loss of generality
  this is  the $k$-th stack.
\smallskip

The conditions \eqref{upperconds}  tell  us   that the NS5-brane  linking numbers are  non-increasing.
If we want them to be positive, we must   impose that
\beq\label{extracond1}
\hat l_{\hat k} =  N_{\hat k -1}  - N_{\hat k}  + M_{\hat k} > 0\ .
\eeq
If we furthermore want  our gauge condition  to respect  mirror symmetry  we must impose  the analog of the first inequality
\eqref{orderedD5}, namely
\beq\label{extracond2}
\hat l_1 =    N_{\hat k} - N_1 + k  \leq k \ .
\eeq
Together  \eqref{extracond1} and \eqref{extracond2} imply   \eqref{strict},  but not the other way around.
Fortunately,  these conditions can be always satisfied in  good quivers, for example by choosing  a gauge factor whose  rank
 is locally  minimum along the chain  (i.e.   $N_{\hat k} < N_1, N_{\hat k - 1}$).  With this choice   we
finally   have
  \beq\label{NS5ordered}
 k \geq \hat l_1\geq  \cdots \hat l_j \geq \hat l_{j+1} \cdots \geq \hat l_{\hat k} >0\ ,
\eeq
 so that the NS5-brane   and the D5-brane linking numbers are on equal footing. They define two
 partitions, $\hat \rho$ and $\rho$ of the same number $N$.
  \smallskip

  Contrary to the case of linear quivers, here  the partitions do not  fully determine the brane configuration.
  The reason is that the number,  $N_{\hat k} \equiv L >0 $, of D3 branes in the $k$-th stack is  still free to vary. We can change it,
  without changing the linking numbers of the five-branes,  by adding or removing D3 branes that wrap
   the  circle (thus increasing or decreasing uniformly  all gauge-group ranks). It follows from
   \eqref{circularlinking} that  the condition for  all gauge-group factors
 to  have positive rank  now reads
     \beq
  L+ \rho^T  >    \hat\rho\     \,  .
  \label{fixedpointcircA}
\eeq
To understand this constraint   intuitively, note that  removing   $L$ winding D3 branes  may convert some stacks of D3 branes to stacks of  anti-D3 branes.
In the case of linear quivers  the inequality $\rho^T  >    \hat\rho$  guarantees  the absence of anti-D3 branes.
  Here anti-D3 branes are tolerated, as long as their number is less than $L$.
  \smallskip

   To any data $(\rho, \hat \rho, L)$ subject to the constraints \eqref{fixedpointcircA},  together with the additional conditions
   $l_1\leq \hat k$ and $\hat l_1 \leq k$, there corresponds a `good' circular-quiver gauge theory, i.e. one conjectured to  flow  to
   an irreducible superconformal theory in the infrared. This description is, however, highly redundant because of the arbitrariness
   in choosing at which  D3-brane stack  to cut open the quiver.  A generic circular quiver will have many gauge-group factors
   for which  \eqref{extracond1} and \eqref{extracond2} are satisfied, so
     many different triplets $(\rho, \hat \rho, L)$ would describe the same  SCFT.

   \smallskip

   To remove  this redundancy, one can impose the extra condition that
  the cut-open segment be  of   minimal rank \emph{globally}, i.e. that
    $L \leq N_j$ for all $j$.\footnote{If there are several gauge factors of globally-minimal rank,
  there will  remain some redundancy in our description of the circular quiver.
  This   is however a non-generic case.}
   This condition is compatible with the earlier ones;  it amounts to further fixing the gauge.  Now removing  $L$
  winding D3-branes does not  create any anti-D3 branes, since $L$ was the absolutely minimal   rank.
  The two partitions thus  obey the stronger inequality
      \beq
   \rho^T \geq \hat\rho\     \,  .
 \eeq
As a bonus, the conditions $l_1\leq \hat k$ and $\hat l_1 \leq k$ are now also automatically satisfied. Note  that
  linear-quiver theories saturating  some of the inequalities $\rho^T \geq \hat\rho$  broke down into smaller decoupled linear
  quivers plus free hypermultiplets. For circular quivers, on the other hand, these disjoint pieces are reconnected by the
  $L>0$ winding D3 branes, giving   irreducible  theories in the infrared.

 \smallskip

Summarizing, the circular-quiver gauge theories conjectured to  flow to irreducible superconformal field theories in the infrared
can be  labeled by a positive integer $L$, and by
two ordered partitions $\rho$ and $\hat\rho$  subject to the condition $\rho^T \geq \hat\rho$.
  An alternative but redundant description is in terms of a triplet $(\rho, \hat\rho, L)$  subject to the looser
  conditions \eqref{fixedpointcircA},  together with the additional constraints $l_1\leq \hat k$ and $\hat l_1 \leq k$.
  Both  descriptions are manifestly  mirror-symmetric.
  As we will later discuss,  in  the dual supergravity theory
  these two descriptions correspond to a complete,  or to a partial gauge fixing of  the 2-form potentials.


\section{Solutions of IIB supergravity}
\label{gravitysolns}

We will now exhibit the solutions of type-IIB supergravity that are holographic duals of  superconformal field theories,
to which the circular-quiver  theories of the previous section (are believed to) flow. The  solutions are
constructed by periodic identification of the  linear-quiver backgrounds  found in \cite{ABEG:2011,Aharony:2011yc}. These latter
 are, in turn,  special
cases of the general local solutions of \cite{DEG1,DEG2}.
 We start  by reviewing very briefly some  key formulae from  these earlier works.

\subsection{Local solutions}
\label{s:localsolutions}

References \cite{DEG1,DEG2} give the general local solutions of type-IIB supergravity preserving the superconformal symmetry $OSp(4|4)$.  This
group is the supergroup of the 3d $\N=4$ SCFTs.  The solutions are parameterized by a choice of a 2-dimensional Riemann surface with boundary, $\Sigma$,
 and by  two harmonic functions,  $h_1$ and $h_2$,  on $\Sigma$.  In terms of
 the auxiliary functions
\begin{align}\label{W}
W = \p \bar \p(h_1 h_2) \ , \qquad N_{j} = 2 h_1 h_2 |\p h_{j}|^2 - h_{j}^2 W,
\end{align}
the metric can be written as
\begin{align}
ds^2 = f_4^2 ds^2_{AdS_4} + f_1^2 ds^2_{S^2_1} + f_2^2 ds^2_{S^2_2} + 4 \rho^2 dz d\bar z\ ,
\end{align}
where the warp factors are given by
\begin{align}
f_4^8 = 16 \frac{N_1 N_2}{W^2} \ ,
\ \ \
f_{1}^8 = 16 h_{1}^8 \frac{N_{2} W^2}{N_{1}^3} \ ,
\ \ \
f_{2}^8 = 16 h_{2}^8 \frac{N_{1} W^2}{N_{2}^3} \ ,
\ \ \
\rho^8 = \frac{N_1 N_2 W^2}{h_1^4 h_2^4} \ .
\end{align}

This geometry is supported by non-vanishing ``matter"  fields, which include the  (in general complex)  dilaton-axion field
\begin{align}
S = \chi + i e^{2 \phi} = i \sqrt{N_2 \over  N_1 }\ ,
\end{align}
in addition to  3-form and 5-form backgrounds. To specify the corresponding gauge potentials
one needs the dual harmonic functions,  defined by
\begin{align}
h_1 = -i ({\cal A}_1 - \bar {\cal A}_1) \qquad &\rightarrow \qquad h_1^D = {\cal A}_1 + \bar {\cal A}_1 \ , \cr
h_2 = {\cal A}_2 + \bar {\cal A}_2 \qquad &\rightarrow \qquad h_2^D = i ({\cal A}_2 - \bar {\cal A}_2) \ .
\end{align}
The constant ambiguity in the definition of the  dual functions is related to changes of the background
fields under large gauge transformations.
The NS-NS and R-R three forms can be written  as
\begin{align}
\label{3forms0}
H_{(3)}  =   \omega^{\, 45}\wedge db_1  \qquad {\rm and} \qquad  F_{(3)} =   \omega^{\, 67}\wedge db_2   \ ,
\end{align}
where $ \omega^{\, 45}$ and $ \omega^{\, 67}$ are the volume forms of the unit-radius  spheres  ${\rm S}_1^{2}$ and ${\rm S}_2^{2}$, while
\begin{align}
\label{3forms1}
b_1 &= 2 i h_1 {h_1 h_2 (\p  h_1\bar  \p  h_2 -\bar \p  h_1 \p  h_2) \over N_1} + 2  h_2^D \ ,  \cr
b_2 &= 2 i h_2 {h_1 h_2 (\p  h_1 \bar\p  h_2 - \bar\p  h_1 \p  h_2) \over N_2} - 2  h_1^D \ .
\end{align}
The expression for the gauge-invariant self-dual  5-form is a little  more involved:
\begin{align}
\label{5form0}
F_{(5)}  = - 4\,  f_4^{\, 4}\,  \omega^{\, 0123} \wedge {\cal F} + 4\, f_1^{\, 2}f_2^{\, 2} \,  \omega^{\, 45}\wedge \omega^{\, 67}
 \wedge (*_2  {\cal F})   \ ,
\end{align}
where $ \omega^{\, 0123}$ is the volume form of the unit-radius ${\rm AdS}_4$,
${\cal F}$ is a 1-form on $\Sigma$ with the property that $f_4^{\, 4} {\cal F}$ is closed,
and $*_2 $ denotes Poincar\' e duality with respect to the $\Sigma$ metric.
The explicit expression for ${\cal F}$  is given by
\begin{align}\label{calF}
f_4^{\, 4} {\cal F} = d j_1\   \qquad {\rm with} \qquad j_1 =
3 {\cal C} + 3 \bar  {\cal C}  - 3 {\cal D}+ i \frac{h_1 h_2}{W}\,   (\p  h_1 \bar\p  h_2 -\bar \p h_1 \p h_2) \ ,
\end{align}
where ${\cal C}$ and ${\cal D}$ are defined by  $\p {\cal C} = {\cal A}_1 \p  {\cal A}_2 - {\cal A}_2 \p  {\cal A}_1$
and  ${\cal D} = \bar {\cal A}_1 {\cal A}_2 + {\cal A}_1 \bar {\cal A}_2$.
\vskip 1mm

 For any choice of $h_1$ and $h_2$,
 equations \eqref{W} to \eqref{calF} give  local  solutions of the supergravity equations which are invariant
 under $OSp(4|4)$. Global consistency puts  severe constraints on these harmonic functions and on the surface $\Sigma$.
  There is  no complete classification of all consistent choices for this data. What has been shown  \cite{DEG1,DEG2}
  is that the most general type-IIB solution  with the $OSp(4|4)$  symmetry can be brought to the above form
   by  an  $SL(2,\mathbb{R})$ transformation. This acts as follows on the dilaton-axion and 3-form fields:
 \begin{align}
 S \to {aS+b\over cS+d}\ , \qquad
  \left(   \begin{array}{c}
H_{(3)}  \\
F_{(3)}
 \end{array}
  \right) \to
   \left(   \begin{array}{cc}
 d & -c \\ -b & a
  \end{array}
  \right)
  \left(   \begin{array}{c}
H_{(3)}  \\
F_{(3)}
 \end{array}
  \right) \ .
 \end{align}
 The Einstein-frame metric and the 5-form $F_{(5)}$ are left unchanged.


\subsection{Admissible singularities}
\label{s:admissible}

  The holomorphic functions ${\cal A}_1$ and ${\cal A}_2$ are analytic in the interior of $\Sigma$,
  but can have singularities on its boundary.  Refs. \cite{DEG1,DEG2} identified  three  kinds of   ``admissible"  singularities, i.e. singularities that
  can be interpreted  as brane sources in   string theory. Two of these  are logarithmic-cut  singularities and correspond to the two
  elementary  kinds of five-brane.
     In local coordinates,  in which the boundary of  $\Sigma$ is the real axis,  these singularities read
     \begin{align}
\underline{\rm D5}:   \qquad {\cal A}_1 =  - i \gamma\,  {\rm log  } w + \cdots  \ , \qquad {\cal A}_2 =  - i   c + \cdots \ ,
\nonumber
 \end{align} \vskip -7mm
 \begin{align} \label{sings}
\underline{\rm NS5}:  \qquad {\cal A}_1 =  -  \hat c + \cdots  \ , \qquad   {\cal A}_2 =   - \hat\gamma\,  {\rm log  } w + \cdots \ .
 \end{align}
 Here $\gamma, \hat\gamma, c, \hat c$ are real parameters  related to the brane charges, and the dots denote
  subleading terms, which are analytic at $w=0$ and have the same reality properties on the boundary as the leading terms.
 These reality properties imply that,   in the case of the D5-brane,    $h_1$ and $h_2$  obey respectively Neumann and Dirichlet boundary conditions,
   i.e.  $(\partial - \bar\partial) h_1 = h_2 = 0$ on the boundary of  $\Sigma$.   For the NS5 brane
 the  roles of  the two harmonic functions are exchanged.
  \vskip 1mm

\begin{figure}
\centering
\includegraphics[height=7cm,width=12cm]{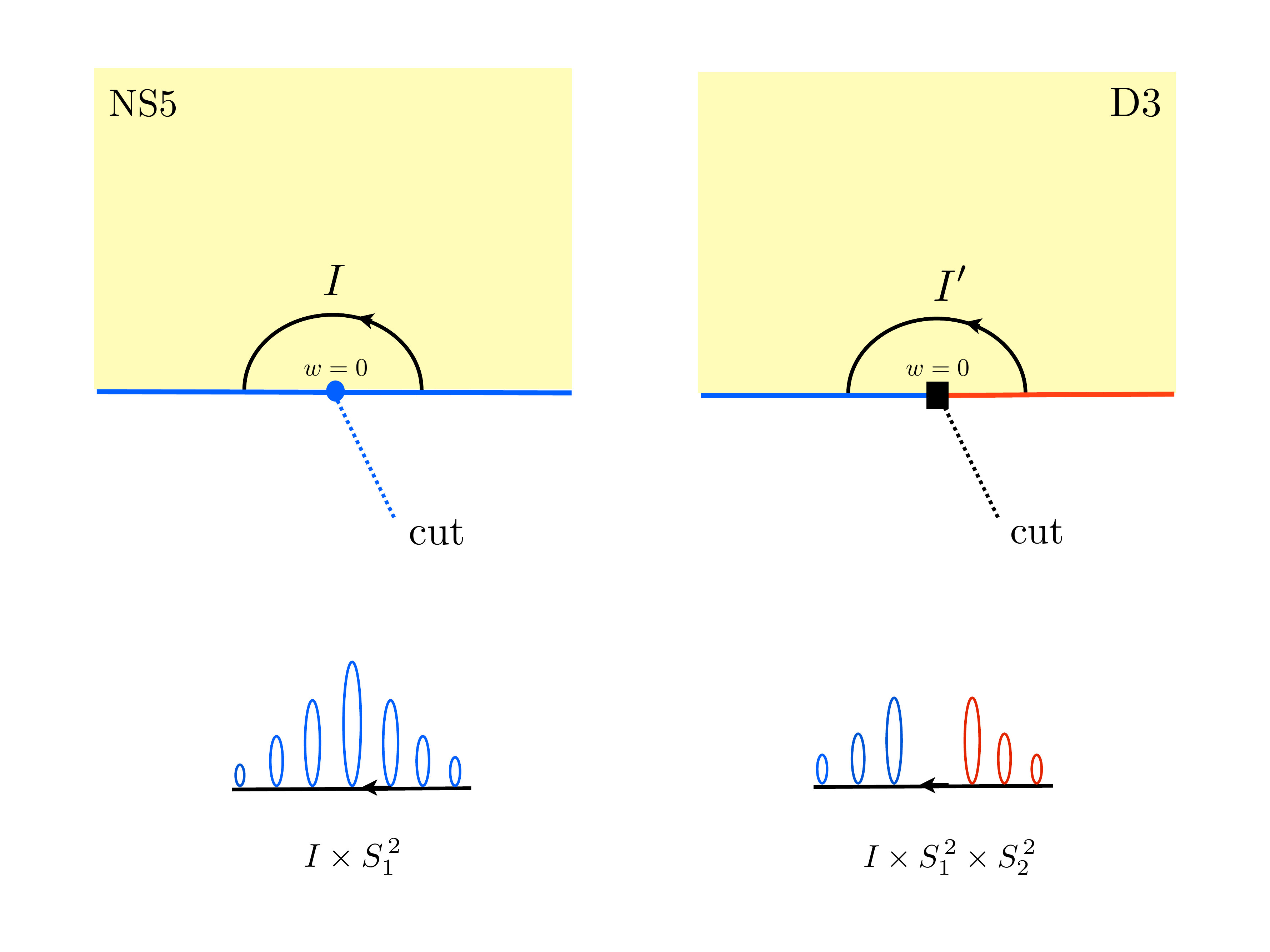}
\vskip -5mm
\caption{\footnotesize  Local singularities corresponding to  a NS5-brane (left) and a D3-brane (right),  as explained in the text.
The boundary of $\Sigma$ is colored  red or blue  according to  which of the two 2-spheres, $S_1^{\, 2}$ or $S_2^{\, 2}$,
shrinks at this part of the boundary to zero.  The  non-contractible cycles
supporting the brane charges are  $I\times S_1^{\, 2}$,  $I  \times S_1^{\, 2}\times S_2^{\, 2}$ and
$I^\prime \times S_1^{\, 2}\times S_2^{\, 2}$,  with  $I$ and $I^\prime$ the (oriented) solid semicircles of the figure. These cycles are topologically
equivalent to a 3-sphere, a 3-sphere times a 2-sphere, and a 5-sphere. The broken lines indicate the logarithmic (on the left) and square root (on the right) branch cuts. }
\label{noncontractI}
\end{figure}

The vanishing of the harmonic function $h_j$ implies that the corresponding 2-sphere $S_j^{\, 2}$ shrinks to a point.
This is necessary in order for points on  the boundary of $\Sigma$,  away from the singularities,  to correspond  to regular  interior points
of the ten-dimensional geometry.
Non-contractible cycles,   which
 support non-zero brane charges,  are obtained by the fibration of one or both 2-spheres
over any curve  that (semi)circles the singularity on $\partial \Sigma$.
  For instance in the case of the NS5-brane  $I \times  S_1^{\, 2}$,
with  $I$   the interval shown in figure \ref{noncontractI},  is topologically a non-contractible  3-sphere.  The appropriately
normalized
flux of $H_{(3)}$ through this cycle is the number of NS5-branes:\footnote{The
five-brane charge  is quantized in units of $2 \kappa_0^2 T_5$,  where $2 \kappa_0^2 = (2 \pi)^7 (\alpha^\prime)^4$
  is the gravitational coupling constant, and $T_5 = 1/[(2 \pi)^5 (\alpha^\prime)^{3}]$ is the five-brane tension. Note that since we have
 kept the dilaton arbitrary,  we are free to  set the string coupling $g_s=1$; the tension of the NS5-branes and the D5-branes is thus the same,
 while the D3-brane tension and charge is $T_3 = 1/[(2 \pi)^3 (\alpha^\prime)^{2}]$.}
 \begin{align}\label{N5hat}
 \hat N_{5} \, =\,   {1\over 4\pi^2\alpha^\prime } \int_{I\times S_1^{\, 2}} H_{(3)} = {2\over \pi\alpha^\prime} h_2^D\Bigl\vert_{\partial I}
 \qquad \Longrightarrow \ \   \hat N_{5} \,
 =\  {4\over \alpha^\prime}\,  \hat\gamma\ .
\end{align}
In evaluating the flux we have taken $I$ to be   infinitesimally small, and we used the fact that in
 the expression \eqref{3forms1}   only $h_2^D$ is discontinuous
across the singularity on the real axis.
We also  assumed
that the logarithmic cut lies outside the surface $\Sigma$, so that fields in the interior of $\Sigma$ are all continuous (see figure \ref{noncontractI}).
\vskip 1mm

In addition to 5-brane charge, the singularities \eqref{sings} also carry D3-brane charge. The corresponding flux
threads  the 5-cycle
$I \times  S_1^{\, 2}\times S_2^{\, 2}$, which is topologically the product of a 3-sphere with a 2-sphere.
There is a well-known subtlety in the definition of this charge, because of
the  Chern-Simons term in the IIB supergravity action \cite{ABEG:2011,Page:1984qv,Marolf:2000cb}. In the case at hand
  the conserved flux  is the integral of the gauge-variant 5-form  $F_{(5)} + C_{(2)}\wedge H_{(3)}$, which obeys
 a non-anomalous Bianchi identity.
   The  number of D3-branes   inside the NS5-brane stack
is thus given by
 \begin{align}\label{insideNS5}
  \hat N_{3} \, =\,
{1\over (4\pi^2\alpha^\prime)^2 } \int_{I\times S_1^{\, 2}\times S_2^{\, 2}} [ F_{(5)} + C_{(2)}\wedge H_{(3)} ]
\, = \,
 -  {2\over \pi \alpha^\prime}\, \hat N_{5} \,  h_1^D\Bigl\vert_{w=0}  \ .
\end{align}
It can be checked,  by taking again $I$ arbitrarily small,
 that $F_{(5)}$,  as well as all terms    in the expression for $C_{(2)}$ other than $h_1^D$,
do not contribute to the above flux. This explains the second equality,  leading finally  to
  \begin{align}\label{N3hat}
  \hat N_3 \,
 =\      \left( {4\over \alpha^\prime} \right)^2 \,  \left(  {\hat\gamma\, \hat c \over \pi} \right) \ .
\end{align}
 Note  that $\hat N_3$  depends on the   potential $C_{(2)}$ at the  position of the
5-brane singularity, and may change  under large gauge transformations. This is related to the Hanany-Witten effect \cite{Hanany:1996ie},
an issue  to which we will return in the  next subsection.

 \vskip 1mm

 In principle, using $SL(2, \mathbb{R})$ transformations one can convert the NS5-brane solution to a
 more general $(p,q)$ fivebrane solution.  Such transformations generate, however,  a
 non-trivial Ramond-Ramond axion background, so $(p,q)$ fivebranes cannot coexist with the NS5-brane
 solution for which the axion vanishes.  There is one exception to the rule: the
  S-duality transformation converts the NS5-brane to a D5-brane without generating an axion background.
  Combined with an exchange of the two 2-spheres, S-duality acts as follows on
   the harmonic functions:
         \begin{align}
 \left(
 \begin{array}{c}
i{\cal A}_2  \\
-{\cal A}_1
 \end{array}
  \right)
 \ \xrightarrow{\tilde S} \
 \left(
 \begin{array}{c}
 {\cal A}_1  \\
i{\cal A}_2
 \end{array}
  \right)\ .
 \end{align}
  This gives
  the D5-brane singularity  anticipated already in equation \eqref{sings}.
The  integer D5-brane  and D3-brane charges read
    \begin{align}\label{N5}
 N_5 =   {4\over \alpha^\prime}  \,   \gamma\, , \qquad  N_3 \,
 =\      \left( {  4\over   \alpha^\prime}  \right)^{  2}  \,   \left( {\gamma\,   c\over \pi}\right) \ .
\end{align}
Note that  the D3-brane charge is here  the flux of the  5-form
$F_{(5)} - B_{(2)}\wedge F_{(3)}$, which is the S-duality transform of
 $F_{(5)} + C_{(2)}\wedge H_{(3)}$. This gauge-variant form  is
 well-defined in any  patch around the D5-brane singularity  as long as this patch does not
 contain NS5-brane sources.

  \vskip 1mm

    The last kind of  singularity,  which can coexist with D5- and NS5-brane singularities, is the one   describing free D3-branes,   with no associated
     fivebrane charge.  In this case the holomorphic functions  have  square-root rather than logarithmic cuts \cite{DEG2}
   \begin{align}\label{D3sing}
\underline{\rm D3}:   \qquad {\cal A}_1 =  {1 \over \sqrt{w}} (a_1   + b_1 w + \cdots  ) \ , \qquad
{\cal A}_2 =   {1 \over \sqrt{w}} (a_2   + b_2 w + \cdots  )
\ .
\end{align}
 Such singularities change the boundary condition of  $h_1$ from Neumann to Dirichlet, and the boundary condition  of
 $h_2$ from Dirichlet to Neumann. This is illustrated in the right part of figure \ref{noncontractI}.
  The integer  D3-brane charge is given by
  \begin{align}
   n_{3} \, =\,
{1\over \left(4\pi^2\alpha^\prime\right)^2 } \int_{I^\prime \times S_1^{\, 2}\times S_2^{\, 2}}  F_{(5)}
\ = \  \left( {4\over \alpha^\prime} \right)^2 \,  \, { (a_1b_2 - a_2 b_1) \over 2\pi}\ .
\end{align}
 The ten-dimensional geometry near the D3-brane singularity is an $AdS_5\times S^5$
 throat  with radius $L$ given by $L^4 = 4\pi \alpha^{\prime\, 2} \vert n_3 \vert$.


\subsection{Linear-quiver geometries}
\label{s:linearq}

\begin{figure}
\centering
\includegraphics[height=8cm,width=12cm]{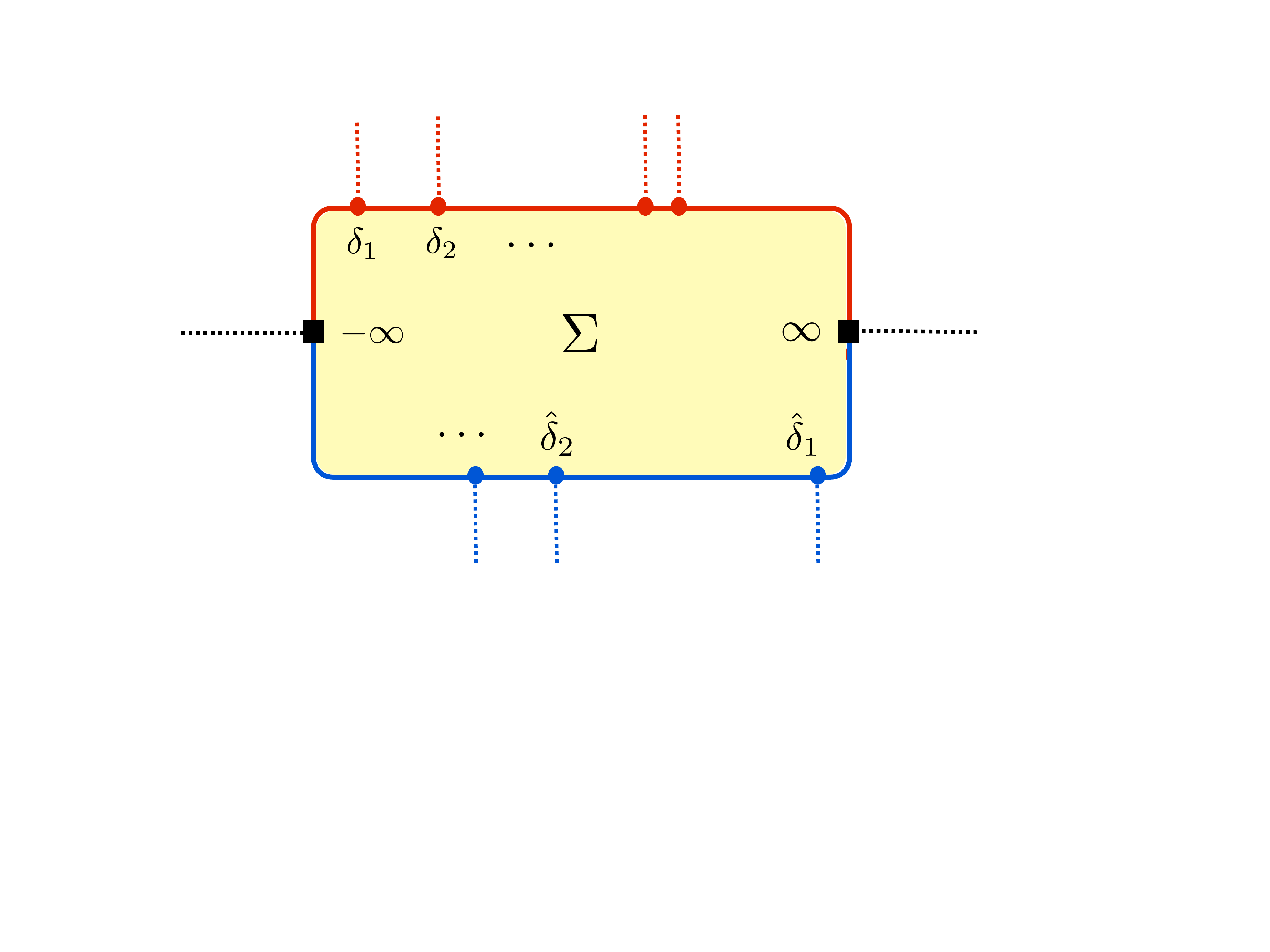}
\vskip -1.9cm
\caption{\footnotesize   Structure of singularities of the harmonic functions for the linear-quiver geometries.
The two boundary-changing singularities at $\pm\infty$, corresponding to $AdS_5\times S^5$ throats, can be capped off by
choosing $a_1=a_2=0$,  as described in the text. They  become regular interior points of the ten-dimensional geometry.}
\label{Disk}
\end{figure}

Consider two  harmonic functions with the singularity structure shown in figure \ref{Disk}.
 The corresponding geometries have the field-theory interpretation of
 superconformal domain walls in ${\cal N} = 4, D=4$ super Yang Mills \cite{Gaiotto:2008ak}.
 If $n_3^\pm$ are the D3-brane charges of the two boundary-changing (black-box) singularities, then
 the domain wall separates two  gauge theories with gauge groups $U(n_3^-)$ and $U(n_3^+)$.
 As pointed out in  \cite{ABEG:2011,Aharony:2011yc},  one may decouple  the three-dimensional
 SCFT that lives on the domain wall from the  bulk  four-dimensional Yang-Mills theories  by setting
 $a_j^\pm = 0$. Equation \eqref{D3sing} shows that  in this case  $n_3^+ = n_3^- =0$.  The square-root
 singularities of the harmonic functions are then simply  coordinate singularities, while the infinite $AdS_5\times S^5$ throats
 are replaced by regular interior points in  ten-dimensions.

\smallskip

Following references \cite{BE:2011, ABEG:2011,Aharony:2011yc}, we choose $\Sigma$ to be the infinite strip and the
harmonic functions to be given by
\begin{align}
\label{harm1}
{\cal A}_1  =  - i \sum_{a=1}^{p} \gamma_{a} \ln\, \tanh\left( \frac{i\pi}{4}-\frac{z-\delta_a}{2} \right)  \ , \ \ \
{\cal A}_2 =  - \sum_{b=1}^{\hat{p}} \hat{\gamma}_{b} \ln \tanh\left(\frac{ z-\hat{\delta}_{b} }{2}\right)    \ .
\end{align}
 Here   $\delta_1 < \delta_2 < ... < \delta_p$ are the positions of the D5-brane singularities on the upper boundary of the
 strip, whereas
   $\hat \delta_1 > \hat \delta_2 > ... > \hat \delta_{\hat p}$ are the positions of the NS5-brane singularities on the lower
   boundary. It can be checked that on these two boundaries $h_1$ obeys, respectively, Neumann and Dirichlet conditions,
   while  $h_2$ has Dirichlet and Neumann conditions. The boundary-changing square-root singularities are at $z= \pm\infty$. In the local
   coordinate $w = e^{\mp z}$ one can verify easily  that $a_j^\pm = 0$,  so these points at infinity correspond to regular interior points
  of the ten-dimensional geometry.
 \smallskip

 To simplify the formulae we will adopt from now on the (non-standard)  convention $\alpha^\prime = 4$.
Equations \eqref{N5} and \eqref{N5hat}   give the numbers of NS5-branes and D5-branes for each   fivebrane
singularity:
\begin{align}
 N_5^{(a)} = \gamma^{(a)}\ , \qquad  \hat N_5^{(b)} = \hat\gamma^{(b)}\ .
   \end{align}
Unbroken supersymmetry requires that there are only branes (or only anti-branes) of each kind. Thus
all the $ \gamma^{(a)}$ must have the same sign, and  likewise for all the
 $ \hat\gamma^{(b)}$. Dirac quantization forces furthermore  these parameters to be integer.
\vskip 1mm

Next let us consider the D3-brane charge.  Inserting the harmonic functions \eqref{harm1} inside the
expressions \eqref{N5} and \eqref{N3hat} gives
\begin{align}
&N_3^{(a)} =  N_5^{(a)} \sum_{b=1}^{\hat p}  \hat N_5^{(b)} \, {2\over\pi} {\rm arctan}(e^{\hat\delta_b -\delta_a}) \ ,
\nonumber  \\
&\hat N_3^{(b)} = - \hat N_5^{(b)} \sum_{a=1}^{p}    N_5^{(a)} \, {2\over\pi} {\rm arctan}(e^{\hat\delta_b -\delta_a})\ ,
\label{ginvN3}
 \end{align}
 where we used the identity $i\,$log\,tanh$({i\pi\over 4} - {x\over 2} )=  -$2\,arctan($e^x$).
As already noted in the previous subsection, this calculation of the  D3-brane charge
 depends on the  2-form potentials $B_{(2)}$ and $C_{(2)}$ and is, a priori, ambiguous.
One may indeed add a real constant
 to ${\cal A}_1$, or an imaginary constant to   ${\cal A}_2$,  thereby changing $h_j^D$ without affecting $h_j$.
This gauge ambiguity is also reflected in the arbitrary choice of Riemann sheet  for the logarithmic functions that enter in
equations  \eqref{harm1}.

\smallskip
Following \cite{ABEG:2011} we fix this ambiguity by placing all logarithmic cuts outside $\Sigma$, as in figure \ref{Disk},
and by choosing the sheet so that   the imaginary part of  (${\rm ln\, tanh} \, z$)   vanishes
when $z$ goes to $+\infty$ on the real axis. This implies that the arctangent functions take values in the interval $[0, \pi/2]$.
Our  choice  of gauge  is continuous in the interior of $\Sigma$ (which is covered by a single patch),  and sets
$B_{(2)}=0$ at $+\infty$  and  $C_{(2)}=0$
at $-\infty$.  With this  choice, D5-branes at $\delta=+\infty$ and NS5-branes at $\hat\delta = -\infty$ do not contribute to
the D3-brane charge. Placing, on the other hand,  one NS5-brane at $\hat\delta = +\infty$ adds one unit of D3-brane charge to
each D5-brane, while placing one D5-brane at $\delta = -\infty$ adds one unit of charge to each NS5-brane.
This is a holographic manifestation of the Hanany-Witten effect.

\smallskip

Since this story will be important to us later, let us explain it a little more.  The 2-form potential $B_{(2)}$ is proportional
to the volume form ($\omega^{45}$)  of the sphere $S_1^{\, 2}$, which shrinks to a point in the lower boundary of the strip
(the blue line in figure \ref{Disk}). When  $B_{(2)}\not= 0$, the corresponding boundary interval corresponds to a Dirac singularity
of codimension 3 in (the 9-dimensional)  space.  This   is unobservable if
\begin{align}
{1\over 2\pi\alpha^\prime} \int_{S_1^{\, 2}} B_{(2)} \in 2\pi \mathbb{Z}\ \Longrightarrow\
B_{(2)}\Bigl \vert_{{\rm Im} z = 0}\  =\  \pi\alpha^\prime\omega^{45} \times ({\rm integer})\ .
\end{align}
With our choice of gauge,
\begin{align}
B_{(2)}\Bigl \vert_{{\rm Im} z = 0}\  =\  \pi\alpha^\prime\omega^{45} \times  \sum_{b=1}^\beta  \hat N_5^{(b)}
\qquad  {\rm for} \ \ \   \hat\delta_{\beta+1} < {\rm Re}z  <  \hat\delta_{\beta }  \ .
\end{align}
Large gauge transformations  change  $B_{(2)}$ everywhere in the strip by a multiple of    $\pi\alpha^\prime\omega^{45}$,
and can remove the Dirac sheet in one of the   intervals of the boundary. For us this was the interval $( \hat\delta_1, \infty)$.
A similar story holds also for the upper (red) boundary and the 2-form $C_{(2)}$. The  D3-brane charges with our  choice of gauge  agree  with
the  invariant linking numbers  defined  in   \S\ref{sec:branes}.

\smallskip

     The brane  engineering of the dual gauge field theories \cite{Hanany:1996ie,Gaiotto:2008ak} involves
     $N$ D3-branes
      suspended between $\hat k$ NS5-branes on the left and $k$ D5-branes on the right.  In the IIB supergravity
      the corresponding numbers are:
   \begin{align}\label{conserv}
   N =  \sum_{a=1}^p N_3^{(a)} = - \sum_{b=1}^{\hat p} \hat N_3^{(b)}\ , \qquad
   k = \sum_{a=1}^p N_5^{(a)}\ , \qquad \hat k = \sum_{b=1}^{\hat p} \hat N_5^{(b)}\ .
   \end{align}
  The way in which the D3-branes are suspended to the five-branes is given by two partitions $\rho$ and $\hat\rho$,
   which define  the linear-quiver gauge  theory. These partitions are given in terms of the linking numbers:
 \begin{align}
\rho &= \Big( \overbrace{l^{(1)},l^{(1)},..,l^{(1)}}^{N_{5}^{(1)}},\ \overbrace{l^{(2)},l^{(2)},..,l^{(2)}}^{N_{5}^{(2)}},\ ...\ ,
\ \overbrace{l^{(p)},l^{(p)},..,l^{(p)}}^{N_{5}^{(p)}} \Big) \ ,   \nonumber \\
\hat \rho &=
 \Big( \overbrace{\hat l^{(1)},\hat l^{(1)},..,\hat l^{(1)}}^{\hat N_{5}^{(1)}},\ \overbrace{\hat l^{(2)},\hat l^{(2)},..,\hat l^{(2)}}^{\hat N_{5}^{(2)}},\ ...\ ,\
 \overbrace{\hat l^{(\hat p)},\hat l^{(\hat p)},..,\hat l^{(\hat p)}}^{\hat N_{5}^{(\hat p)}} \Big) \ ,
\end{align}
where $l^{(a)} = N_3^{(a)}/N_5^{(a)}$  and \,   $\hat l^{(b)} = \hat N_3^{(b)}/\hat N_5^{(b)}$. Here
 $l^{(a)}$ is the number of D3-branes ending on  each D5-brane in the $a$th stack, while   $\hat l^{(b)}$ is the
number of D3-branes emanating from each NS5-brane in the $b$th stack.
Because these numbers must be integers, the parameters $\delta_a$ and $\hat\delta_b$ are quantized.\footnote{The relations
between the integer brane charges and the supergravity parameters are not easily inverted. To express the latter in terms
of the brane charges one must solve a system of  transcendental equations.
}
In all one has $2p + 2\hat p -1$  parameters, since a global translation of all the $\delta_a$ and $\hat\delta_b$ does not
change the solution. The parameters of the quiver are $N_{5}^{(a)}, l^{(a)}, \hat N_{5}^{(b)}, \hat l^{(b)}$ subject
to one constraint  \eqref{conserv},  which expresses the
 conservation of D3-brane charge. The two parameter counts therefore match.

\smallskip

The linking numbers of the supergravity solutions obey the inequalities $\rho^T > \hat\rho$, which
were the conditions for the existence of a
non-trivial infrared fixed point of the quiver gauge theory \cite{Gaiotto:2008ak}, see
\S\ref{sec:branes}. On the supergravity side,  the  inequalities
follow immediately   \cite{ABEG:2011}  from the fact that $0< {\rm arctan}(x) < \pi/2$ for positive $x$.  This is a non-trivial
check of the AdS/CFT correspondence.


\subsection{From   strip to  annulus}
\label{s:striptoannulus}

  The strategy for constructing holographic IIB duals  for the circular quivers is the following:  one starts from
  the  linear-quiver solutions  that we have just described,
 and arranges the  five-branes   in infinite  regular arrays. The holomorphic  functions ${\cal A}_j$   become logarithms of quasi-periodic
 elliptic  functions.
Modding out  by
 discrete translations then converts the  strip domain, $\Sigma$,   to an annulus,
 and the dual  linear-quiver theories to  theories based on circular quivers.
\vskip 1mm

\begin{figure}
\centering
\includegraphics[height=8cm,width=12cm]{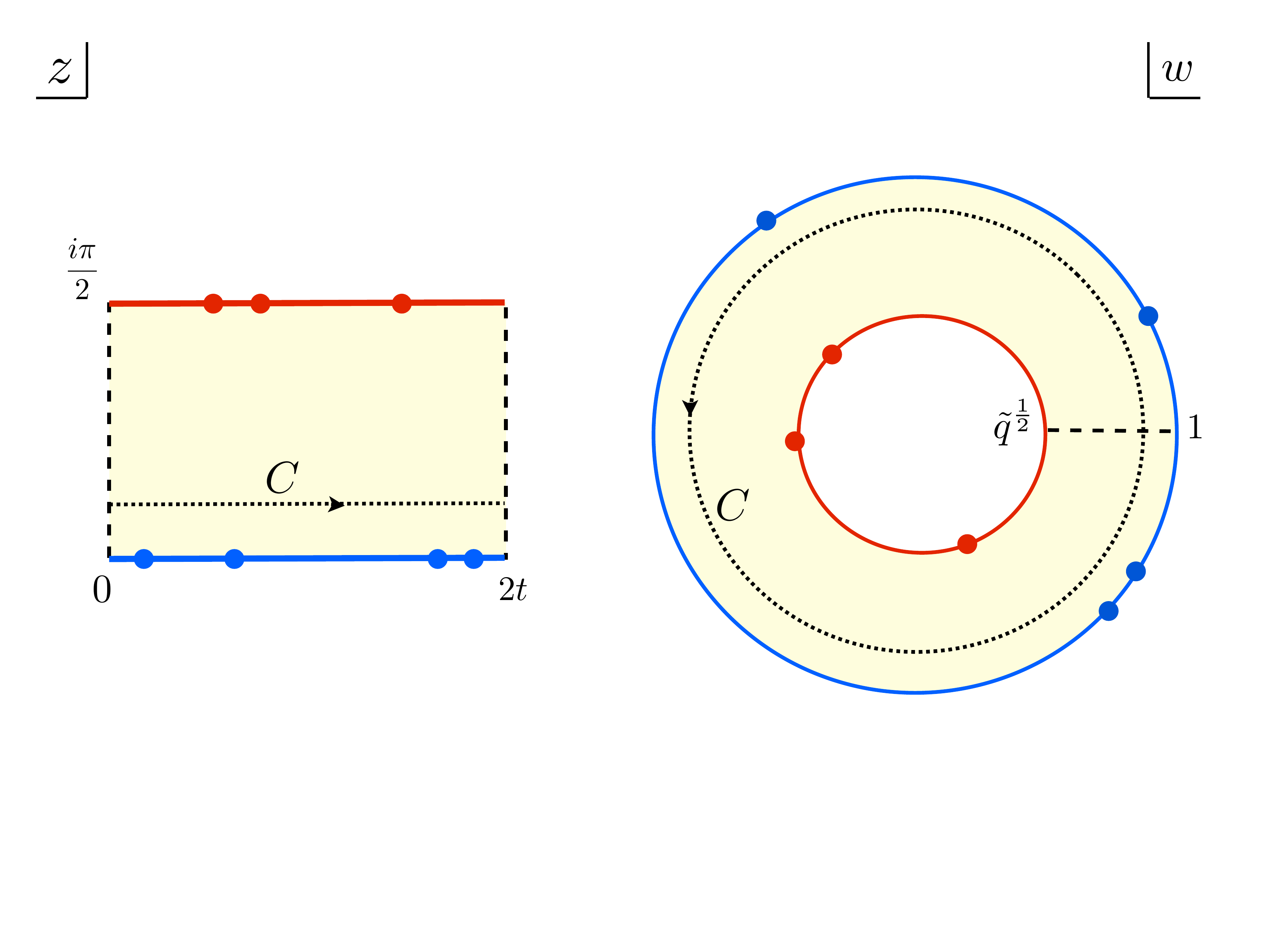}
\vskip -1.6cm
\caption{\footnotesize  The annulus $\Sigma$   for the type-IIB solutions that are
 dual to  $D=3, {\cal N} = 4$  circular-quiver  theories.   $\Sigma$ is the infinite strip in the $z$ plane modulo
 the translations $z\to z+2t$ (left), or the annular domain in the $w = {\rm exp}(i\pi z/t)$ plane (right). The radius
 of the inner boundary of the annulus  is   $\tilde q^{1/2}$ where $\tilde q  =  {\rm exp}(-\pi^2/t)$   is the exponentiated dual modulus
 of the elliptic $\vartheta$-functions. The monodromies of $h_j^D$ around the curve $C$ give the total number
 of NS5 and D5-branes, as explained in the main text. }
\label{Annulus}
\end{figure}

 More explicitly, given a set of fivebrane singularities at $\delta_a$ and $\hat \delta_b$, we may always pick a
 positive parameter $t$ such that, after a rigid translation,    $0  \leq \delta_a \leq 2 t$ and $0 \leq \hat \delta_b \leq 2t$.
Replicating the fivebrane sources with  periodicity $2t$ then leads to the following harmonic functions
\begin{align}
\label{ellipticharm}
h_1  &= -  \sum_{a =1}^p  \gamma_a \ln \bigg[ \prod_{n = -\infty}^{\infty} \tanh \bigg( \frac{i \pi}{4} - \frac{z - (\delta_a + 2 n t)}{2} \bigg) \bigg] + c.c. \ , \cr
h_2 &= - \sum_{b=1}^{\hat p} \hat \gamma_b \ln \bigg[ \prod_{n = -\infty}^{\infty} \tanh \bigg( \frac{z - (\hat \delta_b + 2 n t)}{2} \bigg) \bigg] + c.c. \ .
\end{align}
These functions are  manifestly periodic  under translations by $2t$, so we are free to identify  $z\equiv z+2t$  thereby converting the strip
  $\Sigma$ to  an annulus.
 Figure \ref{Annulus} depicts this annular domain in the $w$-plane, where $w= {\rm exp}({i\pi z/t})$.
 \smallskip

\smallskip
To see that the infinite products in the above expressions converge, we will rewrite them in terms of elliptic
$\vartheta$-functions  (we use the conventions of  reference \cite{Green:1987mn}).  This can be done with the help of the identity
 \begin{align}
\label{thident}
\Bigl\vert \frac{\vartheta_1(\nu\vert \qth)}{\vartheta_2(\nu\vert \qth)} \Bigr\vert \,
=\,   \left\vert   \prod_{n=-\infty}^{\infty}  \tanh(i\pi \nu  +  nt )\,  \right\vert \ ,
 \qquad {\rm where} \quad e^{i \pi \qth} = e^{-t}  \ .
\end{align}
The proof of this identity  follows from  the product formulae for the $\vartheta$-functions
\begin{align}\label{productf}
\vartheta_{1}(\nu\vert \qth) = 2 e^{i \pi \qth /4} \sin (\pi \nu) \prod_{n=1}^\infty (1-e^{2n i \pi \qth})(1 - e^{2n i \pi \qth} e^{2 \pi i \nu})(1 - e^{2n i \pi \qth} e^{-2 \pi i \nu}) \ , \cr
\vartheta_{2}(\nu\vert \qth) = 2 e^{i \pi \qth /4} \cos (\pi \nu) \prod_{n=1}^\infty (1-e^{2n i \pi \qth})(1 + e^{2n i \pi \qth} e^{2 \pi i \nu})(1 + e^{2n i \pi \qth} e^{-2 \pi i \nu}) \ .
\end{align}
Note that the modular parameter is $\tau = it/\pi$,  because the hyperbolic tangents are periodic under  $z\to z+2\pi i$.
Inserting the   identity  \eqref{thident}  in  \eqref{ellipticharm}
leads to the following expressions for $h_1$ and $h_2$:
 \begin{align}
\label{hmany}
h_1 &= -  \sum_{a =1}^p \gamma_a \ln \bigg[ \frac{\vartheta_{1}\left(\nu_a\vert \qth  \right)}{\vartheta_{2}\left(\nu_a\vert \qth  \right)} \bigg]
 + c.c.    \  , \  \qquad {\rm with}\ \ \
  i\, \nu_a = - \frac{z-\delta_a}{2 \pi } + \frac{i}{4} \ , \cr
h_2 &= - \sum_{b=1}^{\hat p} \hat \gamma_b \ln \bigg[ \frac{\vartheta_{1}\left(  \hat \nu_b\vert \qth \right)}{\vartheta_{2}\left(  \hat \nu_b\vert \qth \right)} \bigg]
+ c.c.  \ ,
\ \  \qquad {\rm with}\ \ \    i\,  \hat \nu_b =   \frac{z - \hat \delta_b}{2 \pi } \ .
\end{align}
These harmonic functions are well-defined everywhere inside the annulus. They have logarithmic singularities on the boundaries,
wherever $\nu_a$ or $\hat\nu_b$ vanish.

\smallskip

   Decomposing $h_j$ into holomorphic and anti-holomorphic parts requires, as in the previous subsection, a choice of gauge.
A convenient choice is to make the  ${\cal A}_j$ analytic in the
interior of the covering strip,  {before}  the periodic identification of $z$. This amounts to placing again all logarithmic branch cuts outside the strip.
With this understanding,  and recalling that the Jacobi $\vartheta$-functions are holomorphic,   we have
   \begin{align}
\label{Amany}
{\cal A}_1  = - i \sum_{a =1}^p \gamma_a \ln \bigg( \frac{\vartheta_{1}\left(\nu_a \vert \qth  \right)}{\vartheta_{2}\left( \nu_a\vert \qth \right)} \bigg)
 + \varphi_1   \  , \  \qquad
{\cal A}_2  = - \sum_{b=1}^{\hat p} \hat \gamma_b \ln \bigg( \frac{\vartheta_{1}\left(\hat \nu_b\vert \qth  \right)}{\vartheta_{2}\left(\hat \nu_b\vert \qth\right)} \bigg)
+ i \varphi_2  \
  ,
\end{align}
where the constant phases $\varphi_1$ and $\varphi_2$ are residual quantized gauge degrees of  freedom,
corresponding to large gauge transformations of the 2-form potentials. As in the case of the linear quiver, we may use this residual  freedom to
enforce the absence of Dirac singularities in one interval on each annulus  boundary.
\smallskip

Unlike $h_j$, the above holomorphic functions and  the dual harmonic functions $h_j^D$  are {\it not }
 periodic under $z\to z+2t$.
Their  gauge-invariant holonomies (or Wilson lines)  give  the total fivebrane charges.
  To see why, note that translating  $z\to z+2t$ changes all the arguments $\nu_a$ by $it/\pi$ (and all the $\hat\nu_b$ by $-it/ \pi$).
 From the product formulae \eqref{productf} one finds that
 under these  translations the $\vartheta$-functions are quasi-periodic:
 \begin{align}
\vartheta_1(\nu+ {it\over \pi}\vert \qth ) = -e^{-2\pi i \nu + t} \vartheta_1(\nu\vert \qth)  \ , \qquad
\vartheta_2(\nu+ {it\over \pi}\vert \qth) =  e^{-2\pi i \nu + t} \vartheta_2(\nu\vert \qth) \ .
\end{align}
 The
  ratio  $\vartheta_1/\vartheta_2$ changes  only by a minus sign.
 Thus ln($\vartheta_1/\vartheta_2)\to$ln($\vartheta_1/\vartheta_2) \mp i\pi$ when $\nu \to \nu\pm it/\pi$, from which we conclude
   \begin{align}\label{holonomy}
 {\cal A}_1(z+ 2t) = {\cal A}_1(z) -  \pi \sum_{a =1}^p \gamma_a  \  \, , \qquad
  {\cal A}_2(z+ 2t) = {\cal A}_2(z) - i\pi    \sum_{b=1}^{\hat p} \hat \gamma_b  \ .
 \end{align}
The meaning of these holonomies becomes clear if one integrates the 3-form field strengths over the
3-cycles $C\times S_j^{\, 2}$, where $C$ is the   dotted curve in figure \ref{Annulus}.
Consider for example the $H_{(3)}$ flux through  $C\times S_1^{\, 2}$. From  equations
  \eqref{3forms0} and  \eqref{3forms1} we deduce  that this   is proportional to
  \begin{align}\label{b2hol}
  \oint_C db_1 = 2 \oint_C d h_2^D\,  =\,  4i \, [ {\cal A}_2(z+2t) - {\cal A}_2(z) ]\ ,
  \end{align}
where in the first step we used the fact that  $(db_1 - 2 d h_2^D)$  is an exact differential which, therefore,   integrates to  zero.
Since the  total flux is conserved, the right-hand-side of  \eqref{b2hol}  must be  $z$-independent.
Furthermore, by  deforming the  contour $C$  so that only the  singularities on the outer boundary of the annulus
 contribute, one finds that the integrated flux  is proportional to the total number of NS5-branes.
 This agrees  with the holonomy of ${\cal A}_2$, as computed from the properties of the $\vartheta$-functions.
 The holonomy of ${\cal A}_1$ is likewise determined  by the total number of D5-branes.


\section{The AdS/CFT correspondence}
\label{sec:ads}
We  turn now to a discussion of the dictionary between the type-IIB supergravity solutions of the previous section, and
the circular-quiver theories of  section 2. As explained there, these gauge theories can be parametrized (in a redundant way)
 by the linking numbers of NS5-branes and D5-branes,
and by the number $L$ of  D3-branes that wind the circle.
We will here first relate these numbers  to the brane charges of the supergravity solutions, and then prove the
basic inequalities \eqref{fixedpointcircA}.
 Modulo a few subtleties, this  is a straightforward
extension of the linear-quiver analysis  of \cite{ABEG:2011}.

\subsection{Calculation of D3-brane charges}

\begin{figure}
\centering
\includegraphics[height=8cm,width=12cm]{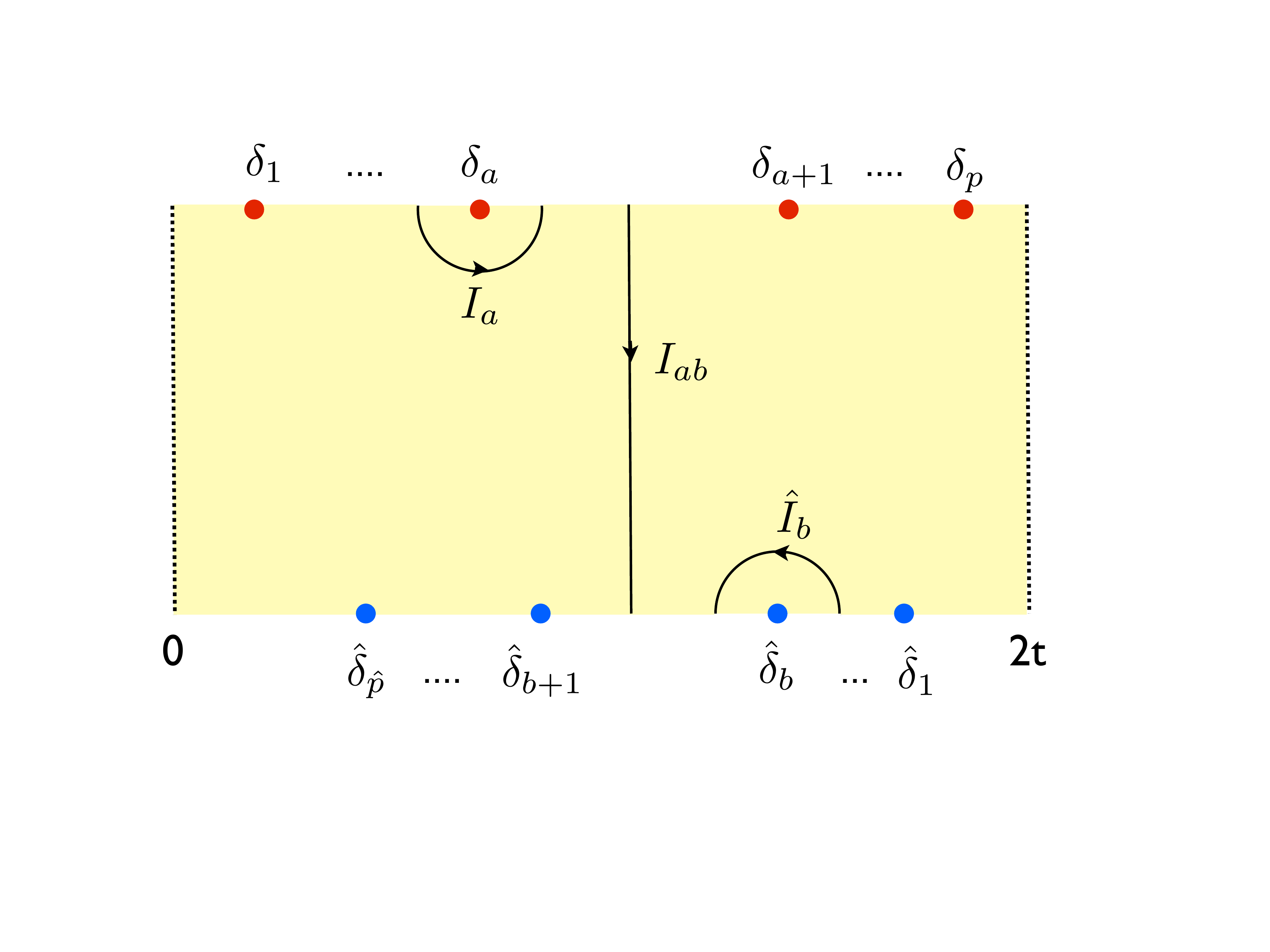}
\vskip -1.6cm
\caption{\footnotesize  The non-contractible 5-cycles in the circular-quiver geometries are fibrations of the two 2-spheres
over the  curves  shown in this figure. $\Sigma$ is an annulus, so the dotted boundaries are identified.
  }
\label{5cycles}
\end{figure}

The ten-dimensional geometries described in \S\ref{s:striptoannulus} have non-contractible three-cycles $I_a\times S^{\, 2}_2$
and $\hat I_b\times S^{\, 2}_1$, where $I_a$ is a semicircular curve around the $a$th singularity of $h_1$ on the upper annulus boundary,
and $\hat I_b$ is  a semicircle around the $b$th singularity of $h_2$ on the lower annulus boundary, see figure \ref{5cycles}.
These three-cycles are threaded respectively  by R-R and NS-NS three-form fluxes,  emanating from $\gamma_a$ D5-branes and from
$\hat\gamma_b$ NS5-branes (in units where $\alpha^\prime = 4$).

\smallskip
 In addition, these geometries have a number of non-contractible five-cycles which can support D3-brane charge.
These  are fibrations of $S_1^2$ and $S_2^2$ over the three types of open curves
$I_a$, $\hat I_b$ and $I_{ab}$ shown in  figure \ref{5cycles}.
Recalling that $S_1^2$ shrinks to a point in the lower boundary,
and $S_2^2$ shrinks to a point in the upper boundary of the annulus, one deduces that   the  topology of these 5-cycles  is as follows:

  \begin{itemize}

\item ${\cal C}_a^5 \equiv  (S_1^2\times S_1^2) \ltimes I_a$ and $\hat {\cal C}_b^5\equiv  (S_1^2\times S_1^2) \ltimes \hat I_b$
are  topologically
 $S^3 \times S^2$;

 \item ${\cal C}_{ab}^5 \equiv  (S_1^2\times S_1^2) \ltimes I_{ab}$ are   topologically    $S^5$ .

\end{itemize}

\noindent Here $I_a$ is a line segment semi-circling the $a$th singularity on the upper boundary, $\hat I_b$ likewise semicircles the $b$th singularity
on the lower boundary, and  $I_{ab}$ is a line segment
 which begins on the upper boundary of the annulus between the points $\delta_a$ and $\delta_{a+1}$
 and ends on
 the lower boundary   between the points $\hat \delta_b$ and $\hat \delta_{b+1}$.
  As shown in the figure, the orientation of the above   segments is chosen  to be counter-clockwise,  or in the case of  $I_{ab}$
from the upper annulus boundary to the lower boundary.

\smallskip

The D3-brane charges emanating from the five-brane singularities can be computed with the help of the
 general formulae  of  \S\ref{s:admissible}. Consider for example the $b$th   NS5-brane stack which corresponds
to  the $z= \hat\delta_b$  singularity on  the lower boundary of the annulus. Using $h_1^D = {\cal A}_1 + \bar {\cal A}_1$ and
the expressions   \eqref{insideNS5}, \eqref{N3hat} and
\eqref{Amany} we find
 \begin{align}
\label{D3charge1}
\hat N^{(b)}_{3} &= - \frac{2}{\pi \alpha'} \hat N^{(b)}_{5} h_1^D|_{z = {\hat \delta_b}}
\cr
&=    \hat N^{(b)}_{5} \, \sum_{a=1}^{p} N^{(a)}_{5} \, \left(  \frac{i}{2\pi}
  \ln \bigg[ \frac{\vartheta_{1}\left(\nu_{ab} \vert \qth \right)}{\vartheta_{1}\left( \bar \nu_{ab}\vert \qth \right)} \frac{\vartheta_{2}\left( \bar \nu_{ab}\vert
 \qth \right)}{\vartheta_{2}\left(\nu_{ab}\vert \qth \right)} \bigg] -   \frac{4}{\pi\alpha^\prime} \varphi_1 \right)\
\end{align}
  where
\bea\label{nuab}
   i \nu_{ab} = \frac{\delta_a - \hat \delta_b}{2 \pi} + \frac{i}{4}  \ , \qquad \qth = e^{-t}\  ,
\eea
and $\bar \nu$ is the complex conjugate of $\nu$.
Likewise, one finds for the $a$th D5-brane:
 \begin{align}
\label{D3charge2}
N^{(a)}_{3} &=  \frac{2}{\pi \alpha'}N^{(a)}_{5}  h_2^D|_{z = {\frac{i \pi}{2} + \delta_a}} \cr
&= N^{(a)}_{5} \, \sum_{b=1}^{\hat p} \hat N^{(b)}_{5} \, \left(
- \frac{i}{2\pi}  \ln \bigg[ \frac{\vartheta_{1}\left(\nu_{ab} \vert \qth\right)}{\vartheta_{1}
\left(\bar \nu_{ab}\vert  \qth  \right)} \frac{\vartheta_{2}\left( \bar \nu_{ab} \vert \qth \right)}{\vartheta_{2}\left(\nu_{ab}\vert \qth \right)} \bigg]
-   \frac{4}{\pi\alpha^\prime} \varphi_2 \right) \ ,
\end{align}
where  the arguments $\nu_{ab}$ are defined again by  \eqref{nuab}.

\smallskip

As has been discussed in  the previous section, the D3-brane (Page) charge suffers from a gauge ambiguity
which corresponds,   in the above expressions,   to the freedom in choosing the  constants $\varphi_1$ and $\varphi_2$.
 In what follows, and until  otherwise specified,
we fix  the  gauge so that the potentials are continuous inside the fundamental domain $0 \leq {\rm Re} z < 2t$, and furthermore
\begin{align}
\label{canongauge}\nonumber
C_{(2)} &= 0 \qquad \rm{in}\ \  \,[0 ,\delta_{1}] \, \ \ \ \textrm{on the upper boundary}, \\
B_{(2)} &= 0 \qquad \rm{in} \,\ \  [\hat \delta_{1},2t] \,\ \ \  \textrm{on the lower boundary}.
\end{align}
The above  choice can be motivated by considering the pinching limit
 $t \longrightarrow +\infty$ with $\delta_a-t$ and $\hat\delta_b -t$ kept fixed. In this  limit the
 geometry degenerates to that of a linear quiver, and our gauge fixing agrees with the one adopted
    in reference  \cite{ABEG:2011}.

 \smallskip
  Using the infinite-product expressions for the $\vartheta$-functions in \eqref{D3charge1} and \eqref{D3charge2},
  and fixing as  just described $\varphi_1$ and  $\varphi_2$,  leads to the expressions
 \bea\label{emanateD5}
  { N^{(a)}_{3}  }  =   N^{(a)}_{5} \sum_{b=1}^{\hat p} \hat N^{(b)}_{5}  \Big[  \sum_{n=0}^{+\infty} f(\hat \delta_b - \delta_a -2nt ) - \sum_{n=1}^{+\infty} f(-\hat \delta_b + \delta_a -2 nt)
 \Big] \ ,
 \eea
and
\bea\label{emanateNS5}
   {\hat N^{(b)}_{3} } =  \hat N^{(b)}_{5} \sum_{a=1}^{p} N^{(a)}_{5} \Big[   \sum_{n=1}^{+\infty} f(-\hat \delta_b + \delta_a - 2 n t )
-\sum_{n=0}^{+\infty} f(\hat \delta_b - \delta_a - 2 n t )
 \Big]  \ ,
\eea
   where $ { N^{(a)}_{3}  } $ is the  D3-brane charge   in the $a$th stack of D5-branes, $  {\hat N^{(b)}_{3} }$
   is  the  D3-brane charge  in  the $b$th stack of NS5-branes, and
\bea
   f(x)= \frac{2}{\pi} \arctan(e^x) \in [0, 1] \ .
\eea
It   can be easily verified that the above charges obey   the sum rule
\begin{align}\label{sumrule2}
  \sum_{a=1}^p N^{(a)}_{3}  =  - \sum_{b=1}^{\hat p} \hat N^{(b)}_{3}  \ \equiv\  N \ .
\end{align}
   In  the pinching limit, where  only the $n=0$
 terms survive in the sums,  all  the $N^{(a)}_{3}$ are positive
 and all the $ \hat N^{(b)}_{3}$ are negative numbers. For finite $t$, on the other hand,  the numbers in each set  can  have
 either sign.

  \smallskip

 Next we consider the 5-cycles ${\cal C}_{ab}^5$.  To associate to these 5-cycles  a Page charge we must decide which (gauge-variant) 5-form
 to integrate. Take for instance the 5-form $\tilde F_{(5)} \equiv  F_{(5)} +  C_{(2)}\wedge H_{(3)}$, which obeys the
  non-anomalous Bianchi identity $d\tilde F_{(5)}=0$.  This is globally defined only on the cycles ${\cal C}_{0b}^5$, since for all other
  choices of $a$,  the gauge potential $C_{(2)}$ has a Dirac string singularity at the upper endpoint of $I_{ab}$. Put differently,
  $\int \tilde F_{(5)}$
   would depend on the precise location of this upper endpoint   unless
  $C_{(2)} =0$ in the corresponding boundary segment.  By a similar reasoning one concludes that $\tilde F_{(5)}^\prime  \equiv  F_{(5)} - B_{(2)}\wedge F_{(3)}$
 should be only integrated on the 5-cycles ${\cal C}_{a0}^5$. Both of these modified 5-forms can be integrated on the
  5-cycle ${\cal C}_{00}^5$,  which is picked out by our gauge fixing \eqref{canongauge}. Furthermore,
  the  Page charge for this cycle does not depend on the choice of the modified 5-form  since
 \bea
 \int_{{\cal C}_{00}^5} (\tilde F_{(5)} - \tilde F_{(5)}^\prime) =   \int_{{\cal C}_{00}^5}  d\, ( C_{(2)}\wedge B_{(2)}) = 0\ .
 \eea

\begin{figure}
\centering
\includegraphics[height=8cm,width=12cm]{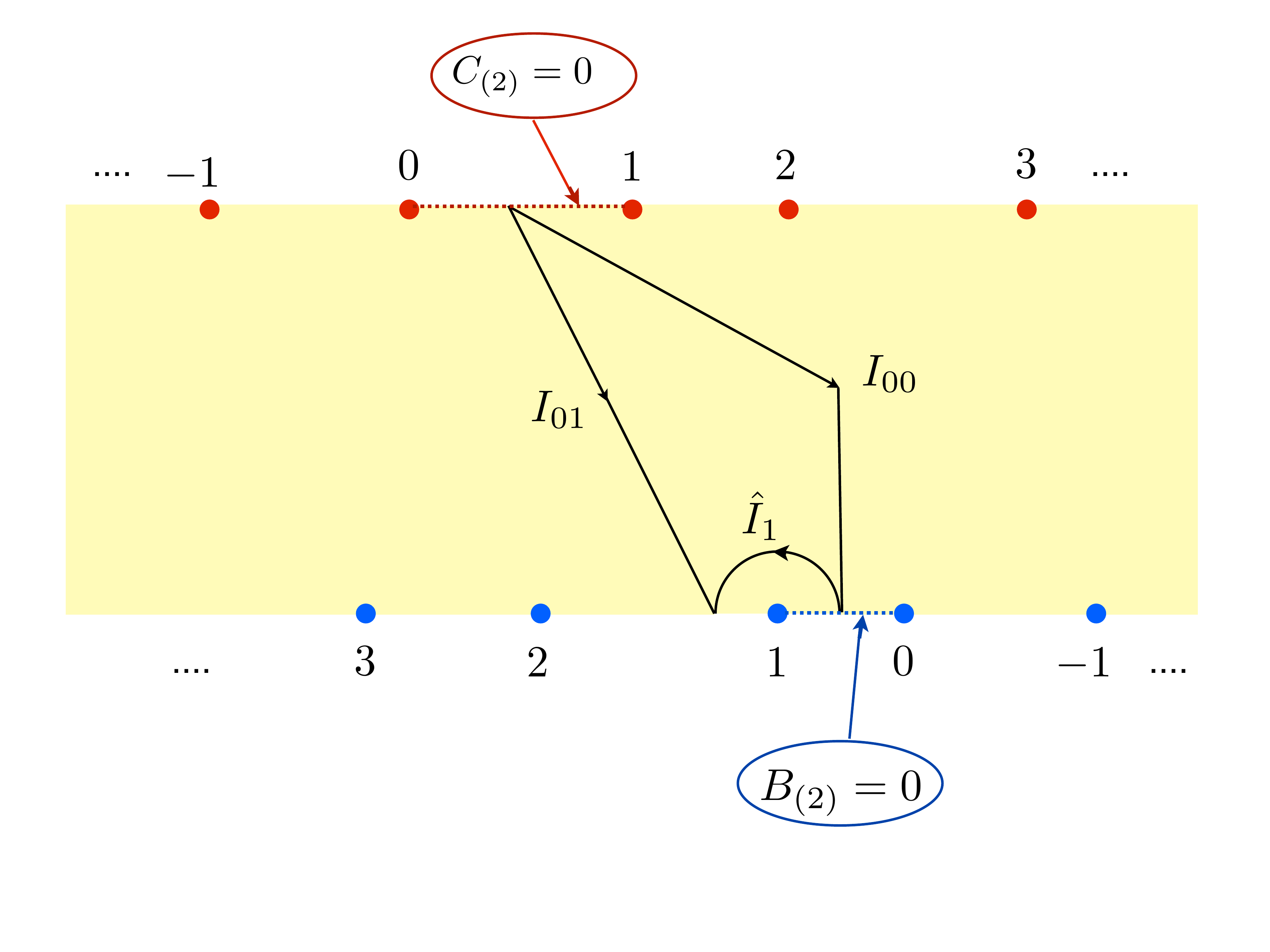}
\vskip -1.cm
\caption{\footnotesize  The infinite array of 5-brane singularities on the universal cover of the annulus.
 The D5-branes on the upper boundary are labelled from left to right, and the NS5-branes on the lower boundary
  from right to left. The choice of gauge
  determines a fundamental domain,  and a  special 5-cycle ${\cal C}_{00}^5 = I_{00}\times S_1^2\times S_2^2$.  The D3-brane charge
  supported by this cycle  is obtained  by  summing over all pairs of singularities with positive labels, and all pairs with non-positive labels, see
 equation \eqref{Mcharge2}.
  }
\label{mnem}
\end{figure}

Let us  denote  the D3-brane charge for this  special 5-cycle  by $M$. If
normalized appropriately, as in equation \eqref{insideNS5},
$M$ must be an  integer charge. We will now argue that this D3-brane charge is given by
the following expression:
 \begin{align}
\label{Mcharge2}
M  = \sum_{a,b >0}  N^{(a)}_{5} \hat N^{(b)}_{5}
  f(\hat \delta_b - \delta_a) + \sum_{a,b \leq 0}  N^{(a)}_{5} \hat N^{(b)}_{5}
  f( \delta_a- \hat \delta_b)\ ,
 \end{align}
where we  here considered  the universal cover of the annulus (i.e. the infinite strip),  and extended the range of  five-brane labels
so that   $-\infty < a< \infty$ is a label for  the infinite
array of D5-brane singularities from left to right, while  $-\infty <b < \infty$ labels  the corresponding  array of NS5-brane singularities from right to left.
Furthermore in this  notation, $\delta_{a+np} \equiv \delta_a + 2nt$ is the position of the $n$th image of the $a$th singularity on the upper
strip boundary; likewise $\hat\delta_{b+m\hat p} \equiv \hat\delta_b - 2mt$ corresponds to the $m$th  image of the $b$th singularity on the
lower strip boundary.
The expression \eqref{Mcharge2} can thus be written more explicitly as follows:
 \begin{align}
\label{Mcharge1}
M &= \sum_{a=1}^p \sum_{b=1}^{\hat p} N^{(a)}_{5} \hat N^{(b)}_{5}
 \Bigl[  \sum_{m,n=0}^{\infty}   f(\hat \delta_b - \delta_a -2nt - 2m t)
  + \sum_{m,n=1}^{\infty}   f(-\hat \delta_b + \delta_a -2nt  - 2 mt) \Bigr]
   \no \\
 &= \sum_{a=1}^{p} \sum_{b=1}^{\hat p} N^{(a)}_{5} \hat N^{(b)}_{5}  \sum_{s=1}^{ \infty} s  \Big[ f (\hat \delta_b - \delta_a -2(s-1) t )
 + f (\delta_a - \hat \delta_b -2(s +1)t ) \Big] \, .
\end{align}
 A  schematic explanation of  the above  expression is given in  Figure \ref{mnem}.
  \smallskip

To see that  \eqref{Mcharge2} is indeed right, let us consider a change of gauge which makes $B_{(2)}$ vanish
on the boundary segment  between the $b=1$ and the $b=2$ singularities.
The privileged 5-cycle is now ${\cal C}_{01}^5$, and the
corresponding D3-brane charge $M^\prime$ reads
\bea
\label{Mcharge3}
M^\prime  = \sum_{a>0 ,b > 1}  N^{(a)}_{5} \hat N^{(b)}_{5}
  f(\hat \delta_b - \delta_a) + \sum_{a  \leq 0, b\leq 1}  N^{(a)}_{5} \hat N^{(b)}_{5}
  f( \delta_a- \hat \delta_b)\ .
\eea
The difference $M^\prime - M$ is equal to $\hat N^{(3)}_1$,   the number of D3-branes in
the first NS5-brane stack, as one can check  with the help of  equation  \eqref{emanateNS5}.
This should be so  since
$I_{01} = I_{00} \oplus \hat I_1$, as illustrated in  Figure \ref{mnem}, and furthermore  the corresponding Page charges,  $M^\prime$
and $M+\hat N^{(3)}_1$,   are given by
 integrals  of  the modified form $F_{(5)} +  C_{(2)}\wedge H_{(3)}$ which does not depend on the choice of $B_{(2)}$ gauge.
\smallskip

This simple consistency check fixes almost uniquely the  expression  \eqref{Mcharge2} for the charge $M$.
To remove all doubts, we have  also verified this formula numerically.

\begin{figure}
\centering
\includegraphics[height=9cm,width=12cm]{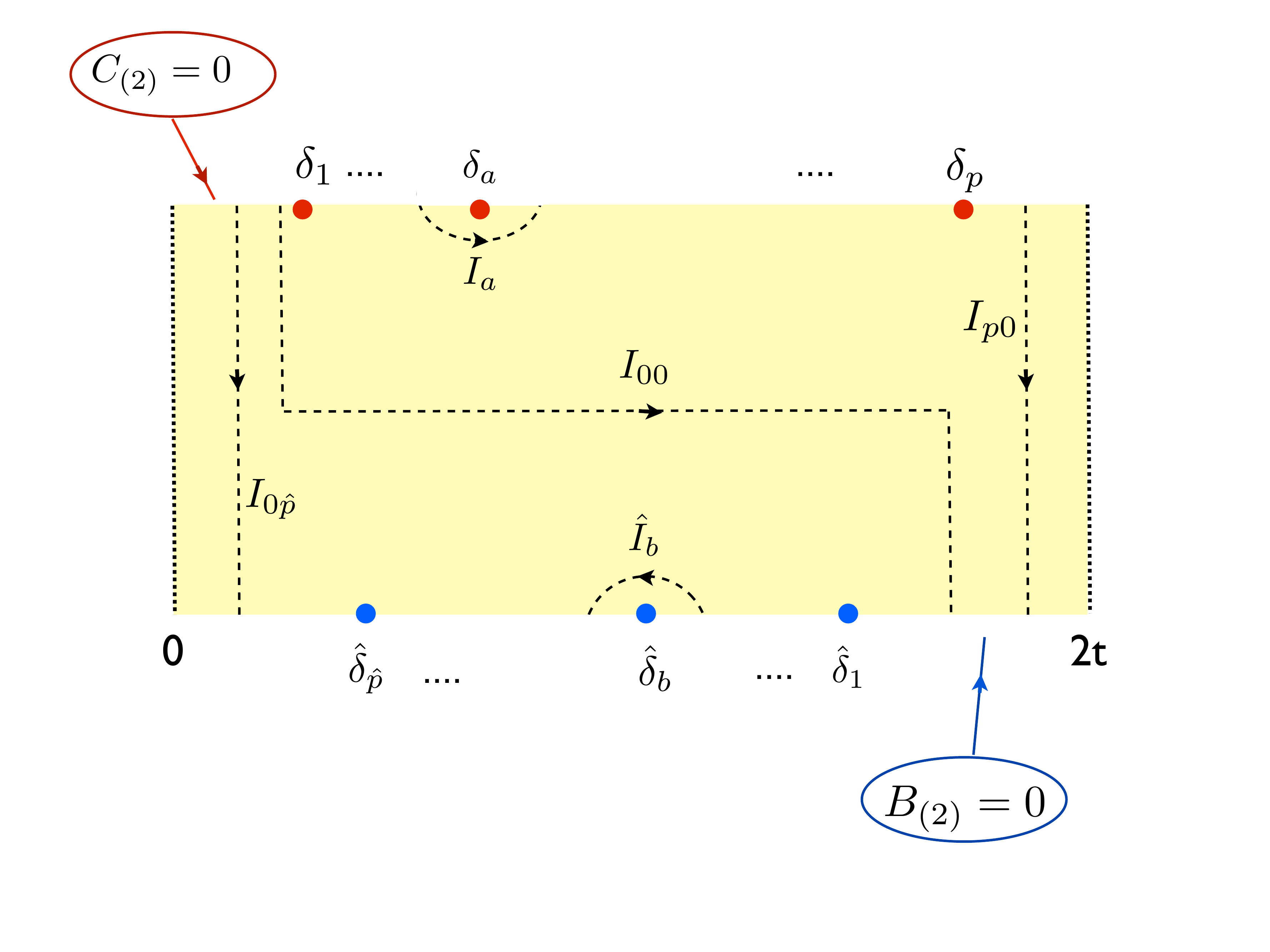}
\vskip -1cm
\caption{\footnotesize
A fundamental domain and the segments  $I_{0\hat p}$ and $I_{p0}$  which correspond to the Page charge $L$.  This
is the number of winding D3-branes, which  vanishes in the (pinching)  limit of a linear quiver.  }
\label{L}
\end{figure}

\smallskip

 Note that  by deforming the open curves as in Figure \ref{mnem}
 one can show   that there are no independent charges besides   $M$ and the Page charges
$\{ N^{(a)}_{5},  \hat N^{(b)}_{5} \}$ of the five-branes.  It will be  convenient for our purposes here to trade $M$ for   $L \equiv M - N$,
where $N$ is the total charge carried by the D5-branes, see  \eqref{sumrule2}.  The charge $L$ corresponds to the 5-form flux through the
 cycle ${\cal C}_{p0}^5$,  or equivalently the  cycle ${\cal C}_{0\hat p}^5$,  depicted  in Figure \ref{L}. Simple manipulations give
   \begin{align}
\label{Lcharge}
L  \ =\  \sum_{a=1}^p \sum_{b=1}^{\hat p} & N^{(a)}_{5} \hat N^{(b)}_{5}  \sum_{n =0}^{\infty} \sum_{m =1}^{\infty}
 \Bigl[   f(\hat \delta_b - \delta_a -2nt - 2m t)
  +  f(-\hat \delta_b + \delta_a -2nt  - 2 mt) \Bigr]
   \no \\
 &= \sum_{a=1}^{p} \sum_{b=1}^{\hat p} N^{(a)}_{5} \hat N^{(b)}_{5}  \sum_{s=1}^{ \infty} s  \Big[ f (\hat \delta_b - \delta_a -2s t )
 + f (\delta_a - \hat \delta_b -2st ) \Big] \, .
\end{align}
 Below, we will identify $L$ with the number of winding D3-branes in a circular quiver.  Consistently  with this interpretation, $L$ can be seen to  vanish  in the
 pinching limit,  $t\to\infty$  with $\delta_a - \hat\delta_b$,  for  all $a=1, \cdots   p$ and $b= 1, \cdots  \hat p$,  held finite and fixed.


\subsection{Parameter match}
\label{s:match}

Following references  \cite{ABEG:2011, Aharony:2011yc} we define the linking numbers of the fivebranes as the Page charge
per five-brane in each given stack:
\begin{align}\label{matchconstraint}
l^{(a)} \equiv \frac{N^{(a)}_{ 3}}{N^{(a)}_{ 5}} \  , \quad \hat l^{(b)} \equiv - \frac{\hat N^{(b)}_{ 3}}{\hat N^{(b)}_{ 5}} \  , \qquad {\rm with}\ \ \
\sum_{a=1}^{p} N^{(a)}_{ 5}l^{(a)} = \sum_{b=1}^{\hat p} \hat N^{(b)}_{ 5}\hat l^{(b)} = N \ .
\end{align}
We here assume that these linking numbers are integer. Strictly-speaking, Dirac's  quantization condition only requires
integrality of the total charge for each five-brane stack, so  solutions with fractional linking numbers
cannot be ruled out a priori as inconsistent.
We will  nevertheless discard this possibility,
 because we  have  no candidate SCFTs on the holographically dual side with fractional linking numbers.
 But the question  deserves   further scrutiny.

  \smallskip

Next let us identify the above liking numbers with those in the brane construction of the circular quivers
 described in    \S\ref{sec:branes},  by defining the following two partitions of $N$:
 \begin{align}\no
\rho &= \Big( \overbrace{l^{(1)},l^{(1)},..,l^{(1)}}^{N_{ 5}^{(1)}},\overbrace{l^{(2)},l^{(2)},..,l^{(2)}}^{N_{ 5}^{(2)}},...,
\overbrace{l^{(p)},l^{(p)},..,l^{(p)}}^{N_{ 5}^{(p)}} \Big) \ , \\
\hat \rho &= \Big( \overbrace{\hat l^{(1)},\hat l^{(1)},..,\hat l^{(1)}}^{\hat N_{ 5}^{(1)}},\overbrace{\hat l^{(2)},\hat l^{(2)},..,\hat l^{(2)}}^{\hat N_{ 5}^{(2)}},...,\overbrace{\hat l^{(\hat p)},\hat l^{(\hat p)},..,\hat l^{(\hat p)}}^{\hat N_{ 5}^{(\hat p)}} \Big) \ .
\end{align}
Together with the additional parameter $L$, we thus have the exact same data that was
used to  define   the circular-quiver gauge theories   $C_{\rho}^{\hat \rho}(SU(N),L)$ .
Put differently,  the supergravity parameters $\{ \gamma_a, \delta_a \}$ can be used to
vary the charges $\{ N_{ 5}^{(a)}, N_{ 3}^{(a)} \}$, the parameters  $\{ \hat\gamma_b, \hat\delta_b\} $ can be used to vary
  $\{ \hat N_{ 5}^{(b)}, \hat N_{ 3}^{(b)} \}$, and the annulus modulus $t$ controls the number $L$ of winding D3-branes.
  One of the charges is not independent because of the sum rule \eqref{matchconstraint}, but this agrees precisely with the fact that
    the supergravity solution is invariant under a common translation of all five-brane singularities on the boundary of the annulus.

  \smallskip

  The parameter counts on the supergravity and gauge-theory sides  therefore  match.
  The quiver data, on the other hand, had  to obey a set of  inequalities in order for the theory to flow to a non-trivial IR fixed point, see section \ref{sec:quivers}.
    We will  show that the same inequalities are also  obeyed on the supergravity side.

\smallskip

Note first that from the expressions \eqref{emanateD5} and \eqref{emanateNS5},  and  the fact that   $f(x)$ is  a monotonic function,
   it follows that the linking numbers of the supergravity solutions are automatically arranged  in decreasing order:
  \begin{align}
\label{ineqauto}
l^{(1)} > l^{(2)} > ... > l^{(p)} \qquad {\rm and}\qquad
\hat l^{(1)} > \hat l^{(2)} > ... > \hat l^{(\hat p)} \ .
\end{align}
From the brane-engineering point of view, it is possible to order the linking numbers by
moving five-branes of the {\it same type}   around each other  in   transverse space (this is
obvious in the configuration of Figure  \ref{separate}).
We have argued in section \S\ref{sec:quivers} that these  moves do not change the infrared limit of the theory,
up to  decoupled free sectors.  Such moves
 should thus be
indistinguishable on the supergravity side.\footnote{Unlike \eqref{orderedD5} and  \eqref{NS5ordered},
the inequalities  \eqref{ineqauto} are strict because they
refer to {\it stacks} of five-branes. Members of a given stack have identical linking numbers, so the linking numbers of individual five-branes are not decreasing
but only  non-increasing.}
 \smallskip

 \begin{figure}
\centering
\includegraphics[height=8.4cm,width=12cm]{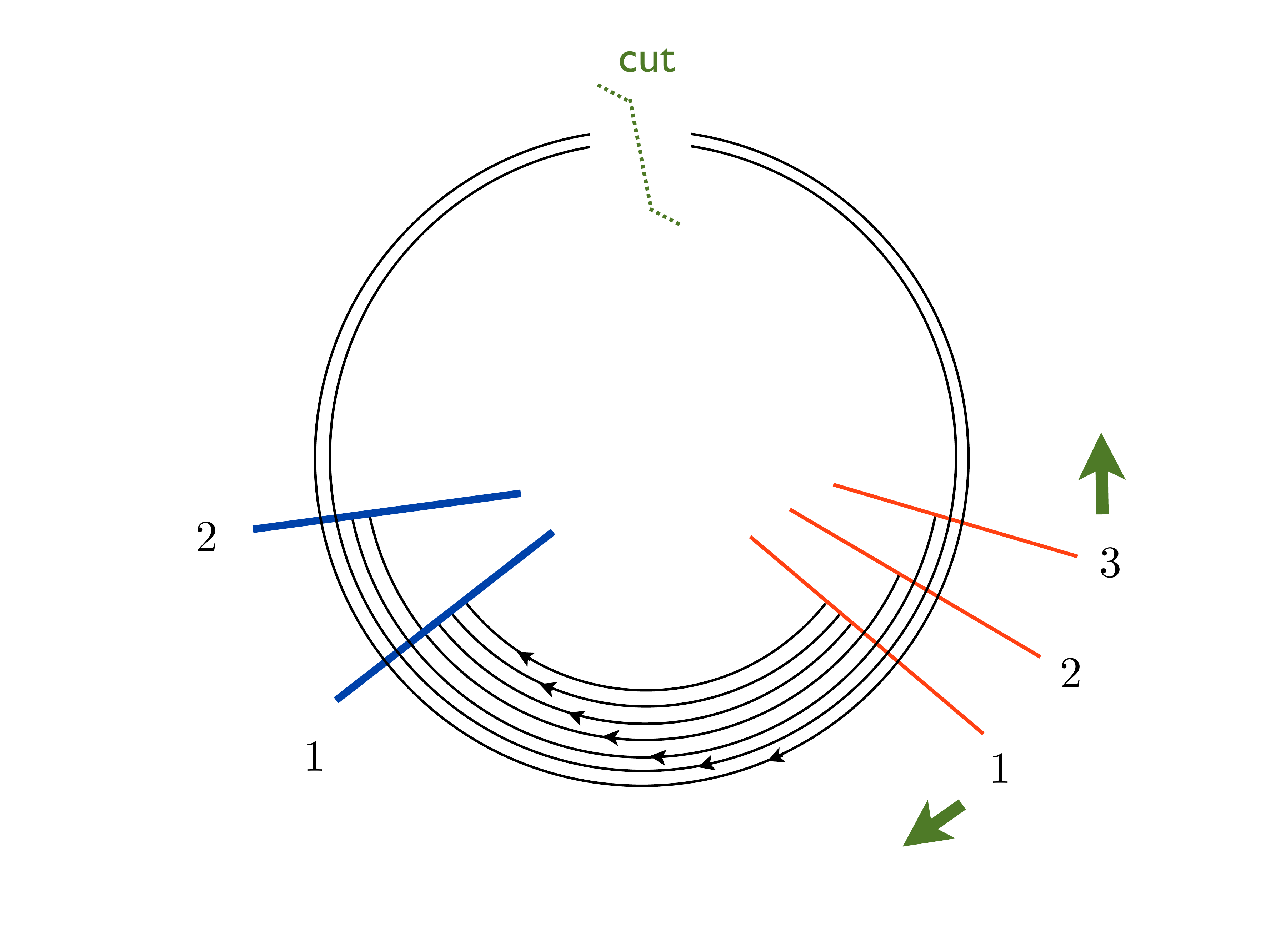}
\vskip -0.7 cm
\caption{\footnotesize  Brane engineering of a circular quiver.
 Cutting open the circle  on its high side leads to the linking-number
assignements $\rho= (3,1,1)$ and $\hat\rho = (3,2)$ with $L=2$; the  corresponding  theory is  $C_\rho^{\hat\rho}(SU(5), 2)$.
The green arrows indicate the elementary D5-brane moves described in the text. For instance,  a rotation of the $3$rd D5-brane
changes these assignments to $\rho^\prime = (3,3,1)$ and $\hat\rho^\prime = (4,3)$ with $L^\prime=3$.
  }
\label{cycle}
\end{figure}

 Besides being arranged  in decreasing order, the linking numbers of the field-theory side could
be furthermore chosen to lie in the intervals $(0, \hat k]$ and $(0, k]$, with
$k$ and $\hat k$   respectively  the total numbers of D5-branes and NS5-branes, see
  \eqref{orderedD5} and  \eqref{NS5ordered}. As was explained in  \S\ref{sec:branes},  these inequalities
  were automatic if one chose to cut open the circular chain at a link of locally-minimal rank.
  We will now explain why the same argument goes through on the supergravity side.

\smallskip

  To this end, consider the circular quiver
  of  Figure \ref{cycle}. Following the discussion in  \S\ref{sec:branes}, to
    assign   linking numbers to the five-branes we  cut   open the circular chain of D3-branes
    and then use the definitions  \eqref{defnlinking}.  Clearly, the  assignment is not unique since
    we are  free to  move one or several five-branes  around the circle before cutting  the chain.
    Let us focus,  in particular,  on the following two ``elementary"  moves:

    \begin{itemize}
    \item  Move the (right-most) $k$th D5-brane anticlockwise, which produces  the  changes
    \bea\label{move1}
    \Delta l_k = \hat k \ , \ \ \Delta \hat l_j = 1\ \ \ \forall \ j=1, \cdots \hat k \ , \ \  \Delta L = l_k\ ;
    \eea

    \item  Move the (left-most) $1$rst  D5-brane  clockwise, which leads to  the  changes
    \bea\label{move2}
    \Delta l_1 = - \hat k \ , \ \ \Delta \hat l_j = -1\ \ \ \forall \ j=1, \cdots \hat k\ , \ \  \Delta L = \hat k -  l_1\ .
    \eea

    \end{itemize}
    These formulae translate the well-known  fact that when a D5-brane crosses a NS5-brane it creates or destroys a D3-brane
 \cite{Hanany:1996ie}.\footnote{The
   linking numbers are actually  invariant under such Hanany-Witten moves, but they change in the way  indicated above
    when the D5-brane crosses the  cutting point.}
    Similar formulae clearly hold for the mirror-symmetric moves of NS5-branes.
     The main point for us here is that
    the inequalities $l_k >  0$ and $\hat k \geq   l_1$  imply that $L$ is a ``local" minimum with respect to  elementary D5-brane moves.
    Likewise,  $\hat l_{\hat k} >  0$ and $ k \geq  \hat l_1$ imply that $L$ is a  minimum with respect to  elementary NS5-brane moves.
    One can thus impose the bounds \eqref{orderedD5} and  \eqref{NS5ordered} by choosing to cut the chain  at a minimum of $L$.

 \smallskip

\begin{figure}
\centering
\includegraphics[height=8.4cm,width=12cm]{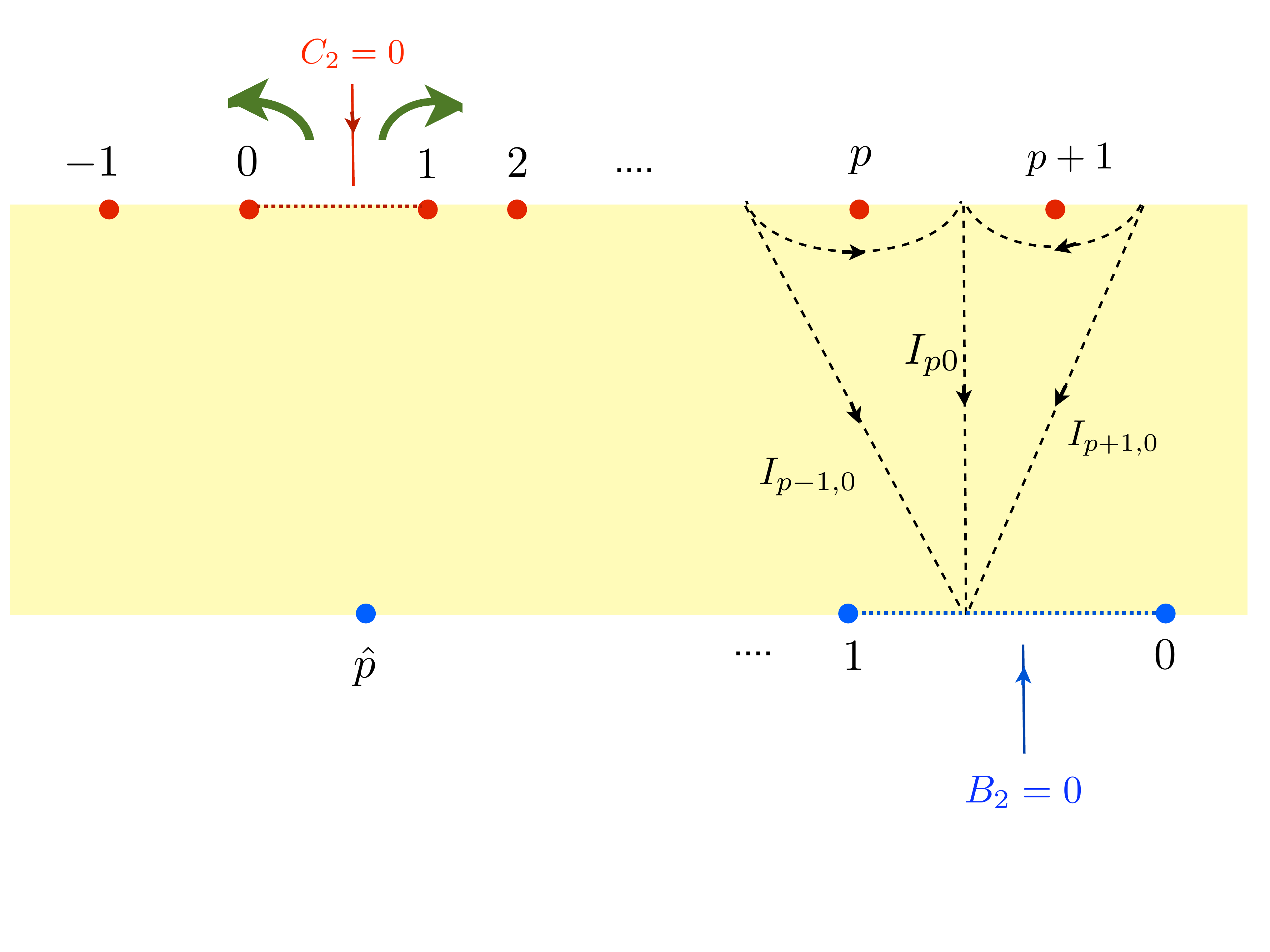}
\vskip -1.cm
\caption{\footnotesize  Global gauge transformations corresponding to the elementary D5-brane moves described in the text.
 Pushing the boundary segment on which $C_{(2)}=0$ one step to the right corresponds to moving the first stack of D5-branes
 around the circular quiver clockwise  once. Pushing this boundary segment to the left corresponds to moving the last D5-brane
 stack once in the anti-clockwise direction.
  }
\label{moves}
\end{figure}

     This same line of argument applies to the supergravity side, where five-brane moves across the cut correspond to large
     gauge transformations.  The  elementary D5-brane moves are illustrated in Figure \ref{moves}.  They correspond to
     shifting the boundary segment on which $C_{(2)}=0$ to a neighboring segment,  on the right or   left.
     Pushing for example this segment  to the left  leads to the following transformations of charges:
        \bea\label{mov1}
    \Delta l^{(p)} = \hat k \ , \ \ \Delta \hat l^{(b)} =  N_5^{(p)}\ \ \ \forall \  b   \ , \ \  \Delta L = N_5^{(p)} l^{(p)} \ .
    \eea
The last two equations follow from the expression for the linking numbers (see \S\ref{s:admissible}) and from the argument
illustrated in Figure \ref{mnem}. As for the first equation, it comes  from the fact the $p$th D5-brane stack is replaced in the fundamental
domain by the $0$th stack. On the universal cover of the annulus linking numbers
(defined as the integrals over the 5-cycles ${\cal C}^5_a$
    and $\hat C_b^5$) obey the  periodicity conditions:
      \bea
      l _{a+np} = l_a -n \hat k\ , \quad  \hat l_{b+n\hat p} = \hat l_b -nk\ .
      \eea
      Thus replacing the $p$th stack by the $0$th stack  changes the associated linking number by
      $\hat k$.\footnote{The notation in \eqref{mov1}
      is slightly abusive, because the change of the fundamental domain should be followed by a relabeling of the D5-branes. Strictly speaking
      $ \Delta l^{(p)} \equiv    l^{(1)}_{\rm new} -  l^{(p)}_{\rm old}$.
      }
    Likewise,  pushing the segment on which $C_{(2)}=0$ one step to the right
leads to the following changes:
     \bea\label{mov2}
    \Delta l^{(1)} = - \hat k \ , \ \ \Delta \hat l^{(b)} =  - N_5^{(1)}\ \ \ \forall \  b   \ , \ \  \Delta L = N_5^{(1)} (\hat k - l^{(1)}) \ .
    \eea
In this case the first D5-brane stack is replaced in the fundamental domain by the $(p+1)$th stack, as   in  Figure \ref{moves}.

\smallskip

Equations \eqref{mov1} and \eqref{mov2} are the same as  \eqref{move1} and \eqref{move2} when $N_5^{(p)} = N_5^{(1)} = 1$.
The  large gauge transformations are in this case the counterpart of the elementary D5-brane moves.
More generally,  they describe the effect of  moving the first and last {\it stacks} of D5-branes around the circular quiver.
Requiring that   $L$ be minimum under these   moves implies that $\hat k - l^{(1)} \geq 0$ and $ l^{(p)} >  0$, as advertized.\footnote{If $l^{(p)}=0$
we push  the selected line segment   to the left until the second inequality becomes strict.} Likewise one shows that
 $  k - \hat l^{(1)} \geq 0$ and $\hat  l^{(p)} >  0$, by requiring minimality under changes of the $B_{(2)}$ gauge.
 That such a minimum exists is guaranteed by the fact that  $L$  is bounded below, and goes to infinity along with the
 separation  $\delta_1 - \hat\delta_1$.
 Note that in general  there are several minima, so different
 triplets of data $(\rho, \hat\rho, L)$ may  correspond to one and the same supergravity solution.

\vskip 1mm

Having established the inequalities  \eqref{orderedD5} and  \eqref{NS5ordered}, we now need to prove the
 inequalities  \eqref{fixedpointcircA} for the associated Young tableaux. In the brane constructions of \S\ref{sec:branes}
these  inequalities guaranteed that all gauge groups have positive rank, i.e. that they are realized on D3-branes
rather than  anti-D3-branes. This is a condition for supersymmetry, so we expect it to be automatically satisfied on the supergravity side.
The proof
is straightforward but tedious, and we relegate  it to   appendix \ref{sec:ineq2}.


\section{Limiting geometries}
\label{sec:lim}

   In this section we discuss the
   solutions  described in   \S\ref{s:striptoannulus},     in regions of the  parameters
   where the annulus with the marked points on the boundary degenerates. Note that taking $(\delta_a - \delta_{a+1})\to 0$ with $t$ held  fixed
     merges the $a$th and $(a+1)$th stack of D5-branes. Modulo the subtle issue of  linking number  quantization,
   this limit is thus rather  dull. The more interesting   limits are those of an infinitely-thin or infinitely-fat annulus, $t\to\infty$ or $t\to 0$.
   We will comment on these two limits   in turn.

\subsection{Pinched annulus and  wormbranes}

     When  taking the limit $t\to\infty$  one must decide what to do with the positions,  $\{ \delta_a\} $ and $\{ \hat\delta_b\}$,  of the  five-brane
     singularities.  If the number of singularities is kept fixed   then, since $\delta_a - \delta_{a+p} = \hat\delta_b - \hat\delta_{b+\hat p} = 2t$,
         at least one of the  intervals  $\delta_a - \delta_{a+1}$ with
     $a\in [1, p]$, and at least one interval
     $\hat\delta_b - \hat\delta_{b+1}$ for some $b\in [0, \hat p -1]$ should become infinite in the limit.
       Without loss of generality, we take these divergent separations  to correspond to $a=p$ and $b=0$.
     From the expression \eqref{Lcharge} we conclude that $L \to 0$ in this limit, so that  the circular quiver  degenerates to
     a linear quiver.  If more than one interval diverges, the linear quiver breaks up further into disjoint linear quivers.

    \smallskip

\begin{figure}
\centering
\includegraphics[height=9.5cm,width=12cm]{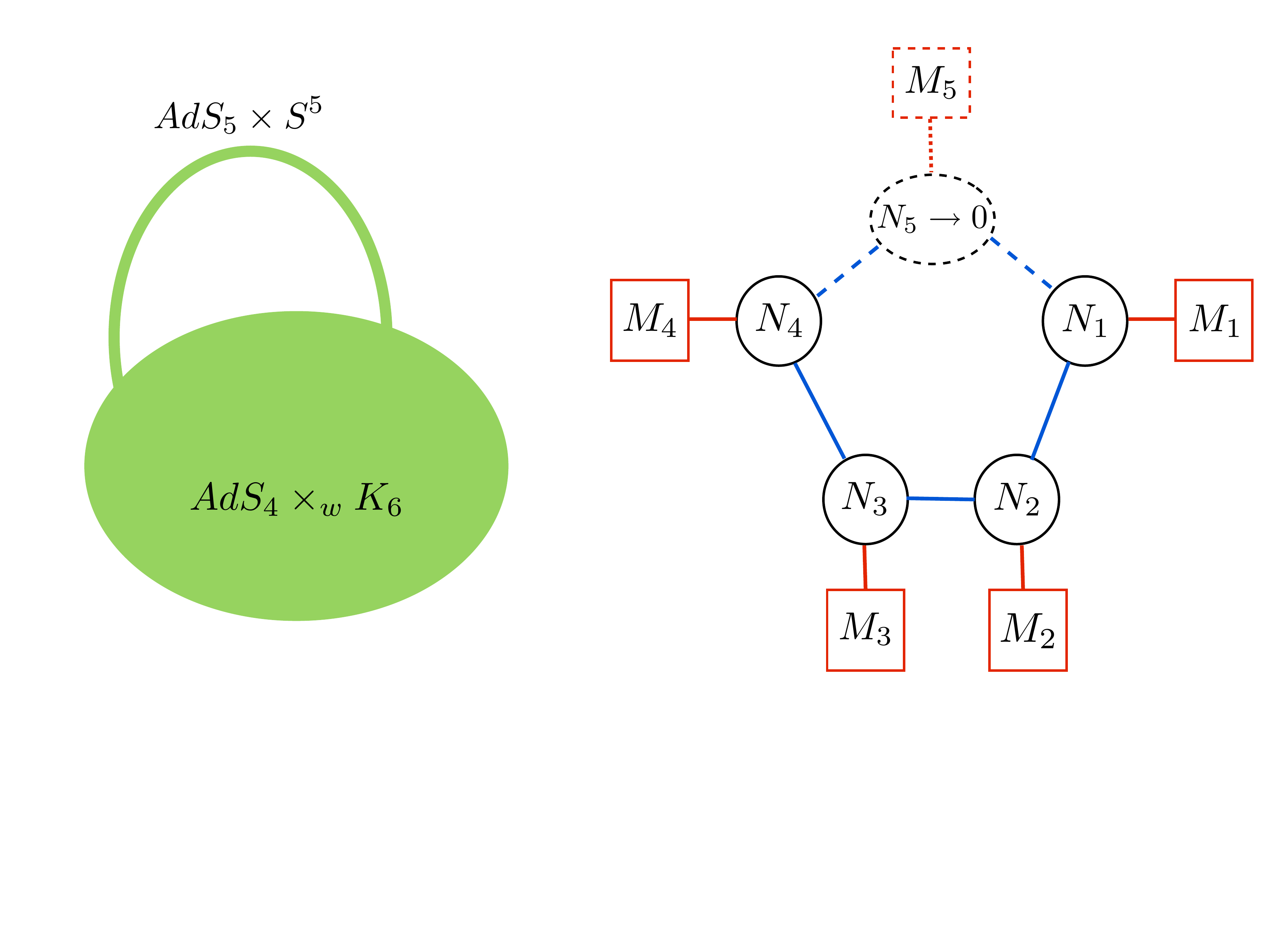}
\vskip -1.9cm
\caption{\footnotesize
Breaking up of a circular quiver as described in the text. On the left, the ten-dimensional geometry describes  a wormhole
 whose entrances are extremal D3-branes. An example of a  dual gauge theory is illustrated on the right:
 a gauge-group factor with vanishing rank opens up
 the chain into a linear quiver.
 }
\label{degenerate}
\end{figure}

         The linear-quiver geometries   were analyzed in  \cite{ABEG:2011}. They are
           warped products $AdS_4 \times_w  K_6$, where $K_6$ is a compact manifold with singularities of co-dimension four
          at the locations of  the five-branes.
          When  $L$  is  small (compared to all  other D3-brane charges) but finite,  the  geometry  describes what one  may call
          a ``worm-brane".  A schematic representation of this space-time  is given in  Figure \ref{degenerate}.
          Two highly-curved $AdS_5\times S_5$  throats\footnote{By scaling up homogeneously all charges, we can keep the curvature small enough so that
           the supergravity approximation stays valid in the $AdS_5\times S_5$ throats (though  of course not near the five-brane singularities).}
           emanate from  two  distinct  points  of  the compact  space $K_6$,  and are  joined together to form a handle.
           The wormhole entrances are three-dimensional extended objects, whence the name ``worm-brane". Note that
          (in theories without  exotic matter)
           point-like wormholes cannot be traversed and, in particular,  they cannot  provide short-cuts  for time travel (see for instance
           \cite{Morris:1988tu, Morris:1988cz,Visser:1995cc}).   Whether these conclusions can change in the case of worm-branes
            is an interesting question to which we may return in future work.

  \smallskip

          From the perspective of the  gauge theory, the pinching-limit   geometries describe circular quivers with a   gauge-group
         factor whose rank is much smaller than all other ranks.   Taking this rank formally to zero opens up the circular chain, and decouples
         the corresponding fundamental hypermultiplets, see Figure \ref{degenerate}.  If several gauge-group ranks are made to vanish,
         the linear quiver  breaks down into disjoint pieces.  In general, the limiting geometries are smooth except when one sends a  set of stacks
         of the same type  infinitely far from all  other  stacks. This corresponds in gauge theory to the decoupling of free hypermultiplets from the
         end-points of a linear quiver.  The geometry with   five-branes of only one type is singular  \cite{ABEG:2011, BE:2011}, consistently with the fact that
   free hypermultiplets should have no smooth supergravity dual.


\subsection{Large-$L$  limit and M2 branes}
\label{secLargeL}

 The second interesting limit of the circular-quiver solutions of section  \S\ref{s:striptoannulus} is the limit  $t\to 0$.
 As we will see, this is the limit of a very large number, $L$,  of winding D3-branes, in which
  the five-branes are effectively smeared, and the solution reduces
 to the near-horizon geometry of M2-branes   at a $Z_k \times Z_{\hat k}$ orbifold singularity.
   \smallskip

 To compute  the geometry in this limit
we use  the asymptotic behavior of the theta functions when $e^{i \pi   \tau} = e^{-t} \to 1$, or equivalently
$e^{i \pi   \tilde \tau} = e^{ -i \pi /  \tau}  =  e^{-\pi^2/t} \to 0$. One finds in this limit
\bea\label{thetaasymptotic}
{\vartheta_1(\nu \vert  \qth)  \over \vartheta_2(\nu \vert  \qth) } = -i {\vartheta_1(\nu \tilde \tau \vert  \tilde \qth)  \over \vartheta_4(\nu \tilde \tau \vert  \tilde \qth) } \,
= \   -2 \,    e^{-\pi^2/4t}\,  {\rm sinh} ({  \pi^2 \nu /  t })  + {\cal O}(e^{-9\pi^2 /4t})  \ ,
\eea
where the second equality  follows from the  expressions of the theta functions as infinite sums.
The formula   simplifies further if ${\rm Re} (\nu)\not= 0$, in which case the hyperbolic sine  can be replaced by
an exponential.  Inserting  \eqref{thetaasymptotic}
  in \eqref{Amany}, and recalling that
 $2\pi {\rm Re} (\nu_a) = \pi/2 - {\rm Im} (z)$  and $2\pi {\rm Re}( \hat \nu_b)= {\rm Im} (z)$, finally   gives
     \begin{align}
\label{Alimit}
{\cal A}_1  \simeq   \sum_{a =1}^p \gamma_a  \,  {\pi \over 2t}  z
 + \varphi_1   \  , \  \quad
{\cal A}_2  =  i \sum_{b=1}^{\hat p} \hat \gamma_b \,    {\pi \over 2t} (z- {i \pi\over 2} )
+ i \varphi_2  \
 ,
\end{align}
where we have absorbed some irrelevant constants
in the arbitrary phases $\varphi_1$ and $\varphi_2$.
 This approximation breaks down at a distance   $\sim t$  from the annulus boundaries, where the linear dependence
is  replaced by the rapidly-oscillating $\log  \, {\rm sinh}$ function.
 \smallskip

    The first thing to note is that, away from the boundaries, the harmonic functions depend only on three parameters: $t$ and
    the total numbers of five-branes, $k = \sum \gamma_a $ and $\hat k = \sum \hat\gamma_b$. The precise locations of
    the five-brane singularities do not matter, as if these   were smeared.  It is convenient to scale out the $t$-dependence
    by redefining the annulus coordinate as follows: $2\pi z= 2t  x +i \pi^2  y$, so that $x\in [0, 2\pi)$ and
    $y\in [0,1]$. In terms of these coordinates, the holomorphic functions read
      \begin{align}
\label{Alimitxy}
{\cal A}_1  \simeq  k  \, (   {x \over 2 } + i {\pi^2 y \over 4t} )
   \  , \  \quad
{\cal A}_2  =  i \hat k   \,    \left(   {x \over 2 } - \pi  + i {\pi^2 (y-1)  \over 4t} \right)
   \
 ,
\end{align}
where we have here chosen $\varphi_1$ and $\varphi_2$  so as to impose the canonical gauge condition \eqref{canongauge}.
Inserting these functions in the general form of the solution, see  \S\ref{s:localsolutions},  gives  the
   Einstein-frame metric  (we recall that $\alpha^\prime =4$):
   \begin{align}
ds^2 =  R^2 g(y)^{1\over 4}  \left[   ds^2_{AdS_4} +  y\,  ds^2_{S^2_1} +   (1-y) ds^2_{S^2_2} \right] +
 R^2 g(y)^{-{3\over  4} } \left[{4t^2\over \pi^4}  dx^2 +  dy^2 \right]  \ , \no
\end{align}
\vskip -8mm
 \bea
 {\rm with} \quad R^4 =  \pi^4  \, { {k \hat k}\over t^2}  \ , \qquad {\rm and}\quad g(y) = y(1-y)\ .
 \label{IIBlargeL1}
 \eea
 Furthermore, the dilaton  and the non-vanishing gauge fields read:
 \begin{align}
e^{2 \phi} =  \frac{\hat k}{k} \sqrt{ \frac{1 - y}{y}}  \ ,   \qquad
C_{(4)} = R^4 \left( {6t x \over \pi^2}  \,  \omega^{0123} + y^2(y- {3\over 2})\,  \omega^{4567} \right) \ , \no
\end{align}
 \begin{align}
B_{(2)}  = 2\hat k(2\pi -x)\,  \omega^{45} \ ,
  \qquad
C_{(2)} =  -2kx\,  \omega^{67}  \  .
\label{IIBlargeL2}
\end{align}
 As already noted, this solution only depends on three integer parameters: the numbers $k$ and $\hat k$ of five-branes,
 and the modulus $t$ of the annulus which can be traded for the number of winding D3-branes via the formula \eqref{Lcharge},
 \begin{align}
L\, =\,  \frac{k \hat k}{2t^2} \, \int_{0}^{+\infty} du \, u\,  \frac{2}{\pi} \arctan(e^{-u}) \, = \, \frac{\pi^2 }{32  }\, \frac{k \hat k}{t^2} \ .
\end{align}
One may also compare \eqref{Lcharge} to the formula  \eqref{Mcharge1} for the charge $M= L+N$, where $N$ gives the
number of D3-branes emanating from five-branes. Since the summands in these two expressions differ by terms of order $t^2$,
we conclude that $N \sim k\hat k$ as $t\to 0$. Thus   the
 number of winding D3-branes  far exceeds, in this limit,  the number of D3-branes that  end on the  five-branes.
 \smallskip

 Not surprisingly, after having effectively smeared the five-branes, the  solution has a Killing isometry under translations of  the coordinate $x$. To be sure,  $x$
  enters in the expressions for $B_{(2)}$ and $C_{(2)}$  but this is a gauge artifact since the 3-form field strengths
 are $x$-independent.  One may thus T-dualize the circle parametrized by  $x$,  using Buscher's rules
 \cite{Buscher:1987qj}, to  find a solution of type-IIA supergravity. This can be then  lifted to eleven dimensions --   the  details of these
  calculations  are given  in appendix \ref{Tduality}.
 The final result for the eleven-dimensional metric is
 \begin{align}
\label{AdS4S7}
ds_{{\rm M-theory}}^2 &=  {\bar R^2}  ds^2_{AdS_4} + \bar R^2 \Big[ 4 d\alpha^2 + \sin^2 \alpha \, ds^2_{S^3/\mathbb Z_{\hat k}} + \cos^2 \alpha \, ds^2_{S^3/\mathbb Z_k} \Big] \ , \no\\
ds^2_{S^3/\mathbb Z_{\hat k}} &=   d\theta_1^2 + d\phi_1^2 + 4 dx^2 - 4 \cos \theta_1 dx d\phi_1    \ , \no\\
ds^2_{S^3/\mathbb Z_{k}} &=  d\theta_2^2 + d\phi_2^2 + 4 dv^2 - 4 \cos \theta_2 dv  d\phi_2   \ ,
\end{align}
where $x$ and $v$ are angle coordinates with periodicities $x= x+ 2\pi/ \hat k$ and $v= v+ 2\pi/ k$,  while the radius of $AdS_4$ is
 $\bar R^2 = (2^{5} \pi^2 k\hat k L)^{1/3}$.

  This is the metric of $AdS^4 \times S^7/ (\mathbb{Z}_k\times  \mathbb{Z}_{\hat k})$   with the two orbifolds acting on the two 3-spheres  in $S^7$.
  The solution furthermore carries $L$ units of four-form flux. It  can be recognized as the near-horizon geometry of $L$ M2-branes sitting
  at the fixed point of the orbifold
  $(\mathbb C^2/\mathbb Z_{\hat k} )\times (\mathbb C^2/\mathbb Z_k )$, where the orbifold identifications are
  $$
  (z_1, \bar z_2) = e^{2i\pi/\hat k} (z_1, \bar z_2)\quad {\rm and}\quad
   (z_3, \bar z_4) = e^{2i\pi/k} (z_3, \bar z_4)\ .
  $$
  Note that the two-forms $B_{(2)}$ and $C_{(2)}$ become,   after the T-duality and the lift, part of the metric. This is in
  line with the fact that D5-branes transform to Kaluza-Klein monopoles, while T-duality in a transverse dimension maps the NS5-branes
  to ALE spaces with  singularities of $A_n$ type \cite{Ooguri:1995wj,Tong:2002rq}.

  \smallskip

   The superconformal field theories that are dual to M theory on  $AdS^4 \times S^7/ (\mathbb{Z}_k\times  \mathbb{Z}_{\hat k})$ are
 close relatives of the  ABJM theory \cite{Aharony:2008ug,Aharony:2008gk} that  have been analyzed by many authors,
  see for example  \cite{Hosomichi:2008jd,Benna:2008zy, Imamura:2008nn,Imamura:2008ji,Herzog:2010hf,Dey:2011pt}. We  will discuss them in more detail in the following
  section. Let us here only quote their free energy  $F = - \log |Z|$ on the 3-sphere.  Using the general formula of \cite{Herzog:2010hf} one finds
  \begin{align}\label{herzogfree}
F \, =\,  L^{3/2}\sqrt{\frac{2 \pi^6}{27 \, {\rm Vol}_7}}  \, =  \,   \frac{\pi }{3} \sqrt{2 k  \hat k} \, L^{3/2}  \ ,
\end{align}
where ${\rm Vol}_7$ is the volume of the compact (Sasaki-Einstein)  manifold  whose metric is normalized so that $R_{ij} = 6 g_{ij}$.
In the case at hand, this is the unit-radius seven sphere with orbifold identifications, so that
  ${\rm Vol}_7 =  {\pi^4}/{3k \hat k}$.
  \smallskip

  As a check of our formulae, we may  compute this free energy   on the type-IIB side.
   Following \cite{Assel:2012cp},
   the on-shell IIB action can be computed via a consistent truncation to pure four-dimensional
    gravity with unit $AdS_4$ metric  multiplied by a 6d volume factor.    The explicit formula derived  in \cite{Assel:2012cp}   is
\begin{align}
\label{acteff}
S_{\rm IIB} = -\frac{1}{(2\pi)^7 (\alpha')^4} \left( \frac{4}{3} \pi^2 \right) (-6)  {\rm vol}_6 \ ,
\end{align}
where for the solutions of interest
\begin{align}
{\rm vol}_6 = -16 (4 \pi)^2 t  \int_0^{2\pi}\hskip -1mm dx \int_0^1\hskip -1mm  dy \, h_1 h_2 \, \p_z \p_{\bar z} (h_1 h_2) \ .
\label{vol6}
\end{align}
 Plugging in the harmonic functions $h_1 = -i( {\cal A}_1 - \bar  {\cal A}_1 )$ and $h_2 = {\cal A}_2 + \bar  {\cal A}_2$, and performing
 the integrals  gives
\beq
S_{\rm IIB} = \frac{\pi^4}{48} \frac{k^2 \hat k^2}{(2t)^3} = \frac{\pi}{3}\sqrt{2 k \hat k}\,  L^{3/2}
\label{SIIB}
\eeq
   in perfect agreement with the   result of M theory.
\smallskip

To summarize, we have shown here  that when $L$ is large our solutions approach smeared backgrounds
dual to M theory on $AdS^4 \times S^7/ (\mathbb{Z}_k\times  \mathbb{Z}_{\hat k})$.
 In this limit  the information about the positions of the
  five-branes is lost, and following \cite{Gregory:1997te,Harvey:2005ab} its reinstatement would require non-trivial backgrounds for the
   wrapped-membrane field.  The essential topological features of the background can be,  however,  
   in principle encoded more simply,  as  3-form torsion of the M-theory orbifold
\cite{Imamura:2008ji, Witten:2009xu,Aharony:2009fc,Dey:2011pt}.  Note that,  contrary to the $N>4$  cases studied in \cite{Aharony:2008gk},
 the orbifolds considered here are not freely-acting on S$\hskip -0.5mm \,^7$, and one would need  to resolve their singularities. 
It would be interesting to work out the precise match 
 of the torsion with the quiver data, and see how the constraints  on the triplet $(\rho, \hat\rho, L)$ arise from  the M-theory side.

\section{ $SL(2, \mathbb{Q})$ and orbifold  equivalences}
\label{sec:sl2r}
\smallskip

Classical type-IIB supergravity has a continuous global $SL(2,\mathbb{R})$ symmetry
\cite{Schwarz:1983wa} which transforms the axion-dilaton field,  $S = \chi + i e^{-2 \phi}$,  and the NS-NS and R-R  three-form  field strengths  as follows:
\begin{align}
\label{sl2rtrans}
 S^\prime =  {a S +  b\over  c S + d}\ , \qquad
  \left(   \begin{array}{c}
\, \, H_{(3)}^\prime  \\
F_{(3)}^\prime
 \end{array}
  \right) =
   \left(   \begin{array}{cc}
 d & -c \\ - b & a
  \end{array}
  \right)
  \left(   \begin{array}{c}
\, H_{(3)}  \\
F_{(3)}
 \end{array}
  \right) \ ,
 \end{align}
where $a,b,c,d$ are real numbers with $ad-bc=1$. The transformations leave invariant the
Einstein-frame metric, and the gauge-invariant five-form field strength.
\smallskip

 As is well known, only the integer subgroup $SL(2,\mathbb{Z})$ is a  symmetry of the full string theory \cite{Hull:1994ys}, whereas continuous transformations can be
 used to generate  inequivalent  solutions.
    The authors of  \cite{DEG1,DEG2} have indeed used such   $SL(2,\mathbb{R})$ transformations
     to bring the general  solution  of the Killing-spinor equations
  to the local  form given in  \S\ref{s:localsolutions}.  Conversely, acting with  the transformations \eqref{sl2rtrans}  generates
  new solutions from the ones of  section 3,  with singularities that  correspond to general $(p,q)$  five-branes.\footnote{The symbol $p$, which usually
   indicates the NS5-brane charge of a $(p,q)$ five-brane, was   also used for the
   number of  five-brane singularities in the upper boundary of $\Sigma$. We hope the context will make it clear in which sense this symbol is being used.
   The same comment applies to the lower-case Latin letters  which   label the five-brane stacks;   following standard notation we also use them for the
    elements of the $SL(2,\mathbb{R})$ matrix. }
 We will now discuss  briefly  these
  new solutions.



\subsection{Solutions with $(p,q)$ five-branes}

   The solutions given by the harmonic functions \eqref{harm1} or \eqref{hmany}
      have  singularities on the upper boundary of the infinite strip or the annulus that correspond to D5-branes,  and
      singularities   on the lower boundary that correspond to NS5-branes. The  charges
    are,
    respectively,  $\gamma^{(e)}$ and  $ \hat\gamma^{(f)}$ for the  stacks labeled by $e$ and $f$.   Since the metric
    is invariant, the $ SL(2, \mathbb{R})$ transformations
    do not change the  positions and the  total number of five-brane stacks. It transforms, however, their charges as follows
       \beq
  \gamma^{(e)} (0, 1) \to   \gamma^{(e)} (-c, a)\qquad {\rm and}\qquad
   \hat\gamma^{(f)} (1,0) \to   \hat\gamma^{(f)}(d, -b )\ ,
   \eeq
where the NS5-brane and D5-brane charges are arranged as usual in a doublet.
    Let us write $(-c, a) = w (p, q)$ and $(d, -b) = \hat w (\hat p, \hat q)$, where $p, q$ and $\hat p, \hat q$ are pairs of relatively-prime
    integers. Charge quantization requires that
   \bea\label{D5NS5prime}
   N_5^{(e)\, \prime} =  w \gamma^{(e)}\quad {\rm  and } \quad    \hat N_5^{(f)\, \prime}  = \hat w \hat\gamma^{(f)}
  \eea
     be integer for all
    $e$ and $f$. Since the $\gamma$'s and $\hat\gamma$'s  are arbitrary parameters, this can always be arranged to get
    any desired number of five-branes in each  stack. The only conditions are that all five-branes on the upper boundary are of  the
    same kind, including the sign, that the same is true for all five-branes on the lower boundary, and that furthermore these two
    kinds are different, $p\hat q -  q\hat p \not=  0$. This last constraint follows from  the fact that the $SL(2, \mathbb{R})$ matrix has determinant one.

 \smallskip

 It should be stressed that the  $SL(2, \mathbb{R})$ transformations take us, in general, outside the ansatz of \S\ref{s:localsolutions};
 they generate in particular a non-vanishing R-R axion field.
 The only exception is  S-duality
  ($S \to -1/S$) which interchanges the harmonic functions, and acts as mirror symmetry on the holographically-dual  SCFT.

 \smallskip

   Consider next the D3-brane charges. These are not affected by  $SL(2, \mathbb{R})$   transformations, provided
   one transforms the gauge choice   covariantly. More explicitly, let us consider the D3-brane charge of the $(p,q)$ singularities in the upper
   boundary. The 2-form that has no component on $S_2^2$ [and is therefore well defined on a patch containing the whole upper boundary where this 2-sphere shrinks]  is $ B_{(2)}= a  B_{(2)}^\prime + c C_{(2)}^\prime$.
   The D3-brane charge of a $(p,q)$ five-brane stack  is given therefore by the integral of the following closed five-form
   \bea
   N_3^{(e)\, \prime} = {1\over (4\pi\alpha^\prime)^2}  \int_{{\cal C}_e^5} \left[ F_{(5)} -  (a  B_{(2)}^\prime + c C_{(2)}^\prime)\wedge (b H_{(3)}^\prime + d F_{(3)}^\prime)  \right]\ ,
   \eea
   with the gauge choice $a  B_{(2)}^\prime + c C_{(2)}^\prime=0$ in the lower-boundary segment
   $[\hat \delta_1, 2t]$.  This is identical to the integral   in the non-transformed solution,  so that
   \bea\label{D3prime}
   N_3^{(e)\, \prime} =     \gamma_e
 \sum_{f=1}^{\hat p}  \hat \gamma_f  \, \left(
- \frac{i}{2\pi}  \ln \bigg[ \frac{\vartheta_{1}\left(\nu_{ef} \vert \qth\right)}{\vartheta_{1}
\left(\bar \nu_{ef}\vert  \qth  \right)} \frac{\vartheta_{2}\left( \bar \nu_{ef} \vert \qth \right)}{\vartheta_{2}\left(\nu_{ef}\vert \qth \right)} \bigg]
-   \frac{4}{\pi\alpha^\prime} \varphi_2 \right) \ ,
   \eea
which is the same result as   \eqref{D3charge2}.   The quantization of this charge puts the same constraints on the continuous parameters
 as in the untransformed solution. This is not however  the case  for the quantization of individual linking numbers,  since the number $w \gamma^{(e)}$ of
 $(p,q)$ five-branes    depends,  via  $w$,  on the $SL(2, \mathbb{R})$   transformation.
     \smallskip

 Among all the solutions discussed here, those  related by $SL(2,\mathbb{Z})$ transformations are physically equivalent \cite{Hull:1994ys}.
 To characterize inequivalent solutions, we may  perform a $SL(2,\mathbb{Z})$ transformation that maps $(\hat p, \hat q)$ to $(1,0)$, so that
 the singularities on the lower boundary correspond to pure NS5-branes. Using then the
  shift symmetry $(p, q) \to  (p + ql, q)$, which leaves invariant the NS5 branes,  we can   bring the  second type
of five-branes to a canonical  form $(p, q)$ with  $0\leq p < \vert q\vert$.
 The $SL(2, \mathbb{R})$  transformation from the ansatz of  \S\ref{s:localsolutions}
 to the above  canonical form of the general  solution is effected by the following matrix
\bea
  \left(   \begin{array}{cc}
 \hat w  & -wp  \\ 0  & wq
  \end{array}
  \right)\quad {\rm with} \quad w\hat w q = 1\ .
\eea
 Multiplying \eqref{D3prime} with $w\hat w q$,    using   \eqref{D5NS5prime} and the infinite-product expressions for
 the $\vartheta$-functions gives
 \bea\label{emanateD5p}
  { N^{(a)\, \prime}_{3}  }  =
   q N^{(a)\, \prime}_{5} \sum_{b=1}^{\hat p} \hat N^{(b)\, \prime}_{5}  \Big[  \sum_{n=0}^{+\infty} f(\hat \delta_b - \delta_a -2nt ) - \sum_{n=1}^{+\infty} f(-\hat \delta_b + \delta_a -2 nt)
 \Big] \ ,
 \eea
and likewise
\bea\label{emanateNS5p}
   {\hat N^{(b)\, \prime}_{3} } =  q \hat N^{(b)\, \prime}_{5} \sum_{a=1}^{p} N^{(a)\, \prime}_{5} \Big[   \sum_{n=1}^{+\infty} f(-\hat \delta_b + \delta_a - 2 n t )
-\sum_{n=0}^{+\infty} f(\hat \delta_b - \delta_a - 2 n t )
 \Big]  \ .
\eea
A similar expression can be written for the winding charge $L^\prime$. Integrality of the linking numbers,  $ N^{(a)\, \prime}_{3} /
   N^{(a)\, \prime}_{5} $ and $ \hat N^{(b)\, \prime}_{3} /  \hat N^{(b)\, \prime}_{5}$,  constraints the modulus $t$ and  the positions of the singularities on
   the annulus boundary. When  $q\not= 1$ there are more allowed choices than in the case of pure D5-branes  and NS5-branes.

  \smallskip

   The
   charges \eqref{emanateD5p} and \eqref{emanateNS5p} obey the sum rule \eqref{sumrule2}, and they thus still define two partitions $\rho$ and $\hat\rho$
   of some integer $N$. Furthermore, these partitions still satisfy the basic   inequalities  \eqref{constraintss}. In general, we have no clear argument for why these
   conditions should be obeyed on the gauge-theory side. Indeed, for arbitrary $(p,q)$ there is no known
  Lagrangian description of the  field theory (we refer the reader to section 8 of \cite{Gaiotto:2008sa} for more details).
   Such a description only exists for the configurations involving $(1,k)$ 5-branes  \cite{Gaiotto:2008sa,Gaiotto:2008sd,Hosomichi:2008jd} :
  the  $U(N)$ gauge theory  living on a stack of $N$ D3-branes has level $k$ (or $-k$) Chern-Simons terms
  depending on whether the  D3-branes end on the $(1,k)$ five-brane from  the left (or the right).


\subsection{Orbifold equivalences and free energies}

    An interesting  corollary of the holographic dualities that we have presented  in this work  is the orbifold equivalence
of different    $\N=4$   superconformal      gauge theories in three dimensions.  Orbifold  equivalences translate the fact
that quantities which  are sensitive only to the untwisted sector,  are not affected by an orbifold operation \cite{Kachru:1998ys,
 Lawrence:1998ja, Bershadsky:1998mb}. Such quantities usually exist in the classical limit of string theory,  and in the large-$N_c$ (planar)
limit of gauge theories.\footnote{For a discussion  of when the equivalence is  exact  see \cite{Armoni:2004ub, Kovtun:2004bz}.}
An example of orbifold equivalence  for the ABJM  theory was analyzed
 recently in \cite{Hanada:2011zx, Hanada:2011yz}. Here we will   present some more examples relating  $\N=4$ circular-quiver theories.

\smallskip

  The   theories that we will discuss are related by  $SL(2,\mathbb{R})$ transformations with rational entries, i.e. by elements of
  $SL(2,\mathbb{Q})$.  Two  theories related in this way are clearly  equivalent in the limit where the supergravity approximation is valid, since
$SL(2,\mathbb{R})$ is a symmetry of  type-IIB supergravity. A similar rational extension of the perturbative T-duality group
$O(d,d,\mathbb{Z})$ has been discussed recently in  \cite{Bachas:2012bj}. As explained in this reference,
$O(d,d,\mathbb{Q})$ transformations can be seen as
 orbifold operations\footnote{If $x=x+2\pi$ parametrizes the orbits of a Killing isometry, then the orbifold identification $x \equiv x + 2\pi \kappa$ for rational $\kappa$
 changes the radius of the Killing orbits,  and can thus be viewed as a $O(1,1,\mathbb{Q})$ transformation. Rationality ensures that the orbifold
 group is of finite order. These observations  generalize  to
 $O(d,d,\mathbb{Q})$.}
  which lead to equivalences that are valid at any order  in
the $\alpha^\prime$ expansion. One may likewise  view the $SL(2,\mathbb{Q})$ transformations as orbifold operations on the F-theory torus.
This formal  interpretation does not, however, imply in any obvious way that the equivalences presented here extend beyond the
supergravity approximation.

\smallskip

The simplest example of  ``equivalent"  theories are theories related by the $SL(2,\mathbb{Q})$ transformation
\bea
   \left(   \begin{array}{cc}
  r/s   & 0   \\ 0  &  s/r
  \end{array}
  \right)\qquad {\rm with}\quad (r,s)  \quad {\rm relatively\, prime \ integers}\, . \no
\eea
Such diagonal transformations do not modify the five-brane types, but they change
  the  number of five-branes in each stack. They also transform  their linking numbers,  so as to leave unchanged the  D3-brane charges:
\bea
\hat N^{(b)\, \prime}_{5} = {r\over s}\,  \hat N^{(b) }_{5}\, , \quad
\hat l_j^{\, \prime} = {s\over r}\, \hat l_j\,  , \quad\quad
N^{(a)\, \prime}_{5} = {s\over r}\,   N^{(a) }_{5}\, , \quad
l_i^{\, \prime} = {r\over s}\,  l_i\, .
\eea
Consistency with charge quantization requires of course that    $ \hat N^{(b) }_{5}$ and  $l_i$ be multiples of $s$, and
  that $N^{(a) }_{5}$ and $\hat l_j$  be  multiples of $r$. Since the number,  $L$,  of winding D3-branes does not transform,  whereas
  \bea
k \to {s\over r}\, k\   \qquad {\rm and}\quad  \hat k \to {r\over s}\, \hat k\ ,
\eea
  the supergravity  free energy
\eqref{herzogfree} is invariant, as expected.
Note that even these simple $SL(2,\mathbb{Q})$ transformations act highly non-trivially on the field theory side.
For instance, the number of gauge-group factors is multiplied by $r/s$, while the total number of fundamental hypermultiplets
is  multiplied by $s/r$.

  \smallskip

\begin{figure}
\centering
\includegraphics[height=9.6cm,width=15cm]{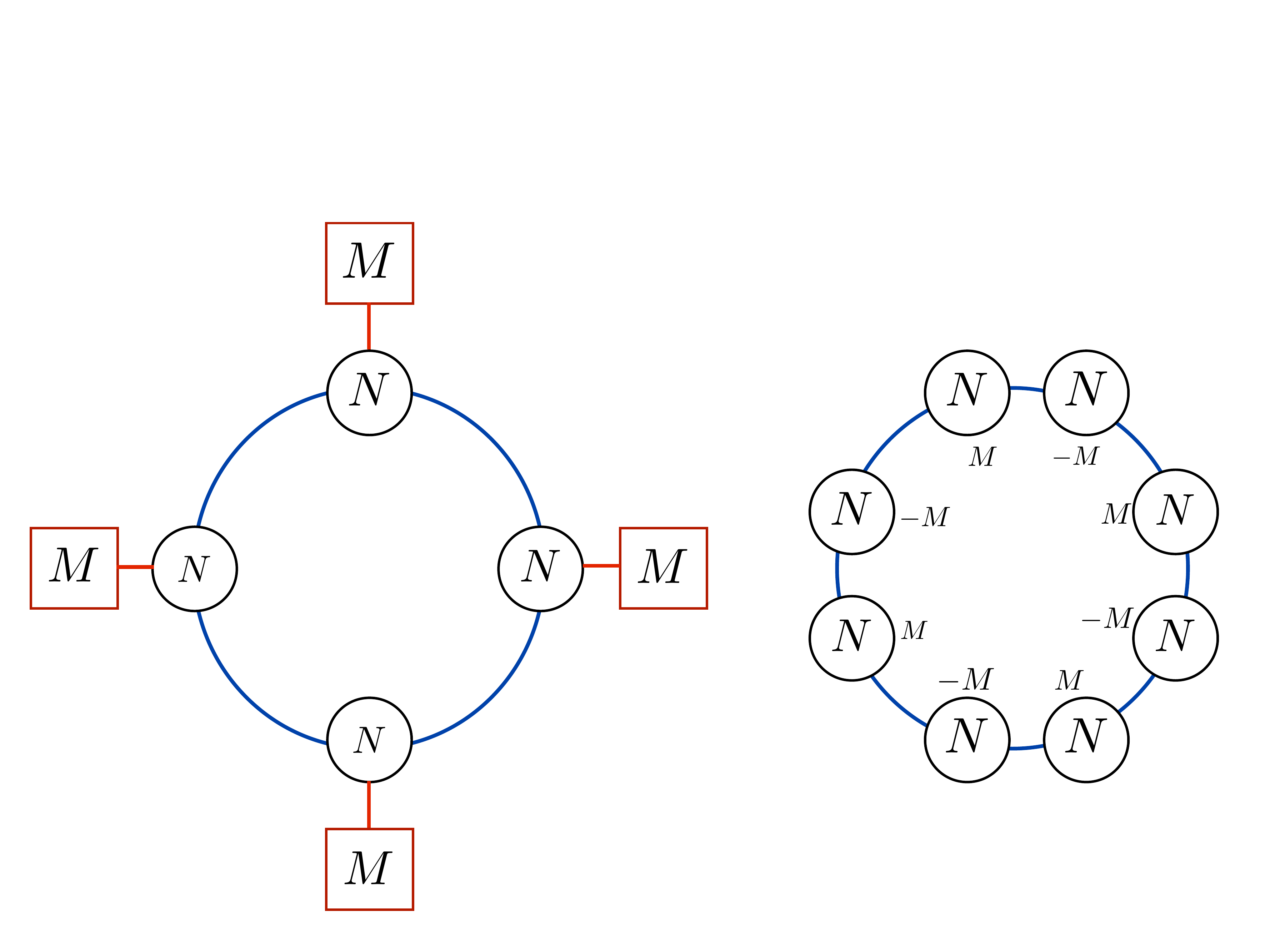}
\caption{ \footnotesize
The two circular-quiver gauge theories related by the $SL(2,\mathbb{Q})$ transformation \eqref{SL2Q}.
  The theory on the right is obtained from the one on the left by doubling the number of gauge-group factors, removing the fundamental hypermultiplets
  and adding   Chern-Simons  terms with alternating sign.}
\label{slQ}
\end{figure}

As another example of $SL(2,\mathbb{Q})$ equivalence, we consider the transformation
\bea\label{SL2Q}
   \left(   \begin{array}{cc}
  1  &    M^{-1}  \\  0   &  1
  \end{array}
  \right)\qquad {\rm with}\quad M\in \mathbb{N} \, .
\eea
This  transformation leaves the NS5-branes invariant, while it  converts a stack of $M$ D5-branes to a
single $(1,M)$ five-brane. Recall that the worldvolume theory of a stack of
$N$ D3-branes intersecting a stack of
 $M$ D5-branes  is a $U(N)$ gauge theory with $M$
  fundamental hypermultiplets. Replacing  the D5-branes by a $(1,M)$ five-brane
     leads to a $U(N)_M \times U(N)_{-M}$ gauge theory with a bifundamental hypermultiplet
     and level $M$ (respectively $-M$) Chern-Simons terms (see e.g. \cite{Aharony:2008ug}).
The  transformation  \eqref{SL2Q} can be used therefore to relate  the following two   theories:\\ \vskip -1.2 mm
\indent (i)  a $U(N)^{\hat k}$ gauge theory, with $M$ fundamental hypermultiplets for every gauge-group factor,
and a bifundamental for each neighboring pair;\\ \vskip -1.2 mm
\indent
(ii) a $U(N)^{2\hat k}$ gauge theory with bifundamentals  for each neighboring pair,  and Chern-Simons terms of alternating
level $\pm M$. \\ \vskip -1.2 mm
\noindent
The corresponding
circular quivers  are illustrated in  Figure \ref{slQ}.
As a test of their  $SL(2,\mathbb{Q})$ equivalence we will
 conclude this section by comparing the free energies of these two gauge field theories in the limit  $N\gg 1$.

\smallskip

Let us first recall the result  \eqref{herzogfree} for the
  free energy on the supergravity side.
  Replacing the number  of winding D3-branes  by $N$,   and the total number
  of D5-branes by $M \hat k$,   leads to  the expression
 \begin{align}
\label{Fsugra}
F_{\rm sugra}  \,  =  \,   \frac{\pi \sqrt{2} }{3}\hat k  M^{1/2} \, N^{3/2}  \ .
\end{align}
This should be compared to the result on the field-theory side.  For
  the necklace quiver of theory  (ii)  the calculation  has been performed
   in  \cite{Herzog:2010hf}. These authors  used the localization techniques of \cite{Kapustin:2009kz}
  to reduce the calculation to a matrix-model  integral, which they then evaluated for large-$N$  by the saddle-point method.
  Their result agrees precisely with \eqref{Fsugra},  confirming the AdS/CFT correspondence. What we need to do  is to
 also  recover  this   result   from the original gauge theory (i).
 \smallskip

  Since     for theories with $\N \geq 4$ supersymmetries the free energy does not run
 \cite{Kapustin:2009kz},  we
 may perform the calculation near  the  (ultraviolet) Gaussian  fixed point.  Using the standard localization techniques, one  reduces the partition function
 of theory (i) to the following matrix-model  integral:
\beq
Z_{(i)} = {1\over (N!)^{\hat k}} \int \prod_{a=1}^{\hat k} \frac{d^N \sigma_a}{(2\pi)^N} \frac{\prod_{i<j} 4\sinh^2 \big(\frac{\sigma_{a}^i - \sigma_{a}^j}{2} \big) }{\prod_{i,j} 2\cosh \big(\frac{\sigma_{a+1}^i - \sigma_{a}^j}{2} \big) } \frac{1}{\big[\prod_{j} 2\cosh \big(\frac{\sigma_{a}^j}{2} \big) \big]^M}\ ,
\eeq
where $i,j$ run from $1$ to $N$. This can be written as  $Z_{(i)} = \int e^{-F (\sigma_a)}$ with
\begin{align}
F (\sigma_a)\,  =\,  - 2 \sum_{a\,;\,i<j} \log\Big[ 2\sinh \big(\frac{\sigma_{a}^i - \sigma_{a}^j}{2} \big) \Big] + \sum_{a\,;\,i,j} \log\Big[ 2\cosh \big(\frac{\sigma_{a+1}^i - \sigma_{a}^j}{2} \big) \Big] \no\\
+ \sum_{a\,;\,j} M \log\Big[ 2\cosh \big(\frac{\sigma_{a}^j}{2} \big) \Big] + \hat k \log(N!) + \hat k N \log(2\pi)\ .
\label{F_A}
\end{align}
\smallskip
Following reference \cite{Herzog:2010hf},  we let  $\sigma_a^j = N^{\beta} x_a^j$, and fix $\beta$ so that
at the saddle point
 the $x_a^j$ are of order one.  Contrary to  this reference,  we do not introduce an imaginary part for the $x_a^j$.
 Indeed,   the saddle point equations are invariant under complex  conjugation, so we are entitled to look   for real solutions.

\smallskip

In the limit $N\gg 1$, we may replace the variables $x_a^i$ by a continuous density $\rho_a(x) $ normalized so that   $\int dx \rho_a(x) = 1$.
The expression \ref{F_A} can be written as
 \bea
F(\rho_a) =\sum_{a=1}^{\hat k} &{1\over 2} \Big[ {\pi^2}  N^{2-\beta} \int dx_a   \rho_a(x_a)^2 +  {M}  N^{1+\beta} \, \int dx_a |x_a| \rho_a(x_a) \Big] \no  \\
&  + O(N^{2-2\beta},N \log N)\ .
\label{F_continuum}
\eea
The details of the computation  are subtle and can be found in appendix A of \cite{Herzog:2010hf}.
\smallskip

The saddle-point equations are non-trivial when the
two terms in this expression are of the same order, so that  $\beta = \half$. Furthermore, thanks to the  symmetries of the problem, we may look
for saddle points with $\rho_a(x) = \rho(x)$ for all $a$,\footnote{The authors of \cite{Herzog:2010hf} arrive to this same ansatz after some approximation
of the saddle point equations.}
 and $\rho(x) = \rho(-x)$.  With these assumptions the above free energy  reduces to
\begin{align}
F(\rho) = \frac{\hat k}{2} N^{\frac{3}{2}} \Big[ \pi^2  \int dx \,  \rho(x)^2 + M  \int dx \, |x| \rho(x) - \gamma \int dx \, \rho(x) +  \gamma  \Big]\ ,
\label{F_A_last}
\end{align}
where the Lagrange multiplier $\gamma$  imposes  the constraint $\int dx \rho(x) = 1$. The ensuing saddle point equation,
\beq
2\pi^2 \rho(x) + M  |x| = \gamma \ ,
\eeq
 is solved by the eigenvalue density
\begin{align}
\rho(x) &= \frac{1}{2\pi^2}\big( \gamma - M  |x| \big) \quad \textrm{for} \quad |x| < x_0 , \no\\
&= 0 \quad \textrm{for} \quad |x| > x_0 .
\end{align}
 The constraint $\int dx \rho(x) = 1$ fixes the Lagrange multiplier
 \bea
 \gamma = \frac{M  x_0}{2} + \frac{\pi^2}{x_0}\ ,
 \eea
whereas the positivity of $\rho$ implies $x_0 \leq \pi\sqrt{\frac{2}{M}}$. Combining all these formulae gives
\begin{align}
F(x_0) = \hat k N^{\frac{3}{2}} \Big[ \frac{\pi^2}{4 x_0} + \frac{M  x_0}{4} - \frac{M^2 x_0^3}{48 \pi^2} \Big]\ .
\end{align}
We now need to minimize this expression with respect to $x_0$ which takes values in $(0, \pi\sqrt{ {2/M}} ]$.
The minimum is achieved at the rightmost endpoint, leading to the final result for the gauge theory (i):
 \begin{align}
F_{(i)}  = \frac{\pi \sqrt{2}}{3}\hat k \sqrt{M} N^{\frac{3}{2}}\ ,
\end{align}
in perfect agreement with both the necklace-quiver and the supergravity calculations.
Note that although the final results agree, the three calculations  differ greatly in their specific details.

\vskip 1cm

{\bf Acknowledgements}: We thank  E. D' Hoker, D. Jafferis, S. Pasquetti, J. Troost and M. Yamazaki  for useful conversations. 
We are also  indebted to the authors of  \cite{Imamura:2009ur} for drawing  our attention to  this reference.
The research leading to these results has received funding from the [European Union] Seventh Framework Programme [FP7-People-2010-IRSES] under grant agreement n 269217.
J.G. thanks the LPTENS for hospitality during  this work.
B.A. thanks the Perimeter Institute for  hospitality during a visit.
J.E. was supported by the FWO - Vlaanderen, Project No. G.0651.11, the ``Federal Office for Scientific, Technical and Cultural Affairs through the Inter-University Attraction Poles Programme,''  Belgian Science Policy P6/11-P, as well as the European Science Foundation Holograv Network, and is currently supported in part by STFC grant ST/J0003533/1.
 Research at the Perimeter Institute is supported in part by the Government of Canada through NSERC and by the Province of Ontario through MRI.
J.G. also acknowledges further support from an NSERC Discovery Grant and from an ERA grant by the Province of Ontario.


\appendix

\section{Mirror symmetry of   inequalities}
\label{sec:ineq}
\setcounter{equation}{0}

\begin{figure}
\centering
\includegraphics[height=8cm,width=11cm]{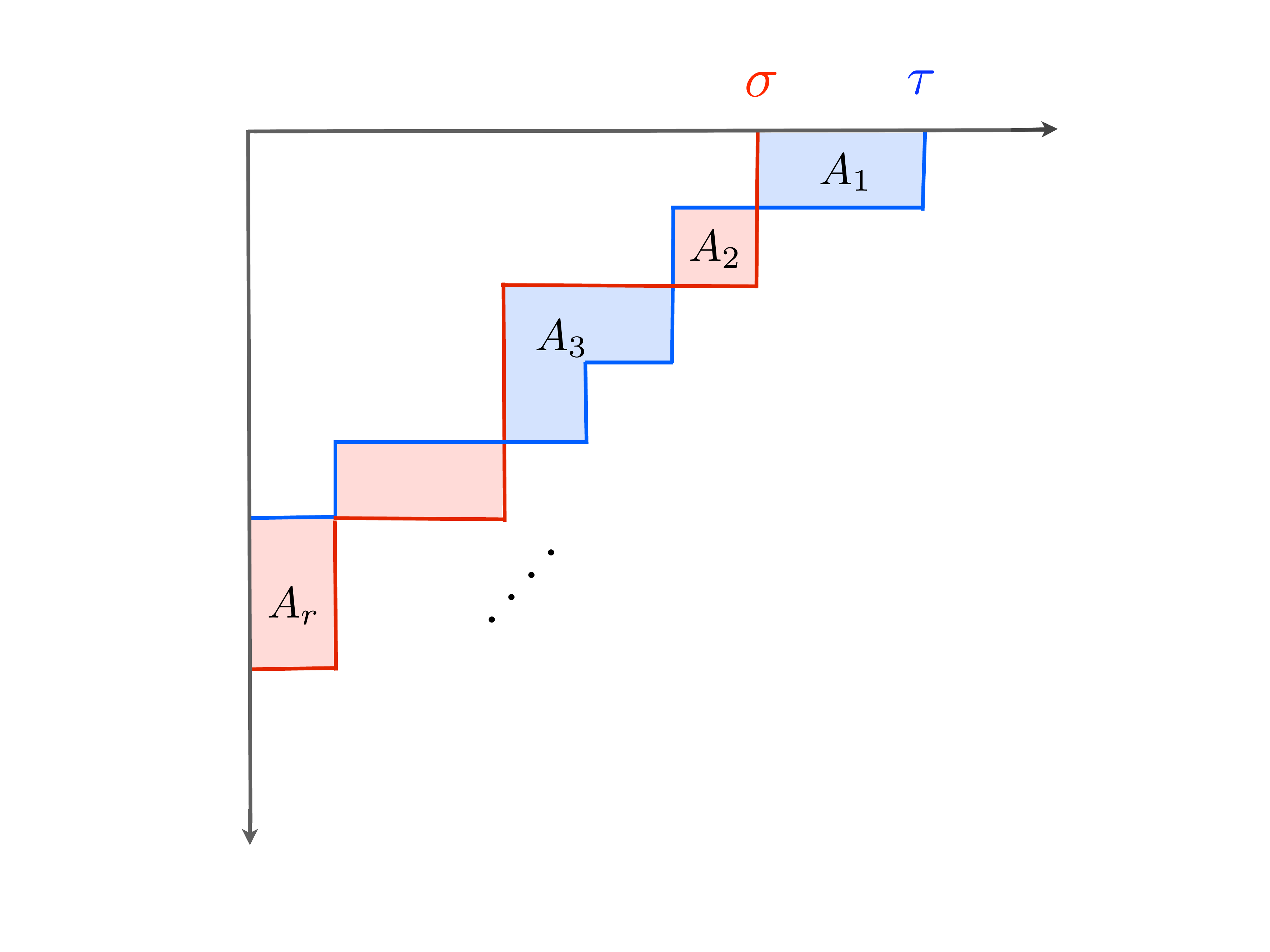}
\vskip -1.cm
\caption{\footnotesize   The difference of two  Young tableaux defines an alternating  sequence  $\{A_1, A_2, \cdots , A_r\}$
where $\vert A_i\vert$  counts  the number of boxes  in the $i$th  region enclosed by  the two histograms of the Young tableaux.
In this  example   $A_1 = 2 , A_2 = -1, A_3 = 3 , \cdots$.
  The difference of the transposed
 tableaux, obtained by rotating the figure by $90^{\rm o}$,   defines  the inverse sequence  $\{A_r,   \cdots , A_2, A_1\}$. }
\label{mirror}
\end{figure}

We will here show that the inequalities
 \eqref{fixedpointcircA} are invariant under the mirror map, i.e. that
 \beq
\label{ineq2a}
L+ \rho^{T}    > \hat \rho\  \ \Longleftrightarrow \ \   L+  \hat \rho^{T}    >   \rho\ .
\eeq
Let us first recall that if $\tau = (a_1,a_2,...,a_t)$ and $\sigma = (b_1,b_2,...,b_s)$ are two partitions of the same number $N$,
expressed as vectors with non-increasing positive components,
then $L+ \tau    > \sigma$  is a shorthand notation for the set of inequalities
\begin{align}
 L+  \sum_{i=1}^{n} a_i > \sum_{i=1}^{n} b_i \quad\quad \textrm{for\ all } \quad n=1,...,\textrm{max}(t,s).
\end{align}
These can be visualized more easily in   the
diagrammatic representation of Figure \ref{mirror}, which defines a sequence $\{A_1, A_2, \cdots , A_r\}$ of
areas with alternating signs. In terms of this sequence, the inequalities read
\bea\label{reverse}
L+ A_1 >0\ , \quad L+(A_1+A_2) >0\ , \quad \cdots \quad , \quad   L >0\ ,
\eea
where the last inequality follows from  the fact that $A_1 + A_2 \cdots  + A_r =0$.
Reversing the order, one may put  these inequalities  in the following form:
\bea\label{reverse2}
  L-A_s  >0\ , \quad  L- (A_s + A_{s-1})>0\ , \cdots \quad , \quad  L  >0\ ,
\eea
or equivalently  $L + \sigma^T > \tau^T$, as is evident if ones rotates by $90^{\rm o}$ Figure \ref{mirror}.
Setting   $\tau\equiv \rho^T$ and $\sigma\equiv \hat \rho$
proves the mirror equivalence  \eqref{ineq2a},   as claimed.

\section{Proof of the inequalities in supergravity}
\label{sec:ineq2}

We have already shown  in \S\ref{s:match} that,  with an appropriate choice of gauge, the linking numbers of the
supergravity solution can be confined to the intervals  $\ l^{(a)}\in (0, \hat k]$ and $\hat l^{(b)}\in (0,  k]$.
In particular, the linking numbers are positive, and we  demand  that they be quantized. Thus the Young tableaux
$\rho$ and $\hat\rho$ are well defined, and the inequalities $L+ \rho^{T}    > \hat \rho$ make  sense. We will now
prove that these inequalities are automatically obeyed on the supergravity side.\footnote{In the graphic form of
Figure \ref{mirror} the inequalities actually make sense for any pair of monotonic functions with equal  definite integral,
and with transposition  of the Young tableau being replaced by   function inversion. This should make it possible to prove the
inequalities without using quantization and the partial gauge fixing that was required to define the Young tableaux.
We will not pursue   this approach further here.
}
\smallskip

Let us recall the explicit expressions of the five-brane  linking numbers and of  $L$:
\begin{align}
\label{formulae}
l^{(a)} &= \sum_{b=1}^{\hat p} \hat N_b  \Big[  \sum_{n=0}^{+\infty} f(\hat \delta_b - \delta_a -2 n t) - \sum_{n=1}^{+\infty} f(-\hat \delta_b + \delta_a - 2 n t)  \Big] \ , \nonumber \\
\hat l^{(b)} &= \sum_{a=1}^{p} N_a \Big[  \sum_{n=0}^{+\infty} f(\hat \delta_b - \delta_a -2 n t ) - \sum_{n=1}^{+\infty} f(-\hat \delta_b + \delta_a -2 n t )  \Big] \ , \\
L  &= \sum_{a=1}^{p} \sum_{b=1}^{\hat p}\sum_{k=1}^{+\infty} k \ N_a \hat N_b \Big[ f(\hat \delta_b - \delta_a -2 k t ) + f(\delta_a - \hat \delta_b - 2k t ) \Big] \ , \nonumber
\end{align}
where  $f(x)= \frac{2}{\pi} \arctan(e^x)$, and we use in this appendix a lighter notation for the five-brane charges,  $N_a \equiv  N^{(a)}_{5}$
and $\hat N_b \equiv  \hat N^{(b)}_{5}$.
In terms of these linking numbers and the five-brane charges the partitions   $\hat \rho$ and $\rho^T$ read:
\begin{align}
 \hat \rho &= (\underbrace{\hat l^{(1)},...,\hat l^{(1)}}_{\hat N_1},...,\underbrace{\hat l^{(b)},...,
\hat l^{(b)}}_{\hat N_b},...,\underbrace{\hat l^{(\hat p)},...,\hat l^{(\hat p)}}_{\hat N_{\hat p}}) \ ,
\end{align}
and
\begin{align}
\rho^T = (\underbrace{\sum_{a=1}^p  N_a,...,\sum_{a=1}^p  N_a}_{l^{(p)}},\underbrace{\sum_{a=1}^{p-1} N_a,...,\sum_{a=1}^{p-1} N_a}_{l^{(p-1)} - l^{(p)}},...,\underbrace{\sum_{a=1}^A N_a,...,\sum_{a=1}^A N_a}_{l^{(A)} - l^{(A+1)}},...,\underbrace{N_1,...,N_1}_{l^{(1)} - l^{(2)}}) \ .
\end{align}
We need now to establish the set of  inequalities
\bea
\label{ineq3}
\sum_{s=1}^r m_s \ + L  \ &>&\  \sum_{s=1}^r \hat l_s\qquad \forall r = 1,\ldots , \max(k,\hat k) \, .
\eea
where $\hat \rho = (\hat l_1,\hat l_2, ..., \hat l_{\hat k})$ and $\rho^T = (m_1,m_2,...,m_k)$ are the above two partitions.

\smallskip

The last  inequality, the one for $r=\max(k,\hat k)$,  implies that
 $L  > 0$. This is obeyed automatically, as seen  from the explicit expression \eqref{formulae}  and the fact that $f$ is strictly positive.
  As in the case of the linear quivers, it is now sufficient to prove the
   inequalities (\ref{ineq3}) for the corners of the histogram $\hat\rho$, i.e. for the values
\begin{align}
\label{defr}
r = \sum_{b=1}^{J} \hat N_b \qquad {\rm where} \qquad J = 1, 2 ,..., \hat p \ .
\end{align}
We refer the reader to  \cite{ABEG:2011} for a detailed explanation of this claim.
The above subset of inequalities takes the following explicit form:
\beq\label{niceform}
\sum_{b=1}^{J}\hat l^{(b)} \hat N_b < L  + \sum_{a=I+1}^{p} l^{(a)} N_a
+ \left(\sum_{a=1}^{I} N_a \right)\left(\sum_{b=1}^{J} \hat N_b\right) \ .
\eeq
 This is the form  that we will  prove using the supergravity calculation of the charges.

 \smallskip

Let us give a name to the infinite sum that enters in the supergravity expressions for the linking numbers:
\bea
F(x, 2t ) \equiv   \sum_{n=0}^{\infty} f(x - 2n t) - \sum_{n=1}^{\infty} f(-x -2 n t )\ .
\eea
 In terms of the function $F$ the inequalities \eqref{niceform}  can be written as
  \begin{align}
\sum_{a = 1}^{p} \sum_{b=1}^{J} N_a \hat N_b \,  F( \hat{\delta}_{b}-\delta_{a}, 2t ) \
&< \ L  + \sum_{a=I+1}^{p} \sum_{b = 1}^{\hat{p}} N_a \hat N_b\,  F ( \hat{\delta}_{b}-\delta_{a}, 2t )  + \sum_{a=1}^{I}\sum_{b=1}^{J} N_a \hat N_b \ . \nonumber
\end{align}
 Splitting the sums,
 simplifying and rearranging terms gives:
\begin{align}
\sum_{a = 1}^{I} \sum_{b=1}^{J} N_a \hat N_b \, F( \hat{\delta}_{b}-\delta_{a}, 2t)
- \sum_{a=I+1}^{p} \sum_{b = J+1}^{\hat{p}} N_a \hat N_b \, F(\hat{\delta}_{b}-\delta_{a}, 2t) \
&< \ L   +  \sum_{a=1}^{I}\sum_{b=1}^{J} N_a \hat N_b \ . \nonumber
\end{align}
We show that this is automatically satisfied by putting the following successive bounds on
  the left hand side:
 \begin{align}
\label{calculus}
& \ \hskip 1.5 cm  \sum_{a = 1}^{I} \sum_{b=1}^{J} N_a \hat N_b \, F( \hat{\delta}_{b}-\delta_{a}, 2t)
- \sum_{a=I+1}^{p} \sum_{b = J+1}^{\hat{p}} N_a \hat N_b \, F(\hat{\delta}_{b}-\delta_{a}, 2t)   \nonumber \\
  & \ \hskip 0.3cm  < \ \sum_{a = 1}^{I} \sum_{b=1}^{J} N_a \hat N_b \, \sum_{n=0}^{\infty} f(\hat \delta_b - \delta_a - 2nt)
+ \sum_{a=I+1}^{p} \sum_{b = J+1}^{\hat{p}} N_a \hat N_b \, \sum_{n=1}^{\infty} f(-\hat \delta_b + \delta_a - 2nt) \nonumber\\
< &  \  \sum_{a = 1}^{I} \sum_{b=1}^{J} N_a \hat N_b \ f(\hat \delta_b - \delta_a) \ + \  \sum_{a = 1}^{p} \sum_{b=1}^{\hat p}
N_a \hat N_b \ \sum_{n=1}^{\infty}  \Big[ f(\hat \delta_b - \delta_a - 2nt) + f(-\hat \delta_b + \delta_a - 2nt) \Big] \nonumber \\
 &   \  \hskip 3.8cm   < \  L + \sum_{a = 1}^{I} \sum_{b=1}^{J} N_a \hat N_b \ .
\end{align}
In the first inequality we have dropped terms that are explicitly negative.
The second inequality is obtained by extension of  the sums.  Finally, in
 the third inequatlity we  used, in addition to  the bound  $0< f(x) <1$,
  the  expression  (\ref{formulae}) for the winding charge $L$.  This completes the proof.

\section{From IIB to M theory for  large $L$}
\label{Tduality}

  We give here the detailed  T-duality transformation of the type-IIB solution for large winding number $L$ to a solution of type-IIA
  supergravity, and the subsequent uplift to eleven dimensions.  We will follow the metric, dilaton and two-form gauge fields,
  which all become part of the metric in eleven dimensions. The  four-form potential of the IIB theory   transforms to the three-form potential of M theory,
  which at leading order  has a field strength  proportional to the ${\rm AdS}_4$ volume form. 
  The way in which the 3-form field may encode the information on the five-brane throats  is a very subtle issue, as already noted in
  the main text. We will not discuss it further in this appendix.    

 \smallskip

   The type-IIB backgrounds in the large-$L$  limit are given by the expressions
  \eqref{IIBlargeL1} and \eqref{IIBlargeL2}.  In order to use the standard Buscher rules, we make a gauge transformation
  that removes the $x$-dependence from the gauge potentials. The new two-form potentials read
     \begin{align}
B_{(2)} &= - 2 \hat k \,  \cos(\theta_1) dx\wedge d\phi_1 \ , \no\qquad
C_{(2)} = - 2  k \,  \cos(\theta_2) dx\wedge d\phi_2  \ ,
\end{align}
where we recall that $x$ is periodic with period $2\pi$.  We also transform the Einstein-frame to the string-frame metric, $G_{MN}  = e^\phi g_{MN}$,
in terms of which  Buscher's rules read  \cite{Buscher:1987qj}:
 \begin{align}
G^\prime_{\mu \nu} &=  G_{\mu \nu} - \frac{G_{x \mu}G_{x \nu}-B_{x \mu}B_{x \nu}}{G_{x x}}  \ , \qquad
G^\prime_{0 \nu} = \frac{B_{x \mu}}{G_{x x}}  \ , \qquad  G^\prime_{x x} =\frac{1}{G_{x x}} \ , \no\\
  B^\prime_{\mu \nu} &= B_{\mu \nu} - \frac{G_{x \mu} B_{x \nu}-B_{x \mu} G_{x \nu}}{G_{x x}}  \ , \qquad
  B^\prime_{0 \nu} = \frac{G_{x \mu}}{G_{x x}} \ , \qquad e^{4   \phi^\prime}  = \frac{e^{4 \phi}}{G_{x x}} \ ,
 \end{align}
 where the prime indicates the type-IIA fields in string frame,  and the lower-case Greek indices $\mu,\nu$ run over all dimensions other than $x$.
  In addition, the 2-form R-R potential transforms to a one-form potential,
\bea
  C^\prime_{(1)\mu} &= C_{(2) x \mu} \ .
  \eea
\noindent Since the IIB metric had no $(x\mu)$ components  $B^\prime$ is zero, while the original 2-form NS-NS gauge field becomes an off-diagonal component of
the IIA  metric. In string-frame this latter reads:
   \begin{align}
dS_{{\rm IIA}}^{\, 2} =  & { { \pi^2 \hat k}\over t }   \sqrt{1-y}  \left[   ds^2_{AdS_4} +  y\,  ds^2_{S^2_1} +   (1-y) ds^2_{S^2_2} \right]    \no \\  &
 +    {4 \pi^2  \over   t }    \sqrt{1-y}  \left[ {y\over \hat k}  (dx - {\hat k\over 2} {\rm cos}\theta_1 d\phi_1)^2 +    \hat k {dy^2\over y(1-y) } \right]     \ ,
\end{align}
  whereas  the R-R gauge field and the transformed dilaton field  are given by
\bea
C_{(1)}^\prime = - 2k\,  \cos \theta_2 d \phi_2 \ , \qquad e^{4 \phi^\prime}  =     {4\pi^2  \over t  } {\hat k\over k^2}  (1-y) ^{3/2} \ . \\ \ \no
\eea
\vskip -3mm
 \indent  Finally we uplift the solution  to M theory,  whose metric  (denoted here by a bar)
is  given in terms of the type-IIA backgrounds by the following  relations \cite{Witten:1995ex}
\bea
 \bar g_{MN}  = e^{- 4    \phi^\prime/3 } ( G^\prime_{MN}  + {1\over 4} e^{4\phi^\prime} C^\prime_M C^\prime_N ) \ , \quad
  \bar g_{Mv} = e^{ 8  \phi^\prime/3 }C^\prime_M\ , \quad
   \bar g_{vv} =  4 e^{ 8  \phi^\prime/3 }\ ,
\eea
where $v = v+2\pi$ parametrizes  the eleventh dimension.
Redefining the coordinates $x \rightarrow \hat k x$ , $v \rightarrow k v$ and $y= \sin^2 \alpha $
gives, after some straightforward algebra,  the $AdS^4 \times S^7/ (\mathbb{Z}_k\times  \mathbb{Z}_{\hat k})$ metric, equation \eqref{AdS4S7}.

\newpage

\bibliography{Circularquiversbib}

\providecommand{\href}[2]{#2}\begingroup\raggedright\begin{thebibliography}{10}

\bibitem{Maldacena:1997re}
J.~M. Maldacena, {\it {The Large N limit of superconformal field theories and
  supergravity}},  {\em Adv.Theor.Math.Phys.} {\bf 2} (1998) 231--252,
  [\href{http://xxx.lanl.gov/abs/hep-th/9711200}{{\tt hep-th/9711200}}].

\bibitem{Gubser:1998bc}
S.~Gubser, I.~R. Klebanov, and A.~M. Polyakov, {\it {Gauge theory correlators
  from noncritical string theory}},  {\em Phys.Lett.} {\bf B428} (1998)
  105--114, [\href{http://xxx.lanl.gov/abs/hep-th/9802109}{{\tt
  hep-th/9802109}}].

\bibitem{Witten:1998qj}
E.~Witten, {\it {Anti-de Sitter space and holography}},  {\em
  Adv.Theor.Math.Phys.} {\bf 2} (1998) 253--291,
  [\href{http://xxx.lanl.gov/abs/hep-th/9802150}{{\tt hep-th/9802150}}].

\bibitem{Hanany:1996ie}
A.~Hanany and E.~Witten, {\it {Type IIB superstrings, BPS monopoles, and
  three-dimensional gauge dynamics}},  {\em Nucl.Phys.} {\bf B492} (1997)
  152--190, [\href{http://xxx.lanl.gov/abs/hep-th/9611230}{{\tt
  hep-th/9611230}}].

\bibitem{ABEG:2011}
B.~Assel, C.~Bachas, J.~Estes, and J.~Gomis, {\it {Holographic Duals of D=3 N=4
  Superconformal Field Theories}},  {\em JHEP} {\bf 1108} (2011) 087,
  [\href{http://xxx.lanl.gov/abs/1106.4253}{{\tt arXiv:1106.4253}}].

\bibitem{Aharony:2011yc}
O.~Aharony, L.~Berdichevsky, M.~Berkooz, and I.~Shamir, {\it {Near-horizon
  solutions for D3-branes ending on 5-branes}},  {\em Phys.Rev.} {\bf D84}
  (2011) 126003, [\href{http://xxx.lanl.gov/abs/1106.1870}{{\tt
  arXiv:1106.1870}}].

\bibitem{Gomis:2006cu}
J.~Gomis and C.~Romelsberger, {\it {Bubbling Defect CFT's}},  {\em JHEP} {\bf
  0608} (2006) 050, [\href{http://xxx.lanl.gov/abs/hep-th/0604155}{{\tt
  hep-th/0604155}}].

\bibitem{Lunin:2006xr}
O.~Lunin, {\it {On gravitational description of Wilson lines}},  {\em JHEP}
  {\bf 0606} (2006) 026, [\href{http://xxx.lanl.gov/abs/hep-th/0604133}{{\tt
  hep-th/0604133}}].

\bibitem{DEG1}
E.~D'Hoker, J.~Estes, and M.~Gutperle, {\it {Exact half-BPS Type IIB interface
  solutions. I. Local solution and supersymmetric Janus}},  {\em JHEP} {\bf
  0706} (2007) 021, [\href{http://xxx.lanl.gov/abs/0705.0022}{{\tt
  arXiv:0705.0022}}].

\bibitem{DEG2}
E.~D'Hoker, J.~Estes, and M.~Gutperle, {\it {Exact half-BPS Type IIB interface
  solutions. II. Flux solutions and multi-Janus}},  {\em JHEP} {\bf 0706}
  (2007) 022, [\href{http://xxx.lanl.gov/abs/0705.0024}{{\tt
  arXiv:0705.0024}}].

\bibitem{Hosomichi:2008jd}
K.~Hosomichi, K.-M. Lee, S.~Lee, S.~Lee, and J.~Park, {\it {N=4 Superconformal
  Chern-Simons Theories with Hyper and Twisted Hyper Multiplets}},  {\em JHEP}
  {\bf 0807} (2008) 091, [\href{http://xxx.lanl.gov/abs/0805.3662}{{\tt
  arXiv:0805.3662}}].

\bibitem{Benna:2008zy}
M.~Benna, I.~Klebanov, T.~Klose, and M.~Smedback, {\it {Superconformal
  Chern-Simons Theories and AdS(4)/CFT(3) Correspondence}},  {\em JHEP} {\bf
  0809} (2008) 072, [\href{http://xxx.lanl.gov/abs/0806.1519}{{\tt
  arXiv:0806.1519}}].

\bibitem{Imamura:2008ji}
Y.~Imamura and S.~Yokoyama, {\it {N=4 Chern-Simons theories and wrapped
  M-branes in their gravity duals}},  {\em Prog.Theor.Phys.} {\bf 121} (2009)
  915--940, [\href{http://xxx.lanl.gov/abs/0812.1331}{{\tt arXiv:0812.1331}}].

\bibitem{Witten:2009xu}
E.~Witten, {\it {Branes, Instantons, And Taub-NUT Spaces}},  {\em JHEP} {\bf
  0906} (2009) 067, [\href{http://xxx.lanl.gov/abs/0902.0948}{{\tt
  arXiv:0902.0948}}].

\bibitem{Aharony:2009fc}
O.~Aharony, A.~Hashimoto, S.~Hirano, and P.~Ouyang, {\it {D-brane Charges in
  Gravitational Duals of 2+1 Dimensional Gauge Theories and Duality Cascades}},
   {\em JHEP} {\bf 1001} (2010) 072,
  [\href{http://xxx.lanl.gov/abs/0906.2390}{{\tt arXiv:0906.2390}}].

\bibitem{Dey:2011pt}
A.~Dey, {\it {On Three-Dimensional Mirror Symmetry}},  {\em JHEP} {\bf 1204}
  (2012) 051, [\href{http://xxx.lanl.gov/abs/1109.0407}{{\tt
  arXiv:1109.0407}}].

\bibitem{Gregory:1997te}
R.~Gregory, J.~A. Harvey, and G.~W. Moore, {\it {Unwinding strings and t
  duality of Kaluza-Klein and h monopoles}},  {\em Adv.Theor.Math.Phys.} {\bf
  1} (1997) 283--297, [\href{http://xxx.lanl.gov/abs/hep-th/9708086}{{\tt
  hep-th/9708086}}].

\bibitem{Tong:2002rq}
D.~Tong, {\it {NS5-branes, T duality and world sheet instantons}},  {\em JHEP}
  {\bf 0207} (2002) 013, [\href{http://xxx.lanl.gov/abs/hep-th/0204186}{{\tt
  hep-th/0204186}}].

\bibitem{Harvey:2005ab}
J.~A. Harvey and S.~Jensen, {\it {Worldsheet instanton corrections to the
  Kaluza-Klein monopole}},  {\em JHEP} {\bf 0510} (2005) 028,
  [\href{http://xxx.lanl.gov/abs/hep-th/0507204}{{\tt hep-th/0507204}}].

\bibitem{Okuyama:2005gx}
K.~Okuyama, {\it {Linear sigma models of H and KK monopoles}},  {\em JHEP} {\bf
  0508} (2005) 089, [\href{http://xxx.lanl.gov/abs/hep-th/0508097}{{\tt
  hep-th/0508097}}].

\bibitem{Bachas:2012bj}
C.~Bachas, I.~Brunner, and D.~Roggenkamp, {\it {A worldsheet extension of
  O(d,d:Z)}},  \href{http://xxx.lanl.gov/abs/1205.4647}{{\tt arXiv:1205.4647}}.

\bibitem{Douglas:1996sw}
M.~R. Douglas and G.~W. Moore, {\it {D-branes, quivers, and ALE instantons}},
  \href{http://xxx.lanl.gov/abs/hep-th/9603167}{{\tt hep-th/9603167}}.

\bibitem{Gaiotto:2008ak}
D.~Gaiotto and E.~Witten, {\it {S-Duality of Boundary Conditions In N=4 Super
  Yang-Mills Theory}},  \href{http://xxx.lanl.gov/abs/0807.3720}{{\tt
  arXiv:0807.3720}}.

\bibitem{BHOOY:1996}
J.~de~Boer, K.~Hori, H.~Ooguri, Y.~Oz, and Z.~Yin, {\it {Mirror symmetry in
  three-dimensional theories, SL(2,Z) and D-brane moduli spaces}},  {\em
  Nucl.Phys.} {\bf B493} (1997) 148--176,
  [\href{http://xxx.lanl.gov/abs/hep-th/9612131}{{\tt hep-th/9612131}}].

\bibitem{Kronheimer:1990ay}
P.~Kronheimer, {\it {Instantons and the geometry of the nilpotent variety}},
  {\em J.Diff.Geom.} {\bf 32} (1990) 473--490.

\bibitem{BHP}
C.~Bachas, J.~Hoppe, and B.~Pioline, {\it {Nahm equations, N=1* domain walls,
  and D strings in AdS(5) x S(5)}},  {\em JHEP} {\bf 0107} (2001) 041,
  [\href{http://xxx.lanl.gov/abs/hep-th/0007067}{{\tt hep-th/0007067}}].

\bibitem{Intriligator:1996ex}
K.~A. Intriligator and N.~Seiberg, {\it {Mirror symmetry in three-dimensional
  gauge theories}},  {\em Phys.Lett.} {\bf B387} (1996) 513--519,
  [\href{http://xxx.lanl.gov/abs/hep-th/9607207}{{\tt hep-th/9607207}}].

\bibitem{Page:1984qv}
D.~N. Page, {\it {Classical Stability of Round and Squashed Seven Spheres in
  Eleven-Dimensional Supergravity}},  {\em Phys.Rev.} {\bf D28} (1983) 2976.

\bibitem{Marolf:2000cb}
D.~Marolf, {\it {Chern-Simons terms and the three notions of charge}},
  \href{http://xxx.lanl.gov/abs/hep-th/0006117}{{\tt hep-th/0006117}}.

\bibitem{BE:2011}
C.~Bachas and J.~Estes, {\it {Spin-2 spectrum of defect theories}},  {\em JHEP}
  {\bf 1106} (2011) 005, [\href{http://xxx.lanl.gov/abs/1103.2800}{{\tt
  arXiv:1103.2800}}].

\bibitem{Green:1987mn}
M.~B. Green, J.~Schwarz, and E.~Witten, {\it {Superstring theory. Vol. 2: Loop
  amplitudes, anomalies and phenomenology}}, .

\bibitem{Morris:1988tu}
M.~Morris, K.~Thorne, and U.~Yurtsever, {\it {Wormholes, Time Machines, and the
  Weak Energy Condition}},  {\em Phys.Rev.Lett.} {\bf 61} (1988) 1446--1449.

\bibitem{Morris:1988cz}
M.~Morris and K.~Thorne, {\it {Wormholes in space-time and their use for
  interstellar travel: A tool for teaching general relativity}},  {\em
  Am.J.Phys.} {\bf 56} (1988) 395--412.

\bibitem{Visser:1995cc}
M.~Visser, {\it {Lorentzian wormholes: From Einstein to Hawking}}, .

\bibitem{Buscher:1987qj}
T.~Buscher, {\it {Path Integral Derivation of Quantum Duality in Nonlinear
  Sigma Models}},  {\em Phys.Lett.} {\bf B201} (1988) 466.

\bibitem{Ooguri:1995wj}
H.~Ooguri and C.~Vafa, {\it {Two-dimensional black hole and singularities of CY
  manifolds}},  {\em Nucl.Phys.} {\bf B463} (1996) 55--72,
  [\href{http://xxx.lanl.gov/abs/hep-th/9511164}{{\tt hep-th/9511164}}].

\bibitem{Aharony:2008ug}
O.~Aharony, O.~Bergman, D.~L. Jafferis, and J.~Maldacena, {\it {N=6
  superconformal Chern-Simons-matter theories, M2-branes and their gravity
  duals}},  {\em JHEP} {\bf 0810} (2008) 091,
  [\href{http://xxx.lanl.gov/abs/0806.1218}{{\tt arXiv:0806.1218}}].

\bibitem{Aharony:2008gk}
O.~Aharony, O.~Bergman, and D.~L. Jafferis, {\it {Fractional M2-branes}},  {\em
  JHEP} {\bf 0811} (2008) 043, [\href{http://xxx.lanl.gov/abs/0807.4924}{{\tt
  arXiv:0807.4924}}].

\bibitem{Imamura:2008nn}
Y.~Imamura and K.~Kimura, {\it {On the moduli space of elliptic
  Maxwell-Chern-Simons theories}},  {\em Prog.Theor.Phys.} {\bf 120} (2008)
  509--523, [\href{http://xxx.lanl.gov/abs/0806.3727}{{\tt arXiv:0806.3727}}].

\bibitem{Herzog:2010hf}
C.~P. Herzog, I.~R. Klebanov, S.~S. Pufu, and T.~Tesileanu, {\it {Multi-Matrix
  Models and Tri-Sasaki Einstein Spaces}},  {\em Phys.Rev.} {\bf D83} (2011)
  046001, [\href{http://xxx.lanl.gov/abs/1011.5487}{{\tt arXiv:1011.5487}}].

\bibitem{Assel:2012cp}
B.~Assel, J.~Estes, and M.~Yamazaki, {\it {Large N Free Energy of 3d N=4 SCFTs
  and $AdS_4/CFT_3$}},  \href{http://xxx.lanl.gov/abs/1206.2920}{{\tt
  arXiv:1206.2920}}.

\bibitem{Schwarz:1983wa}
J.~H. Schwarz and P.~C. West, {\it {Symmetries and Transformations of Chiral
  N=2 D=10 Supergravity}},  {\em Phys.Lett.} {\bf B126} (1983) 301.

\bibitem{Hull:1994ys}
C.~Hull and P.~Townsend, {\it {Unity of superstring dualities}},  {\em
  Nucl.Phys.} {\bf B438} (1995) 109--137,
  [\href{http://xxx.lanl.gov/abs/hep-th/9410167}{{\tt hep-th/9410167}}].

\bibitem{Gaiotto:2008sa}
D.~Gaiotto and E.~Witten, {\it {Supersymmetric Boundary Conditions in N=4 Super
  Yang-Mills Theory}},  \href{http://xxx.lanl.gov/abs/0804.2902}{{\tt
  arXiv:0804.2902}}.

\bibitem{Gaiotto:2008sd}
D.~Gaiotto and E.~Witten, {\it {Janus Configurations, Chern-Simons Couplings,
  And The theta-Angle in N=4 Super Yang-Mills Theory}},  {\em JHEP} {\bf 1006}
  (2010) 097, [\href{http://xxx.lanl.gov/abs/0804.2907}{{\tt
  arXiv:0804.2907}}].

\bibitem{Kachru:1998ys}
S.~Kachru and E.~Silverstein, {\it {4-D conformal theories and strings on
  orbifolds}},  {\em Phys.Rev.Lett.} {\bf 80} (1998) 4855--4858,
  [\href{http://xxx.lanl.gov/abs/hep-th/9802183}{{\tt hep-th/9802183}}].

\bibitem{Lawrence:1998ja}
A.~E. Lawrence, N.~Nekrasov, and C.~Vafa, {\it {On conformal field theories in
  four-dimensions}},  {\em Nucl.Phys.} {\bf B533} (1998) 199--209,
  [\href{http://xxx.lanl.gov/abs/hep-th/9803015}{{\tt hep-th/9803015}}].

\bibitem{Bershadsky:1998mb}
M.~Bershadsky, Z.~Kakushadze, and C.~Vafa, {\it {String expansion as large N
  expansion of gauge theories}},  {\em Nucl.Phys.} {\bf B523} (1998) 59--72,
  [\href{http://xxx.lanl.gov/abs/hep-th/9803076}{{\tt hep-th/9803076}}].

\bibitem{Armoni:2004ub}
A.~Armoni, M.~Shifman, and G.~Veneziano, {\it {Refining the proof of planar
  equivalence}},  {\em Phys.Rev.} {\bf D71} (2005) 045015,
  [\href{http://xxx.lanl.gov/abs/hep-th/0412203}{{\tt hep-th/0412203}}].

\bibitem{Kovtun:2004bz}
P.~Kovtun, M.~Unsal, and L.~G. Yaffe, {\it {Necessary and sufficient conditions
  for non-perturbative equivalences of large N(c) orbifold gauge theories}},
  {\em JHEP} {\bf 0507} (2005) 008,
  [\href{http://xxx.lanl.gov/abs/hep-th/0411177}{{\tt hep-th/0411177}}].

\bibitem{Hanada:2011zx}
M.~Hanada, C.~Hoyos, and A.~Karch, {\it {Generating new dualities through the
  orbifold equivalence: a demonstration in ABJM and four-dimensional quivers}},
   {\em JHEP} {\bf 1201} (2012) 068,
  [\href{http://xxx.lanl.gov/abs/1110.3803}{{\tt arXiv:1110.3803}}].

\bibitem{Hanada:2011yz}
M.~Hanada, C.~Hoyos, and H.~Shimada, {\it {On a new type of orbifold
  equivalence and M-theoretic $AdS_4/CFT_3$ duality}},  {\em Phys.Lett.} {\bf
  B707} (2012) 394--397, [\href{http://xxx.lanl.gov/abs/1109.6127}{{\tt
  arXiv:1109.6127}}].

\bibitem{Kapustin:2009kz}
A.~Kapustin, B.~Willett, and I.~Yaakov, {\it {Exact Results for Wilson Loops in
  Superconformal Chern-Simons Theories with Matter}},  {\em JHEP} {\bf 1003}
  (2010) 089, [\href{http://xxx.lanl.gov/abs/0909.4559}{{\tt
  arXiv:0909.4559}}].

\bibitem{Imamura:2009ur}
Y.~Imamura, {\it {Monopole operators in N=4 Chern-Simons theories and wrapped
  M2-branes}},  {\em Prog.Theor.Phys.} {\bf 121} (2009) 1173--1187,
  [\href{http://xxx.lanl.gov/abs/0902.4173}{{\tt arXiv:0902.4173}}].

\bibitem{Witten:1995ex}
E.~Witten, {\it {String theory dynamics in various dimensions}},  {\em
  Nucl.Phys.} {\bf B443} (1995) 85--126,
  [\href{http://xxx.lanl.gov/abs/hep-th/9503124}{{\tt hep-th/9503124}}].

\end{thebibliography}\endgroup
\bibliographystyle{JHEP}

\end{document}